\documentclass[twocolumn]{aastex61}
\usepackage{graphicx}
\usepackage{amssymb}
\usepackage{amsmath}
\usepackage{multirow}
\usepackage{listings}
\usepackage{hyperref}
\begin{document}

\title{Deep Very Large Array observations of the merging cluster CIZA J2242.8+5301: Continuum and spectral imaging}

\author{G. Di Gennaro}
\affil{Harvard-Smithsonian Center for Astrophysics, 60 Garden Street, Cambridge, MA 02138, USA}
\affil{Leiden Observatory, Leiden University, PO Box 9513, 2300 RA Leiden, The Netherlands}

\author{R.J. van Weeren}
\affil{Leiden Observatory, Leiden University, PO Box 9513, 2300 RA Leiden, The Netherlands}
\affil{Harvard-Smithsonian Center for Astrophysics, 60 Garden Street, Cambridge, MA 02138, USA}

\author{M. Hoeft}
\affil{Th{\"u}ringer Landessternwarte, Sternwarte 5, 07778 Tautenburg, Germany}

\author{H. Kang}
\affil{Department of Earth Sciences, Pusan National University, Busan 46241, Korea}

\author{D. Ryu}
\affil{Department of Physics, School of Natural Sciences, UNIST, Ulsan 44919, Korea}

\author{L. Rudnick}
\affil{Minnesota Institute for Astrophysics, University of Minnesota, 116 Church St. S.E., Minneapolis, MN 55455, USA}

\author{W. Forman}
\affil{Harvard-Smithsonian Center for Astrophysics, 60 Garden Street, Cambridge, MA 02138, USA}

\author{H.J.A. R{\"o}ttgering}
\affil{Leiden Observatory, Leiden University, PO Box 9513, 2300 RA Leiden, The Netherlands}

\author{M. Br{\"u}ggen}
\affil{Hamburger Sternwarte, Universit{\"a}t Hamburg, Gojenbergsweg 112, 21029 Hamburg, Germany}

\author{W.A. Dawson}
\affil{Lawrence Livermore National Lab, 7000 East Avenue, Livermore, CA 94550, USA}

\author{N. Golovich}
\affil{University of California, One Shields Avenue, Davis, CA 95616, USA}

\author{D.N. Hoang}
\affil{Leiden Observatory, Leiden University, PO Box 9513, 2300 RA Leiden, The Netherlands}

\author{H.T. Intema}
\affil{Leiden Observatory, Leiden University, PO Box 9513, 2300 RA Leiden, The Netherlands}

\author{C. Jones}
\affil{Harvard-Smithsonian Center for Astrophysics, 60 Garden Street, Cambridge, MA 02138, USA}

\author{R.P. Kraft}
\affil{Harvard-Smithsonian Center for Astrophysics, 60 Garden Street, Cambridge, MA 02138, USA}

\author{T.W. Shimwell}
\affil{Leiden Observatory, Leiden University, PO Box 9513, 2300 RA Leiden, The Netherlands}

\author{A. Stroe}
\affil{European Southern Observatory, Karl-Schwarzschild-Str. 2, 85748, Garching, Germany}

\correspondingauthor{Gabriella Di Gennaro}
\email{gabriella.di.gennaro@cfa.harvard.edu; digennaro@strw.leidenuniv.nl}

\received{March 27, 2018}
%\revised{??-??-????}
\accepted{July 29, 2018}

\begin{abstract}
Despite progress in understanding radio relics, there are still open questions regarding the underlying particle acceleration mechanisms. In this paper we present deep 1--4 GHz VLA observations of CIZA\,J2242.8+5301 ($z=0.1921$), a double radio relic cluster characterized by small projection on the plane of the sky. Our VLA observations reveal, for the first time, the complex morphology of the diffuse sources and the filamentary structure of the northern relic. We discover new faint diffuse radio emission extending north of the main northern relic. Our Mach number estimates for the northern and southern relics, based on the radio spectral index map obtained using the VLA observations and existing LOFAR and GMRT data, are consistent with previous radio and X-ray studies ($\mathcal{M}_{\rm RN}=2.58\pm0.17$ and $\mathcal{M}_{\rm RS}=2.10\pm0.08$). However, color-color diagrams and modelings suggest a flatter injection spectral index than the one obtained from the spectral index map, indicating that projection effects might be not entirely negligible. The southern relic consists of five ``arms''. Embedded in it, we find a tailed radio galaxy which seems to be connected to the relic. A spectral index flattening, where the radio tail connects to the relic, is also measured. We propose that the southern relic may trace AGN fossil electrons that are re-accelerated at a shock, with an estimated strength of $\mathcal{M}=2.4$. High-resolution mapping of other tailed radio galaxies also supports a scenario where AGN fossil electrons are revived by the merger event and could be related to the formation of some diffuse cluster radio emission.

\end{abstract}

\shorttitle{VLA observations of CIZA\,J2242.8+5301}
\shortauthors{Di Gennaro et al.}

\section{Introduction}\label{sec:intro}
Cluster mergers are the most energetic phenomena in the Universe since the Big Bang, involving kinetic energies of $\sim 10^{63}-10^{65}$ erg, released over a time scale of 1--2 Gyr \citep[e.g.][]{sarazin02} depending on the cluster's mass and on the relative velocity of the merging dark matter halos. In a hierarchical cold dark matter (CDM) scenario, these phenomena are the natural way to form rich cluster of galaxies. Cluster mergers produce shock waves and turbulence in the intracluster medium (ICM). It has been proposed that ICM shocks and turbulence can (re-)accelerate cosmic rays (CRs) which then produce diffuse radio synchrotron emitting sources in the presence of $\mu$Gauss magnetic fields. These diffuse sources are known as {\it radio halos} and {\it radio relics} \citep[see][for a theoretical and an observational review]{brunetti+jones14,feretti+12}.

Radio halos are centrally located unpolarized sources with a smooth morphology, roughly following the X-ray emission from the ICM. The currently favored formation scenario for the formation of radio halos involves the re-acceleration of pre-existing CR electrons \citep[energies of $\sim$ MeV, e.g.][]{brunetti+01, petrosian01} via magneto-hydrodynamical turbulent motions, induced by a merging event. 
Moreover, radio halos are characterized by a steep-spectrum \citep[$\alpha < -1$, with $S_\nu\propto\nu^\alpha$, e.g.][]{brunetti+08}. A correlation between the halo radio power and the cluster X-ray luminosity exists \citep[e.g.][]{cassano+13}, showing that the most X-ray luminous clusters host the most powerful radio halos. Moreover, \cite{cassano+10} showed a clear correlation between the cluster's dynamical state, measured from the X-ray surface brightness distribution, and the presence of giant radio halos, providing strong support that mergers play an important role in the formation of these sources.

As with the radio halos, radio relics are characterized by a steep spectral index ($\rm \alpha\approx-0.8~to~-1.5$). They are generally elongated sources located in the outskirts of clusters, and they are suggested to trace outwards traveling shock fronts, where the ICM and magnetic fields are compressed. \citep[e.g.][]{ensslin+98,finoguenov+10,vanweeren+10}. As a consequence, the magnetic field is amplified and aligned, producing polarized radio emission. One formation scenario for relics is the {\it diffusive shock acceleration} (DSA) mechanism, similar to what occurs in supernova remnants \citep[e.g.][]{blandford+ostriker78,drury83,ensslin+98}. This model involves particles (electrons) that are accelerated from the ICM's thermal pool into CRs at shocks, while the electrons in the downstream region suffer synchrotron and inverse Compton (IC) energy losses. A prediction of the DSA model is a relation between the Mach number of the shock and the radio spectral index at the shock location \citep[e.g.][]{giacintucci+08}. However, recent observations suggest fundamental problems with the standard DSA model: (1) in a few cases, there are cluster merger shocks without corresponding radio relics \citep[e.g. the main shock in the Bullet Cluster,][]{shimwell+14} (2) sometimes, the spectral index derived Mach numbers are significantly higher than those obtained from the X-ray observations \citep[e.g.][]{vanweeren+16}, and (3) some relics require an unrealistic shock acceleration efficiency (for DSA) to explain their observed radio power \citep[e.g.][]{vazza+14,botteon+16}. Another formation scenario is {\it shock re-acceleration} of relativistic fossil electrons \citep[e.g.][]{markevitch+05, macario+11, kang+ryu11, kang+12, bonafede+14, shimwell+15, kang+17, vanweeren+17a}. This mechanism addresses the DSA efficiency problem, and implies a direct connection between radio relics and radio galaxies, which are supposed to be the primary sources of fossil plasma. 

Radio ``phoenices'' are another class of relics that are characterized by their steep curved spectra and often complex toroidal morphologies. They are supposed to trace AGN fossil plasma lobes which have been adiabatically compressed by merger shock waves \citep{ensslin+gopal-krishna01,ensslin+bruggen02,degasperin+15}. 

Relics have a range of sizes and morphologies. So-called {\it double relic} systems, with two relics located on diametrically opposite sides of the cluster center \citep[e.g.][]{rottgering+97, bagchi+06, venturi+07, bonafede+09, vanweeren+09, bonafede+12, degasperin+14} are an important subclass. Numerical simulations \citep[e.g.][]{vanweeren+11} suggest that these systems are the result of binary mergers between two-comparable mass clusters (mass ratio of 1:1 or 1:3), with the line connecting the two relics representing the projected merger axis of the system. 

In this paper we present a radio continuum and spectral analysis of the merging cluster CIZA\,J2242.8+5301, which hosts two opposite radio relics and a faint radio halo, by means of new deep 1--4~GHz Karl G. Jansky Very Large Array (VLA) observations. The paper is organized as follows: in Sect. \ref{sec:sausage} we provide an overview of CIZA\,J2242.8+5301, presenting the results of previous studies performed in the optical, X-ray and radio bands; in Sect. \ref{sec:obs} we describe the radio observations and the data analysis; the radio images, the spectral index maps are presented Sect. \ref{sec:results}. We end with a discussion and conclusions in Sects. \ref{sec:discuss} and \ref{sec:concl}, respectively.

In this paper, we adopt a flat $\Lambda$CDM cosmology with H$_0=70$ km s$^{-1}$ Mpc${-1}$, $\Omega_{\rm m}=0.3$ and $\Omega_\Lambda=0.7$, which implies a conversion factor of 3.22 kpc/$^{\prime\prime}$ and a luminosity distance of $\approx944$ Mpc, at the cluster's redshift \citep[$z =0.192$][]{kocevski+07}.

\section{CIZA J2242.8+5301}\label{sec:sausage}
CIZA\,J2242.8+5301 (hereafter CIZA2242) is a merging galaxy cluster located at $z = 0.1921$ which was first observed in X-ray by {\it ROSAT} \citep{kocevski+07}. 
This merging cluster has been well studied across the electromagnetic spectrum, from radio, optical, to X-ray bands \citep{vanweeren+10, ogrean+13, stroe+13, ogrean+14, stroe+14, akamatsu+15, dawson+15, jee+15, okabe+15,donnert+16, stroe+16, kierdorf+17, rumsey+17}, and it has been used as a textbook example for numerical simulations aimed to model the shock physics \citep{vanweeren+11, kang+12, kang+ryu15, donnert+17, molnar+17,kang+17}.

In the radio band, CIZA2242 shows the presence of a spectacular double relic, located $\sim1.5$ Mpc north and south from the cluster center. The northern relic is composed of an arc-like structure, $\sim2$ Mpc long and $\sim50$ kpc wide, which suggests the relic traces a shock propagating outward, and which gave the cluster the nickname of the ``Sausage'' cluster. The cluster also contains a low surface brightness radio halo \citep{vanweeren+10,stroe+13, hoang+17}. Numerical simulations performed by \cite{vanweeren+11} suggested that the mass ratio of the colliding clusters is between 1.5:1 and 2.5:1, and that the relics are seen close to edge-on (i.e. $|i|\lesssim10^\circ$). \cite{donnert+17} showed that, in the northern shock, the upstream X-ray temperatures and radio properties are consistent with each other, and consistent with weak lensing cluster masses. \cite{dawson+15} found two comparable subcluster masses ($16.1^{+4.6}_{-3.3}\times10^{14}$ M$_\odot$ and $13.0^{+4.0}_{-2.5}\times10^{14}$ M$_\odot$ for the northern and southern clusters, respectively). Their data was consistent with the interpretation of a merger occurring on the plane of the sky. The time since core passage was estimated to be about 0.6~Gyr by \cite{rumsey+17}.

The cluster has an X-ray luminosity of $L_{500}=7.7\pm0.1\times10^{44}$ erg s$^{-1}$ in the 0.1--2.4~keV energy band, within a radius of $R_{500}=1.2$ Mpc \citep{hoang+17}.
Several X-ray surface brightness discontinuities were detected with {\it Chandra} and {\it XMM-Newton} by \cite{ogrean+14}.
Due to the faint X-ray emissivity at the relic location, the temperature measurements across the shock fronts were obtained by means of {\it Suzaku} observations \citep{akamatsu+15}. The estimated shock Mach numbers were $\mathcal{M}_{\rm N}^{\rm X}=2.7^{+0.7}_{-0.4}$ and $\mathcal{M}_{\rm S}^{\rm X}=1.7^{+0.4}_{-0.3}$, for the northern and the southern relics respectively. Despite the initial discrepancy from the Mach numbers found in the radio band \citep[$\mathcal{M}_{\rm N}^{\rm radio}\sim4.6$ and $\mathcal{M}_{\rm S}^{\rm radio}\sim2.8$, for the northern and southern relic respectively,][]{vanweeren+10, stroe+13}, recent observations show an agreement between the X-ray and the radio Mach number ($\mathcal{M}_{\rm N}^{\rm radio}=2.9^{+0.10}_{-0.13}$ \citealp{stroe+14}, $\mathcal{M}_{\rm N}^{\rm radio}=2.7^{+0.6}_{-0.3}$ \citealp{hoang+17} and $\mathcal{M}_{\rm S}^{\rm radio} = 1.9_{-0.2}^{+0.3}$ \citealp{hoang+17}).

The Sausage relic has also been observed at high frequencies ($\gtrsim 2$ GHz), where there is still some debate on the integrated radio spectral shape of the relic. New radio observations (up to 30~GHz) revealed a possible steepening in the integrated radio spectrum from $\alpha\sim-1.0$ to $\alpha\sim-1.6$ at $\nu>2.5$ GHz \citep{stroe+16}, questioning the single power-law spectrum predicted from the DSA model \citep{ensslin+98}. Possible explanations involve a non-negligible contribution from the Sunyaev-Zel'dovich (SZ) effect \citep{basu+16}, the presence of a non-uniform magnetic field in the region \citep{donnert+16} and  evolving shock and re-acceleration models \citep{kang+ryu15}. Contrary to the interferometric observations, no steepening at high frequencies is revealed by single dish observations \citep{kierdorf+17} at 4.85 and 8.35 GHz, with the Effelsberg Telescope, nor by the combined single-dish Sardinia Radio Telescope interferometric measurements from \cite{loi+17}. This result led \cite{loi+17} to suggest that the interferometric observations at very high frequency might lose diffuse flux on large angular scales due to the limitation of the minimum baseline length (although \citealt{stroe+16} did attempt to correct for this effect).

Another interesting characteristic is that the northern relic is strongly polarized at high frequencies (i.e. $\rm\sim GHz$), with an observed polarization fraction in some regions of about 60\% \citep[at 8.35 GHz,][]{kierdorf+17}. Also, from the relic's width, the magnetic field strength was estimated to be between 5 and 7 $\mu$G \citep{vanweeren+10}. Diffuse central emission, classified as a radio halo, was detected at 1.4 GHz with the Westerbork Synthesis Radio Telescope\citep[WSRT;][]{vanweeren+10}, at 325 and 153 MHz with the Giant Metrewave Radio Telescope \citep[GMRT;][]{stroe+13}, and at 145~MHz with LOFAR \citep{hoang+17}. The radio halo has a spectral index of $\alpha=-1.06\pm0.06$, which remains approximately constant across the $\rm\sim1~Mpc^2$ halo region. Moreover, \citeauthor{hoang+17} found that the radio halo power is in agreement with the known correlation between the X-ray luminosity and radio power for giant radio halos, and suggested that the halo traces electrons re-accelerated by turbulence generated by the passing shock wave.

\begin{table*}%[h!]
\caption{Log of the observations.}
\begin{center}
\resizebox{\textwidth}{!}{\begin{minipage}{\textwidth}\renewcommand{\arraystretch}{1}
\begin{tabular}{lccccccc}
\hline
\hline%\noalign{\smallskip}
Band and Array 		&  Obs. date 	& Freq. coverage 	& Channel width& Integration time 	& Time on source 	& $\theta_{\rm FWHM}$$^{\rm a}$ 	& LAS$^{\rm b}$\\
					& 			& (GHz)			& (MHz)		& (s)				& (s)				& $(^{\prime\prime})$			&  $(^{\prime\prime})$\\
\hline%\noalign{\smallskip}
\multirow{4}{*}{\bf L-band D-array}
					& 30 Jan 2013	& 1--2			& 1			& 5				& 3538			& 40							& 970\\
					& 31 Jan 2013	& 1--2			& 1			& 5				& 3538			& 40 							&970\\
					& 27 Jan 2013	& 1--2			& 1			& 5				& 7175			& 40 							&970\\
					& 03 Feb 2013	& 1--2			& 1			& 5				& 7160			& 40 							&970 \\
\noalign{\smallskip}
\multirow{2}{*}{\bf L-band C-array}
					& 03 Jun 2013	& 1--2			& 1			& 5				& 21530			& 11 							&970\\
					& 02 Sep 2013	& 1--2			& 1			& 5				& 21535			& 11 							&970\\
\noalign{\smallskip}
{\bf L-band B-array}		& 30 Oct 2013	& 1--2			& 1			& 3				& 28719			& 4 							&120\\
\noalign{\smallskip}
{\bf L-band A-array}		& 11 May 2014	& 1--2			& 1			& 1				& 28720			& 2 							&36\\
\hline%\noalign{\smallskip}
\multirow{5}{*}{\bf S-band D-array}		
					& 23 Feb 2013	& 2--4			& 2			& 5				& 3580			& 16							&490\\
					& 27-28 Feb 2013& 2--4			& 2			& 5				& 7175			& 16 							&490 \\
					& 18 Feb 2013	& 2--4			& 2			& 5				& 3580			& 16							&490 \\
					& 27 Jan 2013	& 2--4			& 2			& 5				& 7170			& 16							&490 \\
					& 27 Jan 2013	& 2--4			& 2			& 5				& 3505			& 16							&490 \\
\noalign{\smallskip}
{\bf S-band C-array}		& 28 Jun 2013	& 2--4			& 2			& 5				& 28775			& 5 							&490\\
\noalign{\smallskip}
{\bf S-band B-array}		& 21 Oct 2013	& 2--4			& 2			& 3				& 28716			& 1.5 						&58\\
\noalign{\smallskip}
{\bf S-band A-array}		& 13 Apr 2014	& 2--4			& 2			& 1				& 28718			& 0.6 						&18\\

\hline%\noalign{\smallskip}
\end{tabular}
\end{minipage}}
\end{center}
{Note: $^{\rm a}$ synthesized beamwidth; $^{\rm b}$ largest angular scale.}
\label{tab:obs}
\end{table*}

\section{Observations and data reduction}\label{sec:obs}
CIZA2242 was observed by the VLA with all the four array configurations in the L- and S-bands, covering the frequency range from 1 to 4 GHz. The total recorded bandwidth was 1~GHz for the L-band, and 2~GHz for the S-bands, split into 16 spectral windows each having 64 channels (1 and 2~MHz width, for the L- and S-band respectively). Due to the large angular size of the cluster and the S-band field of view (FOV), we  observed three separate pointings. These pointings were located north-west, north-east and south with respect to the cluster's center. An overview of the frequency bands and observations is given in Table \ref{tab:obs}. 

For the primary calibrators we used 3C138 and 3C147, observed $\approx5-10$ minutes each at the end of the observing run. In some configurations 3C48 was also observed ($\approx8$ minutes) near the middle or at the end of the observing run, depending on whether this source was observed in addition to, or in substitution of, the other two primary calibrators. J2202+4216 was included as a secondary calibrator, and observed for $\approx3$ minutes at intervals of 30-40 minutes. 
All four polarization products (RR, RL, LR and LL) were recorded. 

The data were reduced with \texttt{CASA}\footnote{\url{https://casa.nrao.edu}} \citep{mcmullin+07} version 4.7.0, and processed in the same way for all the different observing runs. 
As a first step the data were Hanning smoothed. We removed radio frequency interference (RFI) using the {\it tfcrop} mode from the \texttt{flagdata} task and applied the elevation dependent gain tables and antenna offsets positions. 
We determined complex gain solutions for the central 10 channels of each spectral window. In this way we removed possible time variations of the gains during the calibrator observations. We pre-applied these solutions to find the delay terms (\texttt{gaintype=`K'}) and bandpass calibration. Applying the bandpass and delay solutions, we re-determined the complex gain solutions for the primary calibrators using the full bandwidth. We determined the global cross-hand delay solutions (\texttt{gaintype=`KCROSS'}) from the polarized calibrator 3C138, taking a RL-phase difference of $-10^\circ$ (both L- and S-band) and polarization fractions of 7.5\% and 10.7\% (L- and S-band respectively). We used 3C147 to calibrate the polarization leakage terms (\texttt{poltype=`Df'}), and 3C138 to calibrate the polarization angle (\texttt{poltype=`Xf'}). If not present, we replaced 3C147 with J2355+4950 as polarization leakage calibrator. All relevant solutions tables were applied on the fly to determine the complex gain solution for the secondary calibrator J2202+4216 and to obtain its flux density scale. The absolute flux scale from the primary calibrators was calculated assuming the \cite{perley+butler13} model. The RFI removal was repeated after applying the derived calibration solutions, using the {\it tfcrop} mode first and followed by the {\it rflag} mode.

All the solutions were transferred to the target source, and then the data were averaged by a factor of two in time and a factor of four in frequency (we excluded the first 7 and the last 10 channels). Finally, a last round of RFI flagging was  performed with \texttt{AOFlagger} \citep{offringa+10}, to remove the remaining interference from the dataset.

\begin{table}%[h!]
\caption{Image (\texttt{CASA}) information.}
\begin{center}
\resizebox{0.85\textwidth}{!}{\begin{minipage}{\textwidth}%\resizebox{0.5\textwidth}{!}{%
\begin{tabular}{lccccc}
\hline
\hline%\noalign{\smallskip}
Band		& Resolution							& weighting	& robust	& uv-taper			& $\sigma_{\rm rms}$\\
		& $(^{\prime\prime}\times^{\prime\prime})$	&			&		& $(^{\prime\prime})$& $(\mu{\rm Jy~beam}^{-1})$ \\
\hline
\multirow{5}{*}{L-band}	
		& $1.6\times1.6$						& briggs		& 0		& none			& 5.3 \\
		& $2.5\times2.5$						& uniform		& N/A	& 2.5				& 6.5 \\
		& $6\times6$							& uniform		& N/A	& 5				& 5.8 \\
		& $10\times10$ 						& uniform		& N/A	& 10				& 8.9 \\
		& $25\times25$ 						& uniform 		& N/A	& 25 				& 19 \\
\hline%\noalign{\smallskip}
\multirow{5}{*}{S-band}	
		& $1.3\times1.3$						& briggs		& 0		& none 			& 3.4 \\	
		& $2.5\times2.5$						& uniform		& N/A	& 2.5				& 4.1 \\
		& $6\times6$ 							& uniform 		& N/A 	& 5 				& 4.5 \\
		& $10\times10$ 						& uniform 		& N/A 	& 10 				& 5.6 \\
		& $25\times25$ 						& uniform 		& N/A 	& 25 				& 23 \\
\hline%\noalign{\smallskip}
\end{tabular}
\end{minipage}}
\end{center}
\label{tab:images}
\end{table}

To refine the calibration for the target field, we performed two rounds of phase-only self-calibration and two of amplitude and phase self-calibration on the individual datasets. The only exceptions were the northeastern and northwestern pointing of the A-array data taken in the S-band, for which the low signal-to-noise ratio (SNR) allowed only for phase-only self-calibration, combining both polarizations (\texttt{gaintype=`T'}). All imaging in {\tt CASA} was done with w-projection (\citealp{cornwell+05}, \citeyear{cornwell+08}), which takes the non-coplanar nature of the array into account. We also used briggs weighting \citep{briggs95} with a robust factor of 0. The spectral index was taken into account during the deconvolution, using \texttt{nterms=3} \citep{rau+cornwell11}. To deconvolve a few bright sources outside the main lobe of the primary beam, image sizes of up to $9720^2$ pixels were needed (in A-array configuration). Clean masks were employed during the deconvolution. Initially, the clean masks were automatically made with the \texttt{PyBDSM} source detection package \citep{mohan+rafferty15}, and then updated after each imaging cycle. We manually flagged some additional data during the self-calibration process by visually inspecting the self-calibration solutions.

The final images were obtained by combining the data from all array configurations and performing a final self-calibration with a long solution interval to align the datasets. We produced images over a wide range of resolutions and with different uv-tapers and weighting schemes to emphasize the radio emission on various spatial scales. Moreover, to ensure the recovery of flux on the same spatial scale for the complementary observations (150 MHz LOFAR, \citealp{hoang+17}, and 610 MHz GMRT, \citealp{vanweeren+10}), we kept data on common uv-distances, discarding all the data below 120$\lambda$. The final image resolutions are $1.6^{\prime\prime}$ and $1.3^{\prime\prime}$ (full resolution, L- and S-band respectively), and $2.5^{\prime\prime}$, $5^{\prime\prime}$, $10^{\prime\prime}$ and $25^{\prime\prime}$ (see Table \ref{tab:images}). We used a multiscale \texttt{clean} \citep{cornwell+08}, with scales of $[0, 3, 7, 25, 60]\times$ the pixel size\footnote{We chose the pixel size so that the beam is sampled by $\approx5$ pixels.} to make these images. The images were corrected for the primary beam attenuation, with the frequency dependence of the beam taken into account (\texttt{widebandpbcor} task\footnote{This task was run by means of \texttt{CASA} version 5.0, which has an update beam shape model.}). 

To create a set of ``deep'' images, we employed the {\it w}-Stacking Clean algorithm (\texttt{WSClean}) by \cite{offringa+14} (Table \ref{tab:images_wsc}). We also stacked images from the L- and S-band data. We only use these images for viewing purposes and determinations of source morphologies. No flux density measurements or other quantitative measurements in this paper have been extracted from them, since \texttt{WSClean} just provides an averaged flux density across the bandwidth without taking into account the spectral index during the deconvolution.

\begin{table}%[h!]
\caption{Image (\texttt{WSClean}) information.}
\begin{center}
\resizebox{0.85\textwidth}{!}{\begin{minipage}{\textwidth}%\resizebox{0.5\textwidth}{!}{%
\begin{tabular}{lccccc}
\hline
\hline%\noalign{\smallskip}
Band		& Resolution							& weighting	& robust	& uv-taper			& $\sigma_{\rm rms}$\\
		& $(^{\prime\prime}\times^{\prime\prime})$	&			&		& $(^{\prime\prime})$& $(\mu{\rm Jy~beam}^{-1})$ \\
\hline
L-band	& $2.1\times1.8$ 						& briggs		& 0		& none 			& 3.8 \\
S-band	& $0.8\times0.6$						& briggs		& $-0.5$	& none 			& 2.7 \\
\hline%\noalign{\smallskip}
\multirow{4}{*}{LS-band}	
		& $3.8\times3.8$						& briggs		& 0		& 2.5				& 3.4 \\
		& $6\times6$							& briggs		& 0		& 5				& 4.2 \\
		& $11\times11$ 						& briggs		& 0		& 10				& 6.2 \\
		& $26\times26$ 						& briggs 		& 0		& 25 				& 16.7 \\
\hline%\noalign{\smallskip}
\end{tabular}
\end{minipage}}
\end{center}
%{Note:}		
\label{tab:images_wsc}
\end{table}

\begin{figure*}%[h!]
\centering
\includegraphics[scale=0.8]{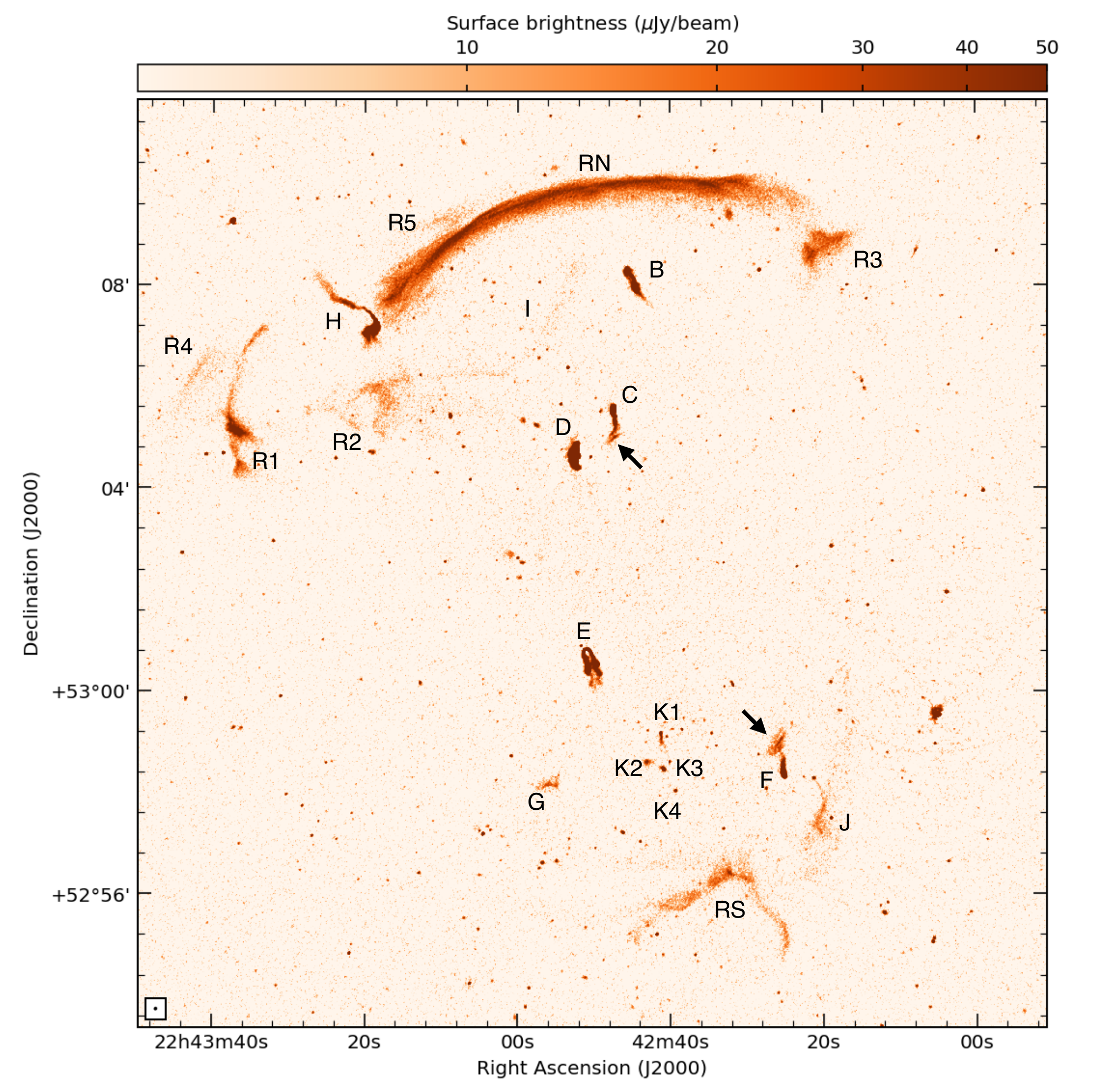}
\caption{L-band (1.5 GHz) VLA high-resolution image of CIZA2242.
The map has a noise of  $\sigma_{\rm rms}=3.8~\mu$Jy~beam$^{-1}$. The black arrows highlight the ``broken'' nature of the radio tails. Sources are labeled following \cite{stroe+13} and \cite{hoang+17} (see Fig. \ref{fig:labels} for a more complete labeling).}\label{fig:deep_fullLband}
\end{figure*}

\begin{figure}[h!]
\centering
{\includegraphics[width=0.47\textwidth]{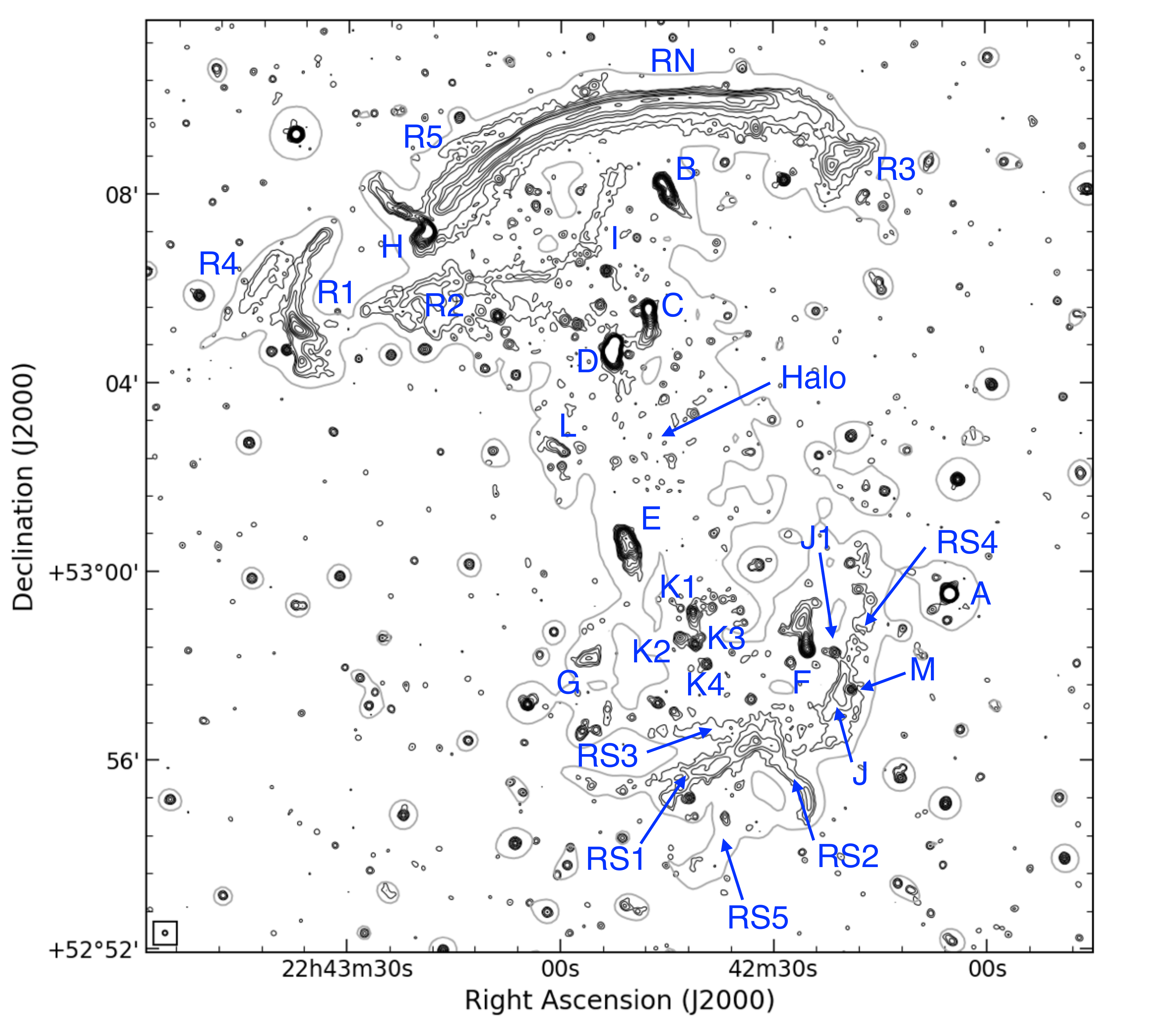}}
\caption{Labeled cluster sources, adapting the scheme from \cite{stroe+13} and \cite{hoang+17}. We used the $5^{\prime\prime}$ resolution (black solid line) to emphasize the diffuse emission. The grey solid line is the $3\sigma$ radio contour at $25^{\prime\prime}$ resolution. The radio contours are the same of the ones in Fig.~\ref{fig:low_res_comb}).}\label{fig:labels}
\end{figure}

\section{Results}\label{sec:results}
\begin{figure*}%[h!]
\centering
{\includegraphics[width=0.47\textwidth]{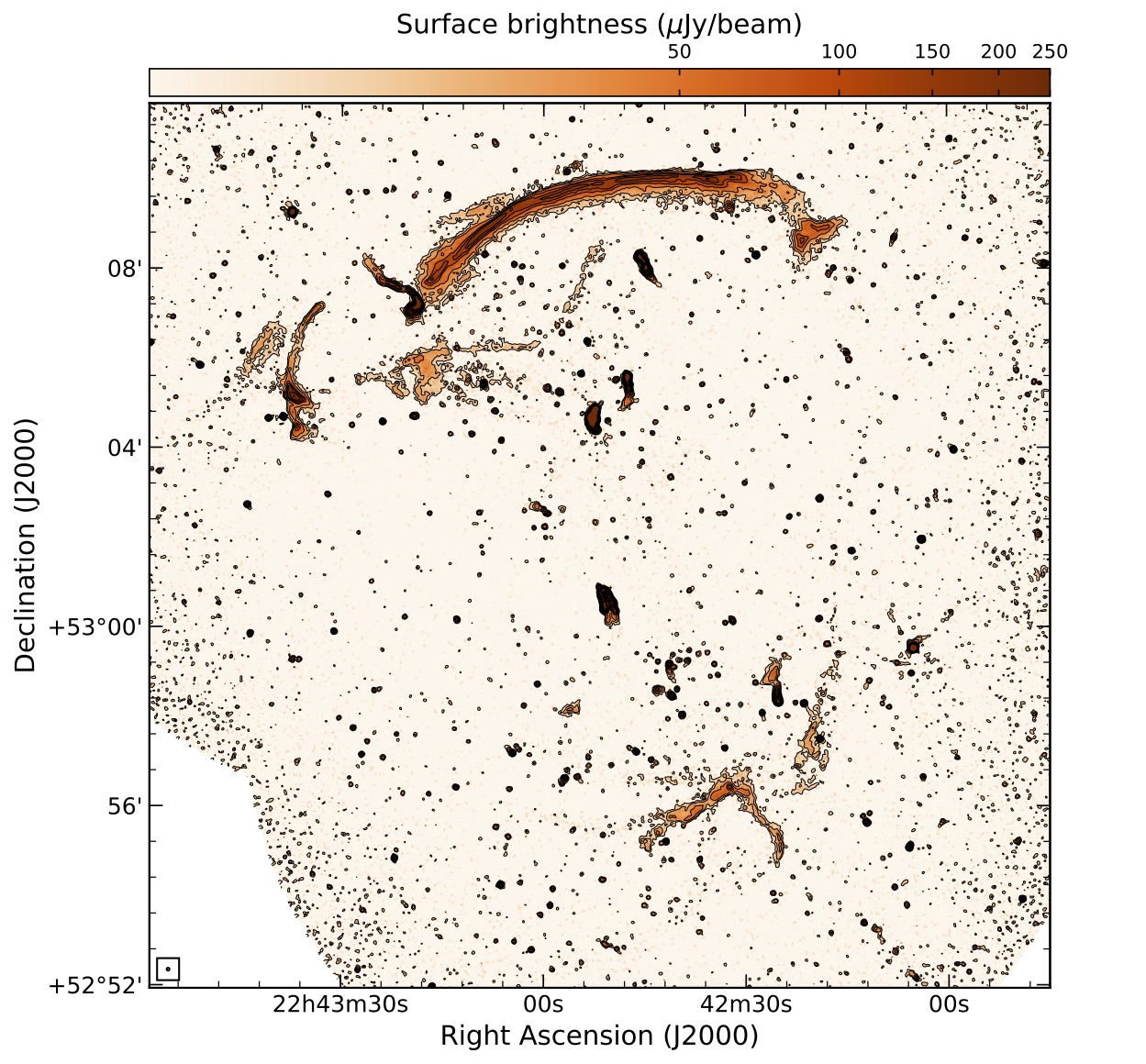}}
\hspace{2mm}
{\includegraphics[width=0.47\textwidth]{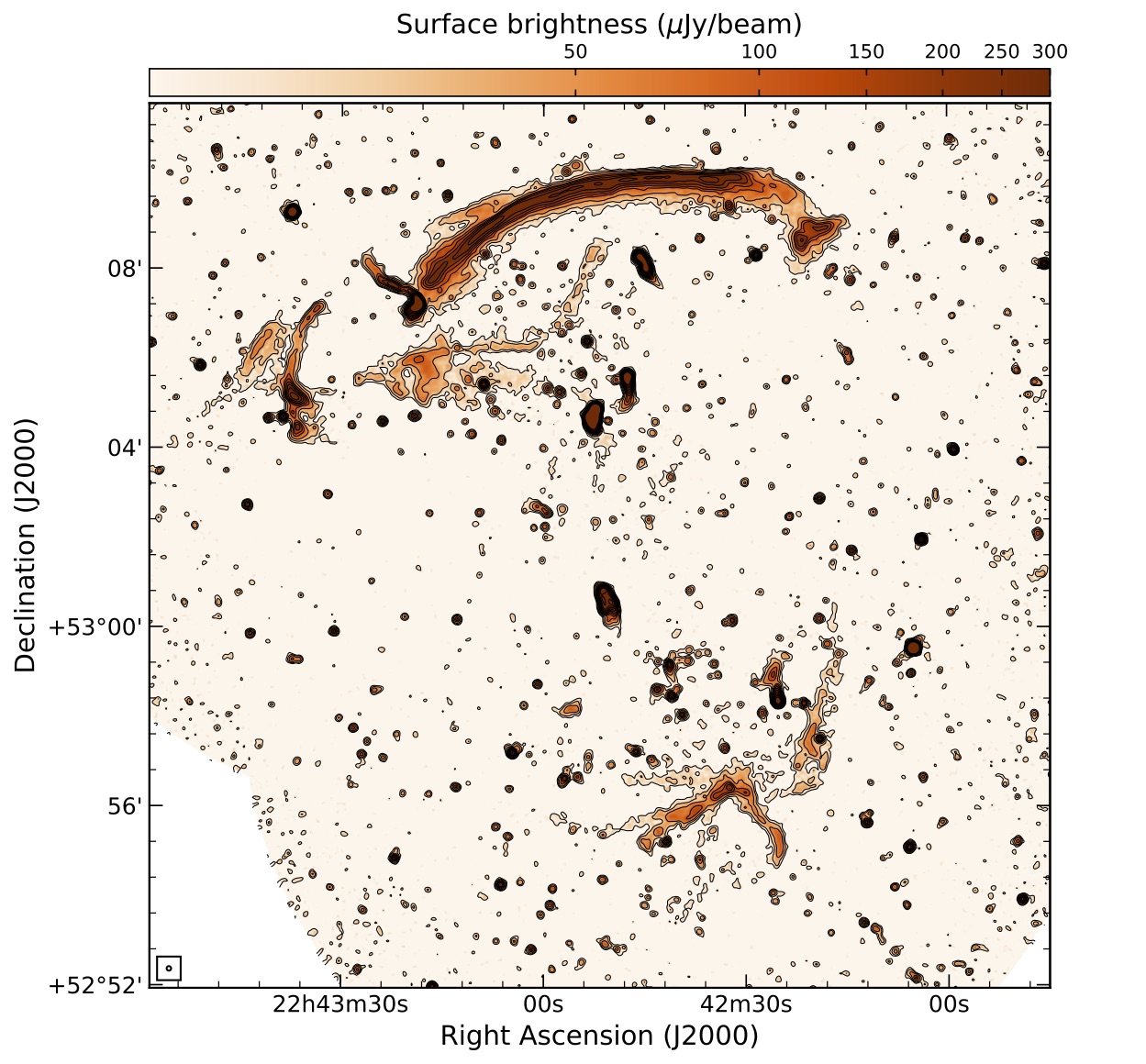}}\\
{\includegraphics[width=0.47\textwidth]{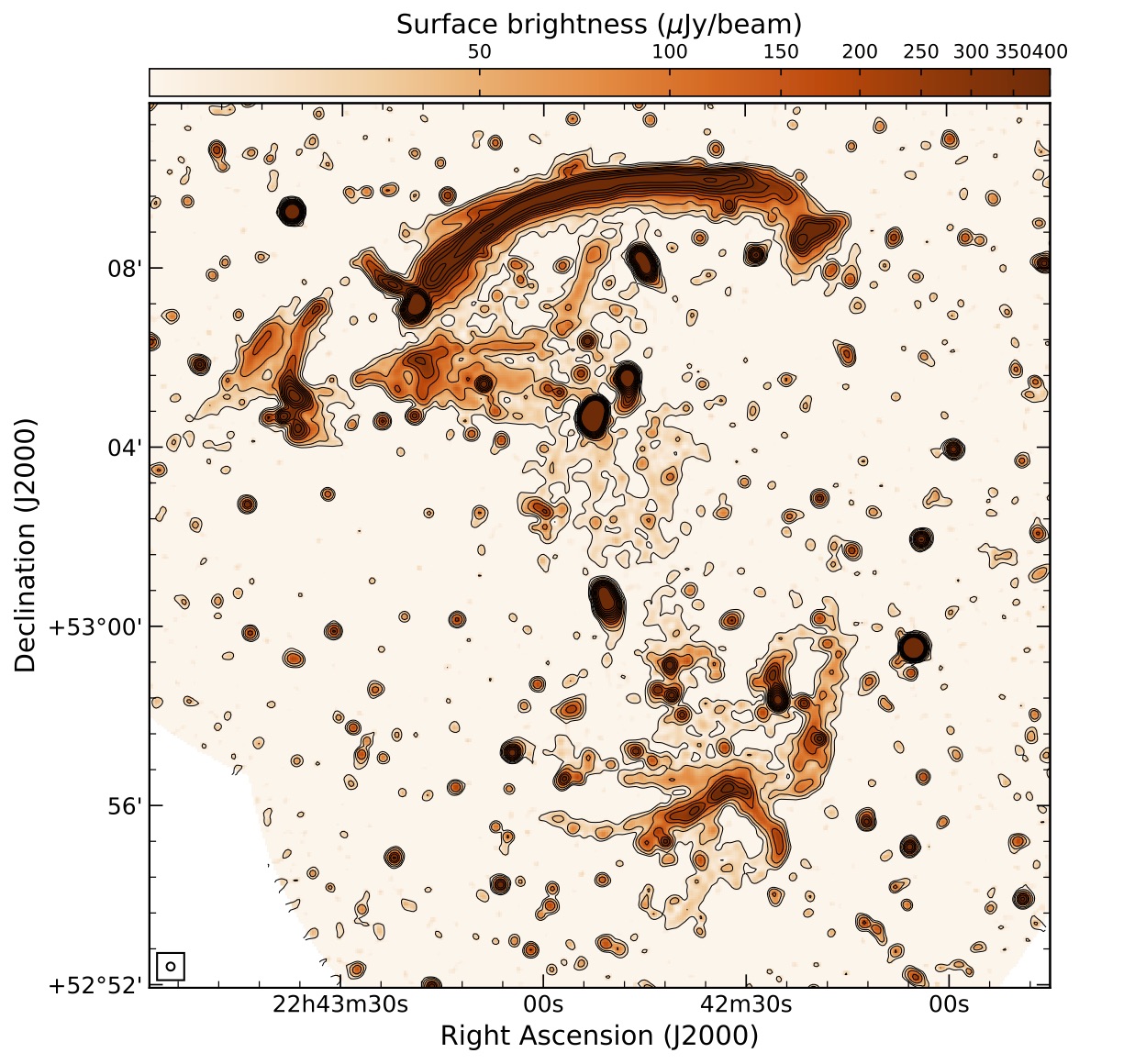}}
\hspace{2mm}
{\includegraphics[width=0.47\textwidth]{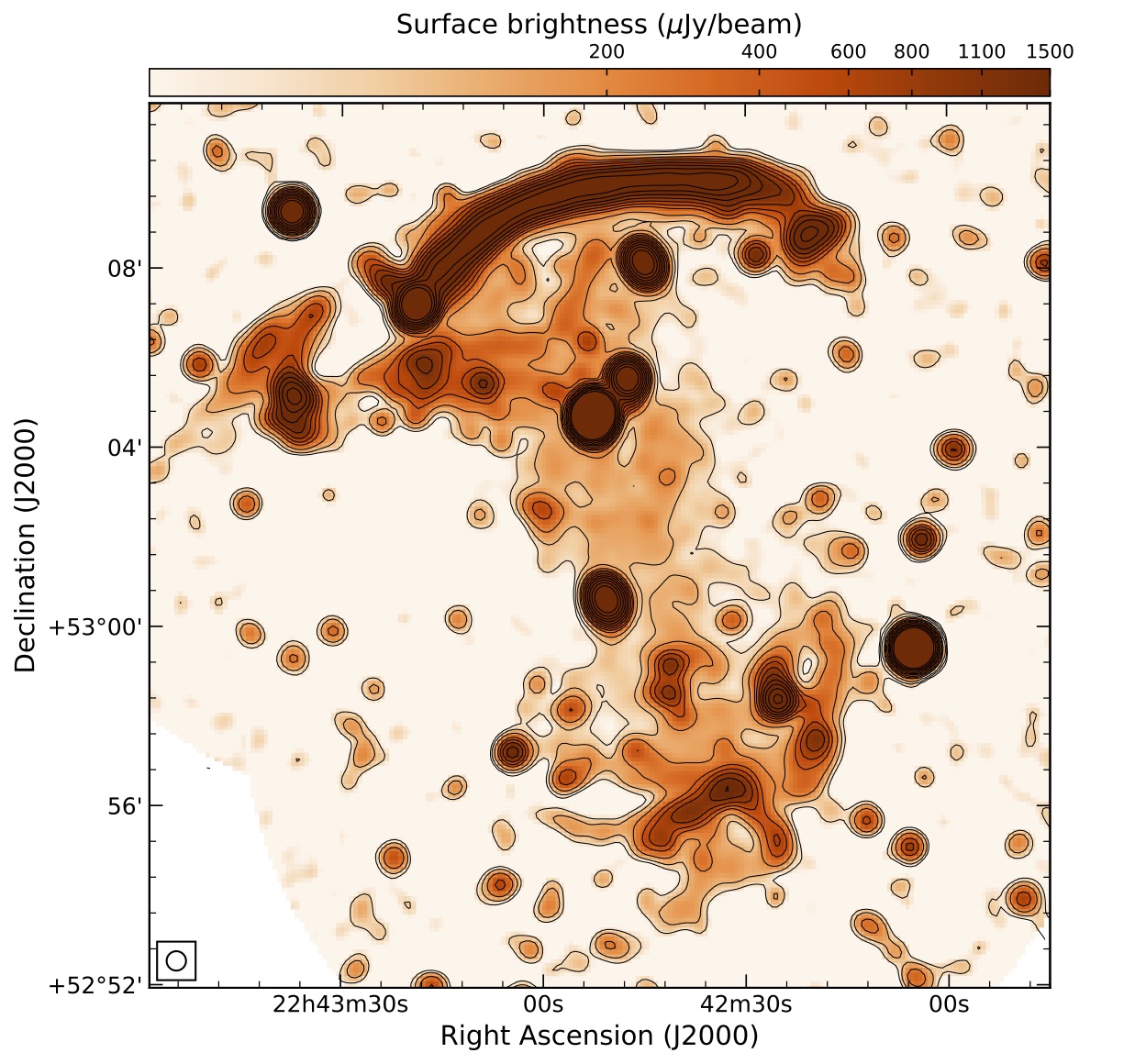}}
\caption{Combined L- and S-band (1--4 GHz) VLA deep images at low resolution ($2.5^{\prime\prime}$, top left; $5^{\prime\prime}$, top right; $10^{\prime\prime}$, bottom left; $25^{\prime\prime}$, bottom right). The beam size is displayed in the bottom left corner of each image. The radio contours are at $3\sigma_{\rm rms}\times\sqrt{[1, 4, 16, 64, \ldots]}$, where $\sigma_{2.5^{\prime\prime}}=3.4~\mu$Jy~beam$^{-1}$, $\sigma_{5^{\prime\prime}}=4.2~\mu$Jy~beam$^{-1}$, $\sigma_{10^{\prime\prime}}=6.2~\mu$Jy~beam$^{-1}$, $\sigma_{25^{\prime\prime}}=16.7~\mu$Jy~beam$^{-1}$.}
\label{fig:low_res_comb}
\end{figure*}

In Figs. \ref{fig:deep_fullLband} and \ref{fig:deep_fullSband} we present our deep, high-resolution VLA images of CIZA2242, in L- (1--2 GHz, $2.1^{\prime\prime}\times1.8^{\prime\prime}$, weighting briggs' robust 0 and uniform) and S-band (2--4 GHz, 0.8$^{\prime\prime}\times0.6^{\prime\prime}$, weighting briggs' robust $-0.5$ and uniform) respectively. The rms noise is $3.8~\mu$Jy~beam$^{-1}$ at 1--2 GHz and  $2.7~\mu$Jy~beam$^{-1}$ at 2--4~GHz (Table \ref{tab:images_wsc}). We also show all the 1--4~GHz lower resolution images produced (Fig.~\ref{fig:low_res_comb}). In all the images we detect the two main relics (north and south), and five other areas/regions of diffuse emission and several tailed radio sources above the $3\sigma_{\rm rms}$ level. We label the cluster sources using the same convention of \cite{stroe+13} and \cite{hoang+17}, see Fig.~\ref{fig:labels}.

At full resolution, the length of the northern relic (RN) remains constant between the two frequencies (i.e. $\sim1.8$ Mpc), while the width decreases from 80 to 40 kpc, in L- and S-band, respectively. The high resolution radio maps show, for the first time, that the relic  structure is not continuous, but broken into six different ``sheets'' or ``filaments'' (Fig.~\ref{fig:zoomRN_fullLband}) with lengths of about $200-600$~kpc. This is visible at both frequencies (Figs.~\ref{fig:deep_fullLband} and \ref{fig:deep_fullSband}). The integrated flux densities of RN, measured within the $3\sigma_{\rm rms}$ region, are $128.1\pm3.2$ mJy and $56.1\pm1.4$ mJy at 1.5 and 3.0 GHz, respectively. We did not included the western emission (R3), for which we measured $12.1\pm0.3$ mJy and $4.5\pm0.2$ mJy, at 1.5 and 3.0~GHz respectively\footnote{\cite{hoang+17} measured one single integrated flux value for RN and R3 ($1548.2\pm4.6$ mJy at 145 MHz)}. We also note the presence of additional faint emission at the $3\sigma_{\rm rms}$ level located northeastward of the northern relic ($4.2\pm0.2$ mJy and $2.1\pm0.1$ mJy, in L- and S-band respectively). We label this emission as R5 (see Fig.~\ref{fig:labels}) and, as for R3, it was not included into the RN surface brightness measurement. R5 has an extent of 215~kpc in the full resolution image, but its size increases up to 660~kpc in the $10^{\prime\prime}$ resolution image (Fig. \ref{fig:low_res_comb}). On the eastern side, at a distance of about $\sim 30$ kpc, a wide-angle tailed source (H in Fig.~\ref{fig:labels}) is located, which has a much higher surface brightness than the relic at both frequencies. Comparing the optical and the radio contours of this source (right panel in Fig. \ref{fig:spix_tails}(d)), we note that the northern lobe breaks $\sim50$ kpc from the AGN and then proceeds quite straight, while the southern lobe bends immediately, in projection. We also note that the surface brightness of the northern lobe increases by a factor $\sim4$ about 85 kpc northeast of the host galaxy. The southern lobe, on the other hand, is $\sim7$ times brighter than the northern one ($F_{\rm lobe~S}=21.5$ mJy and $F_{\rm lobe~N}=3.2$ mJy at 1.5 GHz). The zoom on the the northern relic (Fig. \ref{fig:zoomRN_fullLband}) shows that the flux boost in the northern lobe coincides with the RN6 filament, while the brighter southern lobe is approximately located along the extent of the RN1 filament.

At a distance of about 600~kpc east of source H, there is another arc-like patch of diffuse emission, labeled R1. This relic extends in the north-south direction for a length of $\approx610$ kpc, while the width changes from about 20~kpc to 100~kpc, going from north to south. The flux density of this diffuse source is about one order of magnitude lower than RN: we measure $16.0\pm0.5$~mJy and $7.8\pm0.3$~mJy, at 1.5 and 3.0~GHz respectively. We also note that the southern part of R1 is brighter than the rest of this relic. East of R1, faint emission, labeled as R4, extends for $\approx315$ kpc in NW-SE direction. The size of R4 increases up to 640~kpc, at $10^{\prime\prime}$ resolution. This emission was also detected by \cite{hoang+17}, who labeled it as R1 east (and R1 as R1 west). South of RN/H, we detect a patch of extended emission with a toroidal morphology, labeled R2. Although the $3\sigma_{\rm rms}$ emission is seen only within a region of $180\times210$ kpc, some residuals cover a broader region, which is better detected at lower resolutions (Fig.~\ref{fig:low_res_comb}). At $5^{\prime\prime}$ resolution, the size increases to $710\times250$ kpc. Towards the west, we detect source I, which was also reported by \cite{stroe+13} and \cite{hoang+17}. It is barely detected in our high-resolution images, but it is clearly visible in our low resolution images (Fig.~\ref{fig:low_res_comb}). Our images reveal that source I is a $\approx825$~kpc long filament that connects with R2.

At full resolution, the southern relic (RS) is well detected only in the L-band image. It is located $\approx2.5$~Mpc south of RN, and formed by two ``arms'', labeled RS1 and RS2 in Fig.~\ref{fig:labels}. They measure $530\times70$ and $25\times380$~kpc, respectively.
We measure a total flux density for these two ``arms'' of $16.7\pm0.5$~mJy and $7.7\pm0.3$~mJy, at 1.5 and 3.0~GHz respectively. Moreover, at lower resolutions we also detect three additional ``arms'', labeled RS3, RS4 and RS5, extending eastwards and westwards (Fig.~\ref{fig:labels}). The southern relic has a total size of $\approx1.5$ Mpc at $25^{\prime\prime}$ resolution.

A patch of faint emission is seen to the west of RS1 and RS2 (source J). Our deep VLA observations reveal for the first time a connection between source J and an AGN core located northwards, and labeled as J1 ($\rm RA=22^h42^m21^s.35$ and $\rm DEC=+52^\circ58^\prime17^{\prime\prime}.92$, J2000). Hereafter, we will refer to the whole radio source as J, and to the core as J1. We measure a flux density for its radio lobe of $6.3\pm1.9$ mJy and $1.7\pm0.2$ mJy, at 1.5 and 3.0 GHz respectively.
Moreover, at low-resolution source J seems to be completely embedded into the southern relic (bottom right panel in Fig.~\ref{fig:low_res_comb}). 

Towards the north of RS, the patchy extended emission detected by \cite{stroe+13} and \cite{hoang+17} is now resolved into four different sources, which are now labeled in Fig.~\ref{fig:labels} as K1 (tailed radio source), K2, K3 and K4 (compact radio sources). All these sources have and optical counterpart (see Fig. \ref{fig:opt_galaxies} and the zoom in right panel in Fig. \ref{fig:spix_tails}(g)). A patch of diffuse emission is detected on the east of these radio galaxies, namely source G. No clear optical counterpart has been found to be associated to this source from previous optical studies \citep[Fig. \ref{fig:opt_galaxies} and the zoom in right panel in Fig. \ref{fig:spix_tails}(f)]{dawson+15}.

The radio halo, already detected by other authors \citep[i.e.][]{vanweeren+10,stroe+13,hoang+17} becomes visible at $10^{\prime\prime}$ resolution (bottom left panel in Fig.~\ref{fig:low_res_comb}), at a level of $3\sigma_{\rm rms}$ ($\sigma_{\rm rms}=6.2~\mu$Jy~beam$^{-1}$). The halo is visible up to $6\sigma_{\rm rms}$ ($\sigma_{\rm rms}=16.7~\mu$Jy~beam$^{-1}$) level at $25^{\prime\prime}$ resolution (bottom right panel in Fig.~\ref{fig:low_res_comb}). It extends the entire distance between RN and RS, with a size of about 2.1~Mpc$\times$0.8~Mpc (N-S and E-W directions, respectively), at this resolution.
To estimate the halo radio flux density, we need to remove the contribution of radio galaxies and other sources. This is not trivial, since some these sources are also characterized by some extended emission.

To construct a model for the emission of the extended and compact sources (e.g. radio galaxies, R2 and source I), we imaged the data with the same settings as for the $5^{\prime\prime}$ resolution image (uniform weighting and a uv-taper of $5^{\prime\prime}$), since it catches properly the diffuse emission from the tails. To avoid the emission of the halo, we include an inner uv-cut of 0.86k$\lambda$, which filters out the emission scales larger than $4^\prime$ ($\sim770$ kpc at cluster's redshift). The model of the radio galaxies was then subtracted from the uv data by means of the task \texttt{uvsub}. 
Finally, the new visibilities were re-imaged at low resolution (i.e. a uv-taper of $35^{\prime\prime}$ and uniform weighting) and the final image was primary-beam corrected.

The measured halo radio flux density at 1.4~GHz is $25.2\pm4.1$ mJy. This flux density measurement agrees with the previous result found by \cite{hoang+17}. The flux density results in a 1.4~GHz radio power\footnote{$P_{\rm 1.4~GHz}=4\pi D_{\rm L}^2S_{\rm 1.4~GHz}(1+z)^{-(\alpha+1)}$ W Hz$^{-1}$, where $D_{\rm L}=944$ Mpc is the luminosity distance and $(1+z)^{-(\alpha+1)}$ the $k$-correction; we use a spectral index of $\alpha=-1.03\pm0.09$ \citep{hoang+17}} of $P_{\rm 1.5~GHz} = (2.69\pm0.37)\times10^{24}$ W Hz$^{-1}$. 
The total error on the halo flux density has been estimated including the flux scale uncertainty (i.e. 2.5\%), image noise (i.e. $\sigma_{35^{\prime\prime}}=36.3~\mu$Jy~beam$^{-1}$) scaled to halo area, and the uncertainty due to the subtraction of the discrete radio sources in the same region \citep[$\sigma_{\rm sub}$, see Eq.~1 in][]{cassano+13}. We used $\sigma_{\rm sub}=2.5\%$ of the total flux of the subtracted radio galaxies, given by the ratio of the post-source subtraction residuals to the pre-source subtraction flux of a nearby compact source (${\rm RA=22^h41^m33^s.02}$ and ${\rm DEC=+53^\circ11^\prime05^{\prime\prime}.63}$, J2000). Additional uncertainties come from the estimation of the halo size and from the flux recovered during the imaging \citep{bonafede+17}.
Therefore, we consider this value a lower limit.
Unfortunately, the LAS (largest angular scale) of the halo does not allow us to properly recover all the halo flux at 3.0~GHz, hence we do not report a flux density measurement at this frequency.

\begin{table}[h!]
\caption{Flux densities and integrated spectral index of the diffuse cluster sources measured from the full resolution VLA images.}
\begin{center}
\begin{tabular}{lccc}
\hline
\hline%\noalign{\smallskip}
Source & $S_{\rm 1.5~GHz}$ & $S_{\rm 3.0~GHz}$ & $\alpha_{\rm int}$ \\
	    & (mJy)	& (mJy)  &	  \\
\hline%\noalign{\smallskip}
RN & $128.1\pm3.2$ & $56.1\pm1.4$ & $-1.19\pm0.05$ \\
RS & $16.7\pm0.5$ & $7.7\pm0.3$ & $-1.12\pm0.07$ \\
R1 & $16.0\pm0.5$ & $7.8\pm0.3$ & $-1.03\pm0.06$ \\
R2 & $10.3\pm0.4$ & $4.5\pm0.2$ & $-1.19\pm0.09$ \\
R3 & $12.1\pm0.3$ & $4.3\pm0.2$ & $-1.43\pm0.07$ \\
R4 & $4.0\pm0.2$ & $2.3\pm0.1$ & $-0.93\pm0.11$ \\
R5 & $4.2\pm0.2$ & $2.1\pm0.1$ & $-0.93\pm0.12$ \\
%F & $2.2\pm0.1$ & $0.4\pm0.1$ & $-2.46\pm0.23$ \\
I & $3.4\pm0.2$ & $1.5\pm0.2$ & $-1.18\pm0.17$ \\
%J (lobe) & $6.3\pm0.3$ & $1.7\pm0.2$ & $-1.89\pm0.16$\\
Halo & $25.2\pm4.1$ & $\dots$ & $\dots$ \\
\hline%\noalign{\smallskip}
\end{tabular}
\end{center}
%{Note:}
\label{tab:fluxes}
\end{table}

A study of the tailed radio galaxies was performed in detail by \cite{stroe+13}. Similar to this previous work, we find that the majority of the radio tails
are stretched in the north-south direction, tracing the merger direction.
Interestingly, our deep high-resolution images reveal that source C and F have ``broken'' tails (see black arrows in Fig.~\ref{fig:deep_fullLband} and \ref{fig:deep_fullSband}), although they are located at different places in the cluster. In contrast with the previous classification as a head-tail radio galaxy \citep{stroe+13}, our highest resolution images reveal that source B is a double-lobe source, but where some emission from the lobes has been stripped backwards in the north-south direction, suggesting possible stripping of lobe plasma related to the merger event. We also underline the proximity between this tailed radio galaxy and source I (Figs. \ref{fig:labels} and \ref{fig:low_res_comb}), although no direct connection is visible in our images. Classical radio tail shapes are seen in sources E and K1, with a tail extension of 160 and 48 kpc, respectively, based on our 1.5 GHz image. 
The integrated flux densities of the diffuse sources at 1.5 and 3.0~GHz\footnote{In the flux density estimation we used identical regions at both frequencies.} and the spectral index between these two frequencies, are reported in Table \ref{tab:fluxes}. All the spectral index values are in agreement with previous works, i.e. \cite{stroe+13} and \cite{hoang+17}. We estimated the spectral index uncertainties by taking into account the map noise ($\sigma_{\rm rms}$) and a flux scale uncertainty of 2.5\%, using:

\begin{figure*}%[h!]
\centering
\includegraphics[scale=0.8]{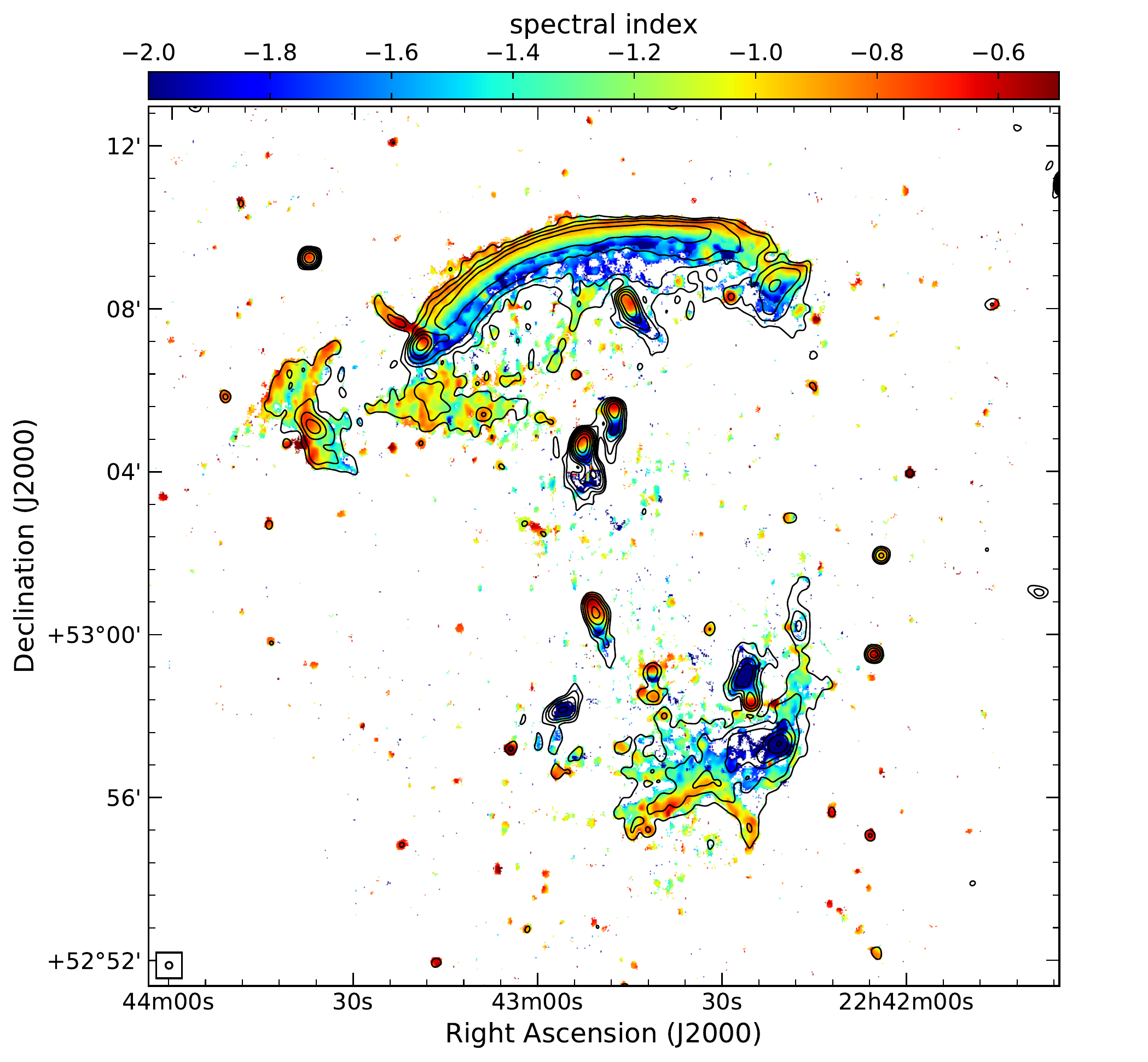}
\caption{10$^{\prime\prime}$ spectral index map of CIZA2242 obtained between 145~MHz, 610~MHz, 1.5~GHz, and 3.0~GHz. The radio contours are from the LOFAR image, with contours drawn at levels of $3\sigma_{\rm rms}\times\sqrt{[1, 4, 16, 64, 256,\ldots]}$, with $\sigma_{\rm rms}=230~\mu$Jy~beam$^{-1}$. The most important cluster sources have been labeled as Fig. \ref{fig:labels}.}\label{fig:spix10}
\end{figure*}

\begin{equation}\label{eq:err_spix}
\sigma_\alpha=\frac{1}{\ln \frac{S_{\rm 1.5\,GHz}}{S_{\rm 3.0\,GHz}}}\sqrt{ \left ( \frac{\Delta S_{\rm 1.5\,GHz}}{S_{\rm 1.5\,GHz}} \right )^2+ \left ( \frac{\Delta S_{\rm 3.0\,GHz}}{S_{\rm 3.0\,GHz}} \right )^2}
\end{equation}
where $\Delta S_\nu= \sqrt{\sigma_{\rm rms}^2 \bigl (\frac{A_{\rm source}}{A_{\rm beam}} \bigr ) + (0.025 \times S_\nu)^2}$ is the total uncertainty on $S_\nu$, while $A_{\rm source}$ and $A_{\rm beam}$ are the area of the source and beam respectively.

\begin{figure*}%[h!]
\centering
{\includegraphics[width=0.47\textwidth]{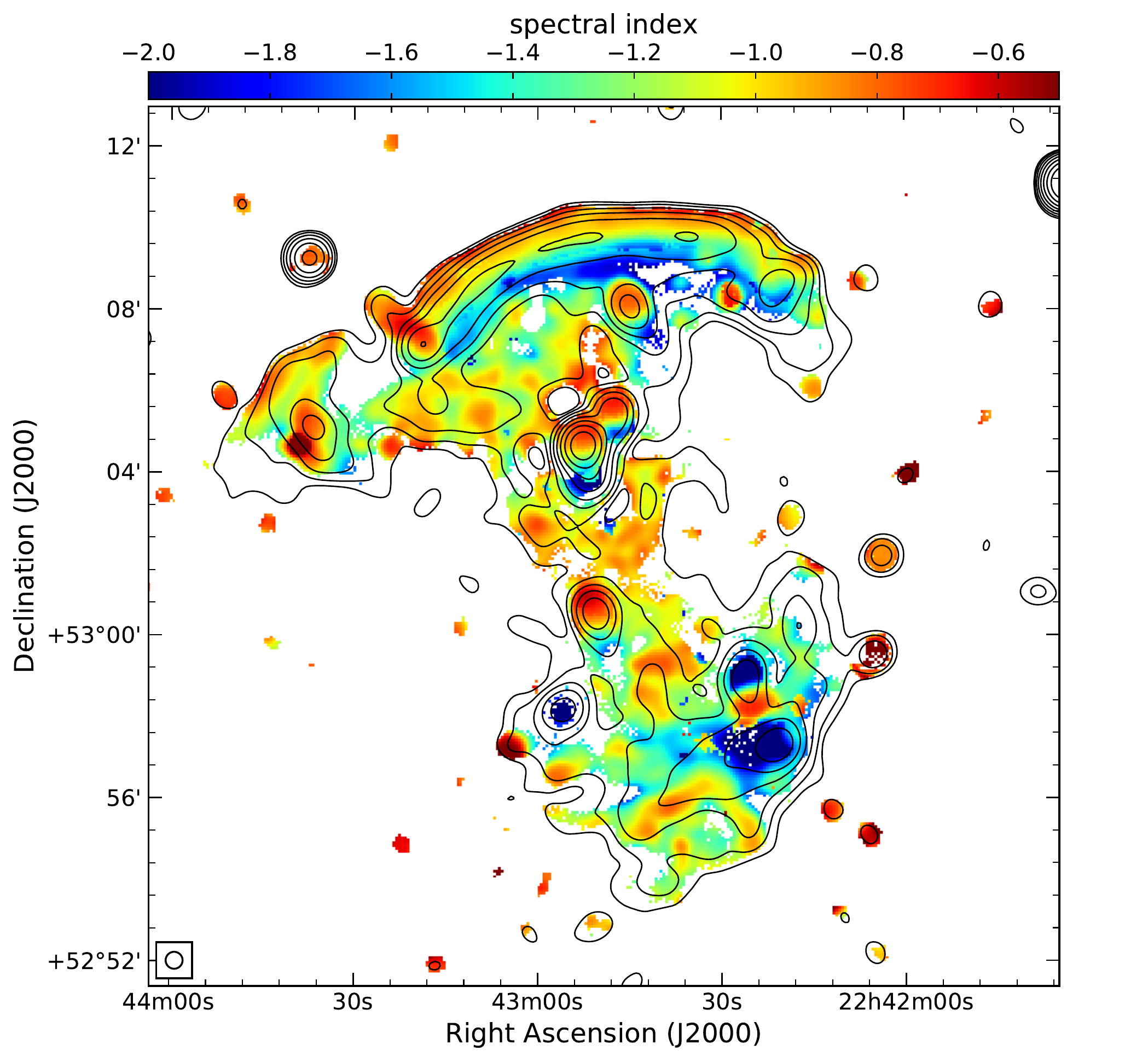}}
\hspace{0mm}
{\includegraphics[width=0.47\textwidth]{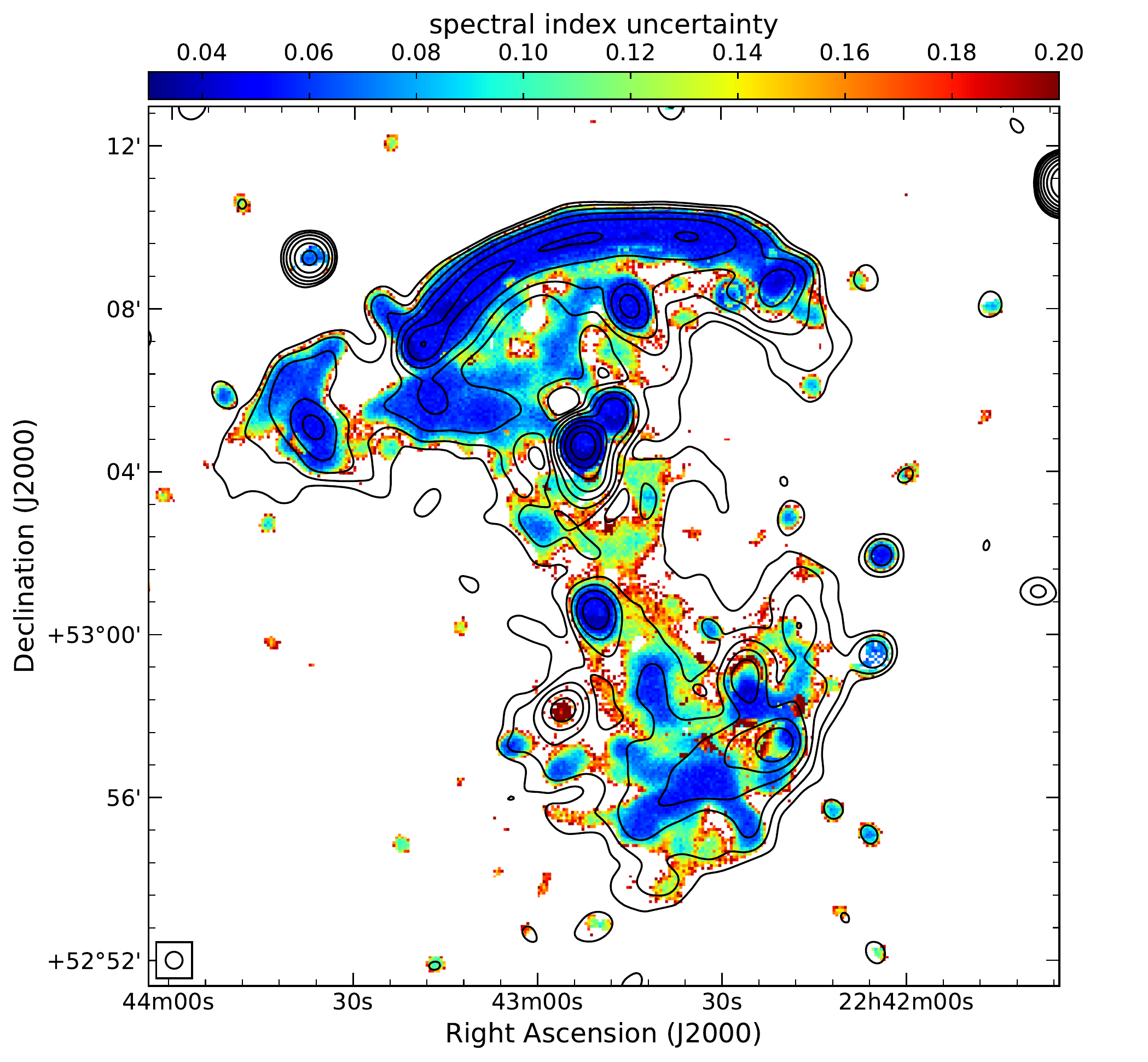}}
\caption{25$^{\prime\prime}$ spectral index (left panel) and spectral index uncertainty (right panel) maps of CIZA2242 obtained between 145~MHz, 610~MHz, 1.5~GHz, and 3.0~GHz. The radio contours are from the LOFAR image, contours are drawn at levels of $3\sigma_{\rm rms}\times\sqrt{[1, 4, 16, 64, 256,\ldots]}$, with $\sigma_{\rm rms}=340~\mu$Jy~beam$^{-1}$.}\label{fig:spix25}
\end{figure*}

\begin{figure*}
\centering
\gridline{
	\fig{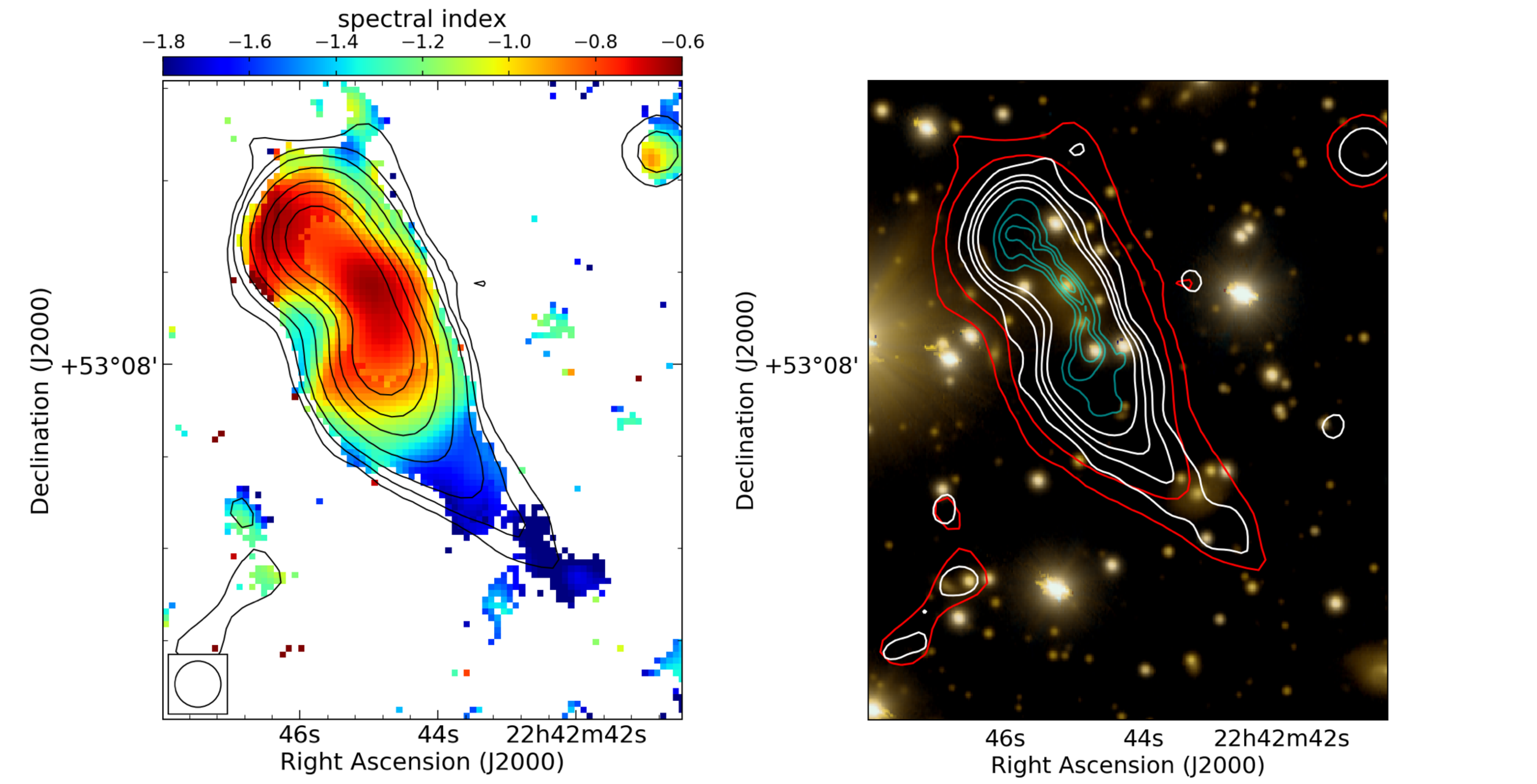}{0.5\textwidth}{(a) source B}
    \fig{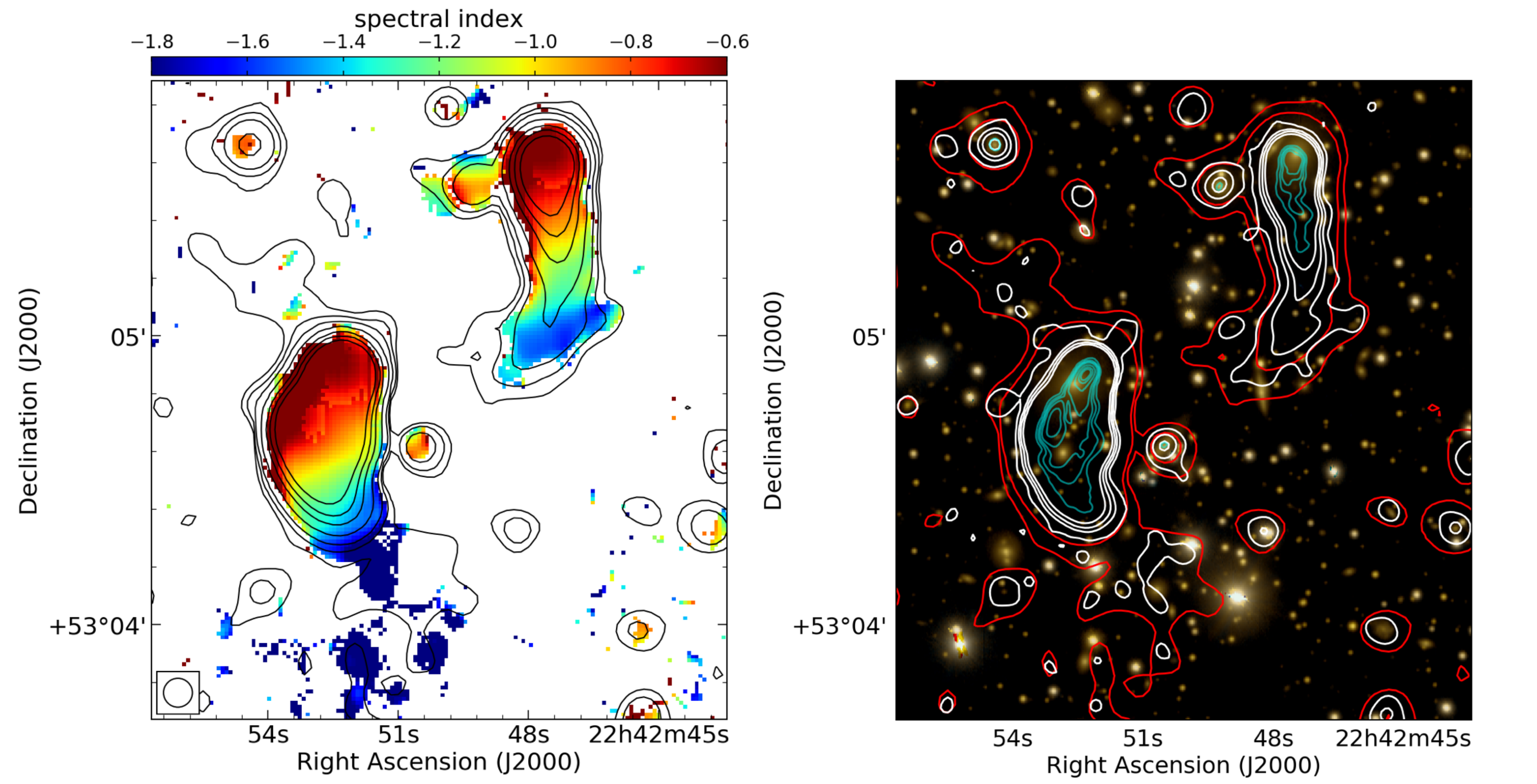}{0.5\textwidth}{(b) sources D (bottom left) and C (top right)}
	}
\medskip
\gridline{
	\fig{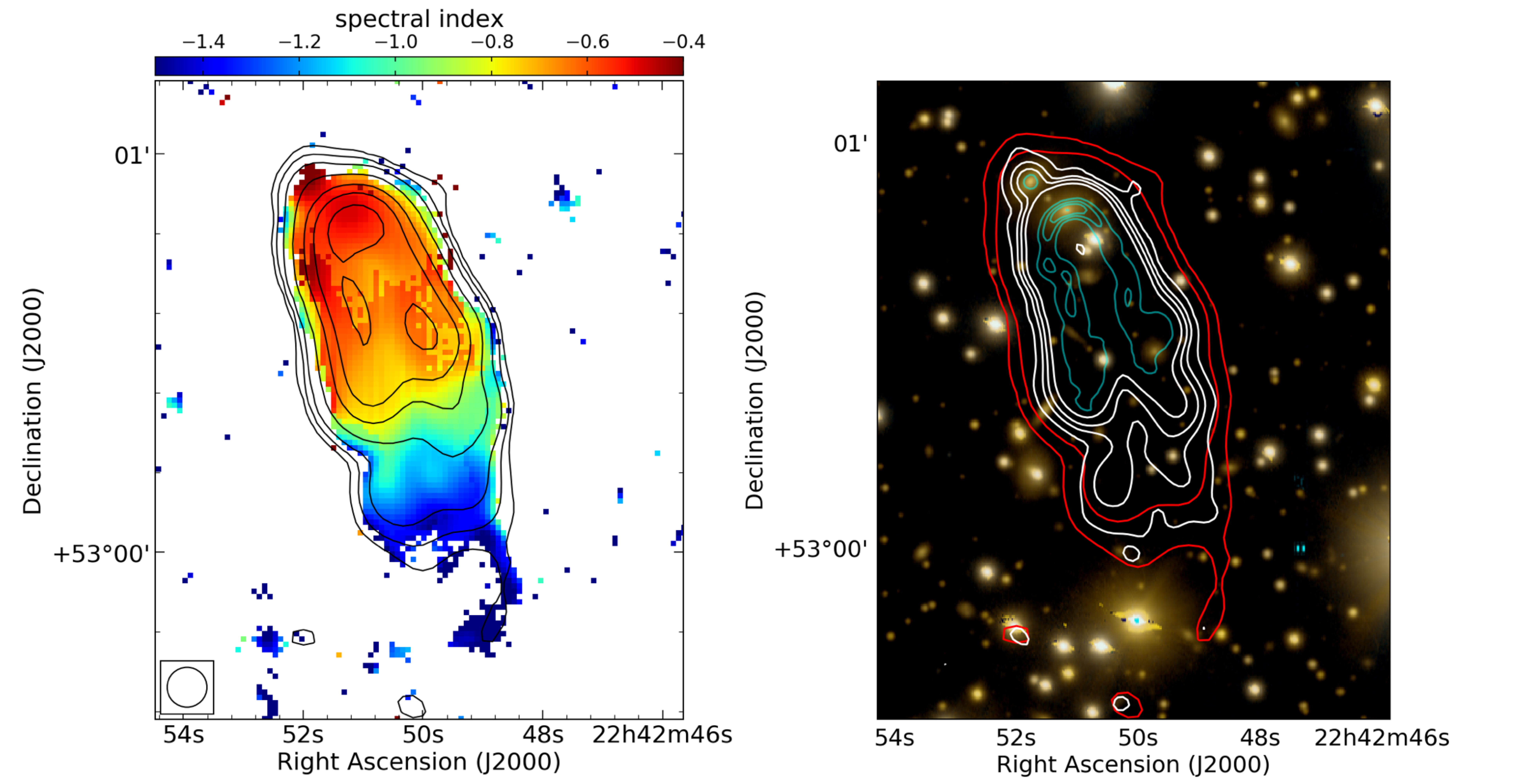}{0.5\textwidth}{(c) source E}
    \fig{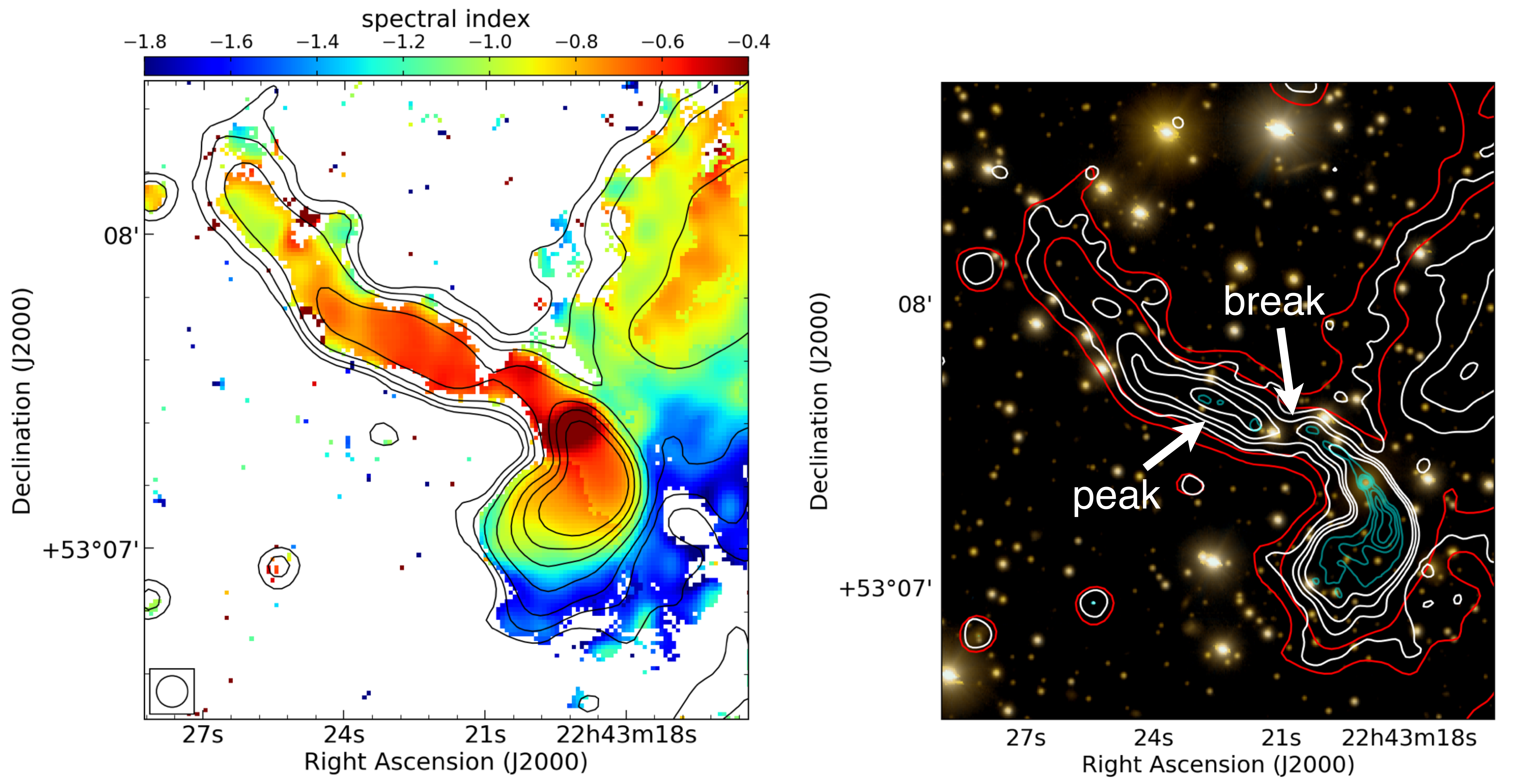}{0.5\textwidth}{(d) source H}
	}
\medskip
\gridline{
	\fig{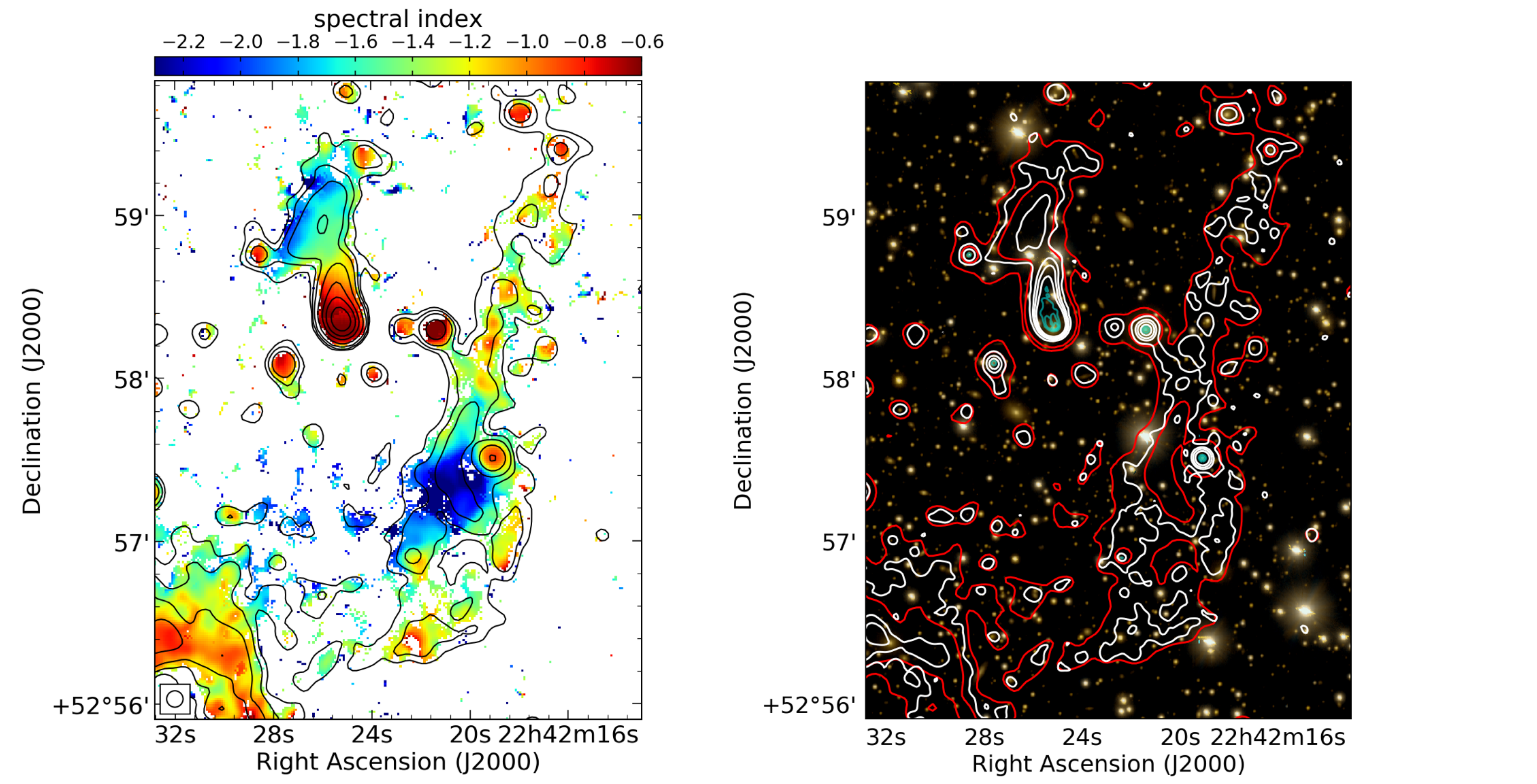}{0.5\textwidth}{(e) sources F (top left) and J (bottom right)}
    \fig{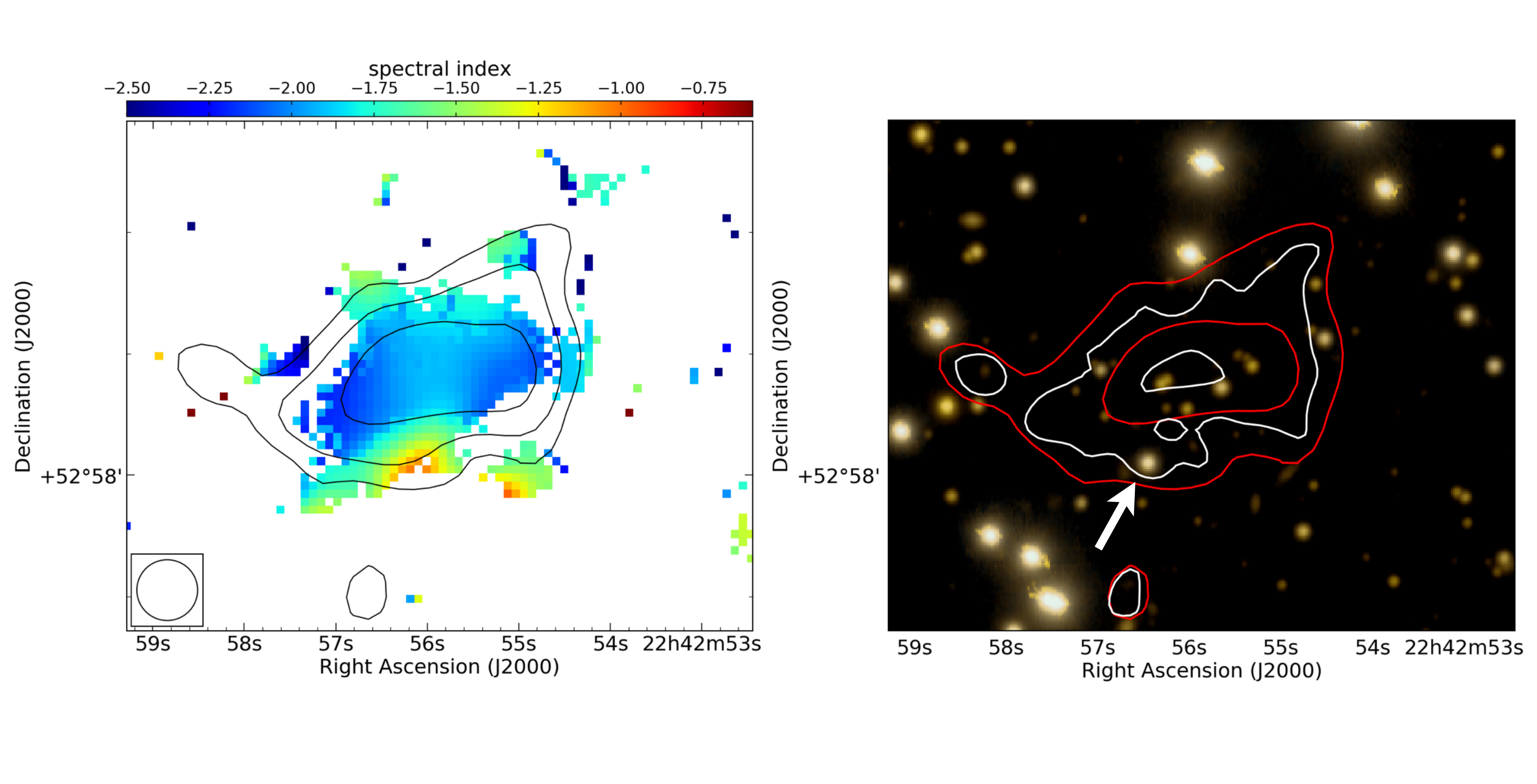}{0.5\textwidth}{(f) source G}
	}
\medskip
\gridline{
	\fig{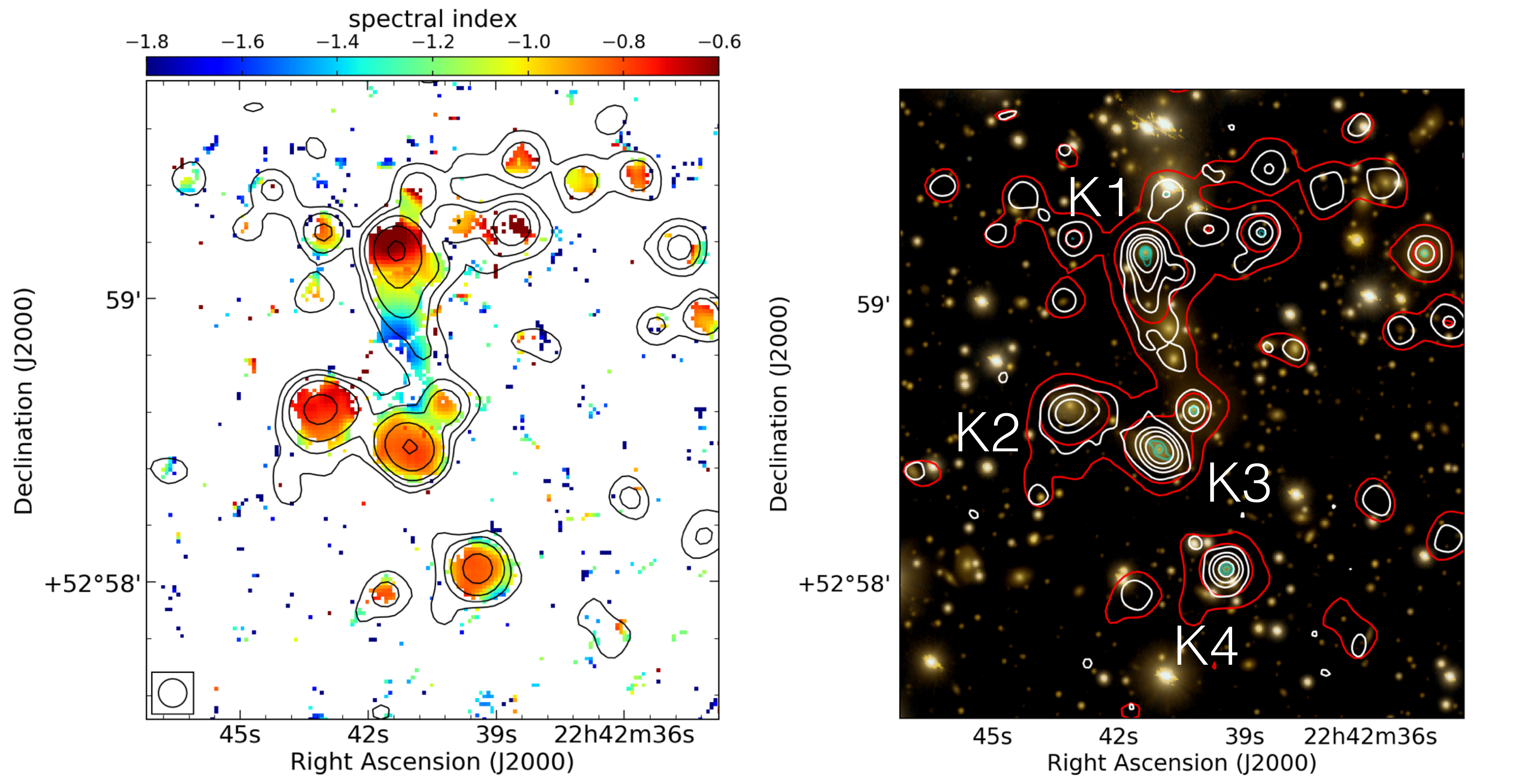}{0.5\textwidth}{(g) sources K1 (top), K2 (left), K3 (right) and K4 (bottom)}
    }
\caption{Tailed radio galaxies in CIZA2242 (for the labels see Fig.~\ref{fig:low_res_comb}). Left panels: $5^{\prime\prime}$ resolution spectral index map with the 1--4 GHz radio contours at the same resolution. Right panels: Subaru $gri$ composite optical images \citep{dawson+15,jee+15} with the $0.8^{\prime\prime}\times0.6^{\prime\prime}$ (cyan) 2--4 GHz, and $2.5^{\prime\prime}$ (white) and $5^{\prime\prime}$ (red) 1--4 GHz radio contours. For both panels, the radio contours are drawn at $3\sigma\times\sqrt{(1, 4, 16, \ldots)}$ levels.}\label{fig:spix_tails}
\end{figure*}

\subsection{Spectral index maps}\label{sec:spix}
We combine our 1--2 and 2--4 GHz VLA images with previous observations performed at 610 \citep{vanweeren+10} and 145 MHz \citep{hoang+17} to obtain  spectral index maps over a wide frequency range. To take into account the differences of each dataset (e.g. interferometer uv-coverage), we produced new radio images for the previous datasets, cutting at a common uv-distance (120$\lambda$). We used uniform weighting to compensate for differences in the uv-plane sampling. The images of each dataset were then convolved to the same resolution, and re-gridded to the same pixel grid (i.e. the LOFAR image). The effective final resolutions and the noise levels are listed in Table \ref{tab:images}.

We obtained the spectral index maps by fitting both first and second order polynomials through the flux measurements. Although the maximum signal to noise ratio (SNR) is given by the former case, the latter one allows us to take into account possible non-negligible deviation from a straight spectrum. We proceeded as follow:
\begin{enumerate}
\item fit with a second order polynomial;
\smallskip
\item determine if the spectral index shows significant curvature (i.e. above the $2\times\sigma$ threshold, with $\sigma$ the uncertainty associated with the second order term);
\smallskip
\item if the second order term is consistent with zero, we redo the fit and use a first order polynomial; otherwise we keep the second order result.
\end{enumerate}

For the spectral index maps,
we accepted the $\alpha$ values only when the associated uncertainty was above a given threshold, i.e. $\sigma_{\alpha}=0.2$ and $\sigma_{\alpha}=0.4$, for the first and second order polynomial fits respectively. In this way we prevent a noisy measurement at a certain frequency from rejecting a possible good fit.

For a first order polynomial fit (i.e. $y=a_0+a_1x$), the spectral index $\alpha$ is defined as the slope of the fit, i.e. $\alpha=a_1$, while the spectral curvature $C$ is defined as:

\begin{equation}
C = -\alpha_{\nu_2}^{\nu_1} + \alpha_{\nu_3}^{\nu_2} 
\end{equation} 
where $\nu_1$ is the lowest of the three frequencies,  $\nu_2$ the middle one and  $\nu_3$ the highest, as described by \cite{leahy+roger98}. In this convention, the curvature is negative for a convex spectrum. Since this curvature definition works only when we have three frequencies, we define the curvature $C$ in the following way \citep{stroe+13}:

\begin{equation}\label{eq:curv}
C =  \alpha_{\rm high} -\alpha_{\rm low}
\end{equation} 
where, in our case, the low frequency spectral index was calculated using the LOFAR and the GMRT observations, and the high low frequency spectral index was calculated using the L- and S-band VLA observations. The uncertainty associated with this measure is:

\begin{equation}
\sigma_C = \sqrt{(\sigma_{\alpha_{\rm low}})^2+ (\sigma_{\alpha_{\rm high}})^2 }
\end{equation} 
where the spectral index uncertainties $\sigma_{\alpha_{\rm low}}$ and $\sigma_{\alpha_{\rm high}}$ were obtained similarly to Eq. \ref{eq:err_spix}.

When the second order polynomial fit (i.e. $y=a_0+a_1x+a_2x^2$) was used, we defined the curvature, $C$, and the spectral index, $\alpha$, as
\begin{align}
C &= a_2 \\
\alpha &=\left . \frac{dy}{dx} \large \right |_{x\equiv\nu={\rm 608~MHz}} = a_1 + 2a_2x \, ,\label{eq:second_order}
\end{align}

respectively. By using this convention, convex spectra are characterized by more negative values of $C$. The uncertainties associated with $\alpha$ and $C$ were obtained via Monte Carlo simulation.

The spectral index maps at $10^{\prime\prime}$ and $25^{\prime\prime}$ of the entire cluster are shown in Figs.~\ref{fig:spix10} and~\ref{fig:spix25}, respectively. The corresponding spectral index uncertainty map at $10^{\prime\prime}$ resolution is presented in Fig.~\ref{fig:spix_error}. 

In the $10^{\prime\prime}$ map (Fig.~\ref{fig:spix10}) a spectral index gradient across the two main relics is seen from the outskirts toward the cluster center. In the northern relic (RN) and in R3, the spectral index values range between $-0.8$ to $-1.7$, across their width. Two flatter spots, detected in the southern part of RN, are associated with two point-like sources. A weaker gradient is also seen in R1 and R3, from east to west (from $-0.7$ to $-1.2$), and in the southern relic (RS), from south to north. A spectral index of about $-0.8$ is measured both on the outskirts of RS1 and RS2, while an average spectral indices of $\approx -1.2$ and $\approx -1.5$ are measured for RS3 and RS4, respectively. No strong gradient is seen for R2, for which we measure steep ($\alpha\approx-1.2$) spectral index values. Relic R5 is barely detected in our spectral index map, because of its low surface brightness in the S-band. For these relics, the spectral index values are about $-0.8$. 

All the radio galaxies show a spectral index gradient, as expected from cluster galaxies, with a steepening from $\alpha\approx-0.6$ in the nuclei to $\alpha\approx-2.0$ in the tails (sources E and K1, left panel in Figs. \ref{fig:spix_tails}(c) and (g), respectively). An exception is  source B, where the spectral index gradient ($\alpha\sim2$) is seen in the south-west direction and not along the nucleus-lobe axis (left panel in Fig. \ref{fig:spix_tails}(a)), supporting a scenario of plasma stripping as consequence of the cluster merger.
A remarkable steep gradient is detected in sources F and J (left panel in Fig. \ref{fig:spix_tails}(e)), the spectral index reaching values lower than $-2.5$. Moreover, for the first time we detect spectral index values 
across the region connecting J1 ($\alpha\approx-0.6$) and the lobe of source J, with values of about $-1.6$ (left panel Fig.~\ref{fig:spix_RS5}). 
Particularly interesting is source H, where the spectral index has a gradient (from $\alpha\approx-0.6$ to $\alpha\approx-1.9$) only in the southern lobe, while it remains quite constant in the northern lobe (from $\alpha\approx-0.6$ to $\alpha\approx-0.9$, left panel in Fig. \ref{fig:spix_tails}(d)). We also note that the spectral index of source G is rather uniform and steep, with an average value of $\sim -1.7$ (left panel in Fig. \ref{fig:spix_tails}(f)). A suggestion of spectral index flattening is seen at the southern boundary of the source, in correspondence of a cluster galaxy which could be a candidate 
optical counterpart of this diffuse emission, although no spectroscopic confirmation has been found so far \citep[Fig. \ref{fig:opt_galaxies}]{dawson+15}.

We made use of the $25^{\prime\prime}$ resolution images (Fig.~\ref{fig:spix25}) to determine the spectral properties of the radio halo, which are consistent with the ones obtained by \cite{hoang+17}. No gradient is seen across the halo, and the spectral indices range between $-1.2$ and $-1.0$.

\section{Discussion}\label{sec:discuss}
The Sausage cluster is a well known double-relic system. Due to its size, regularity, and radio brightness, it offers a unique opportunity to study the particle \mbox{(re-)acceleration} mechanisms at shocks, and the particle aging in the shock downstream region, with a relatively small contribution from projection effects. Indeed, numerical simulation by \cite{vanweeren+11} has shown that CIZA2242 is a binary cluster merger seen very close to edge-on (i.e. $|i| \lesssim10$). 
Here we discuss the relics' morphologies, formation scenarios, and underlying particle (re-)acceleration mechanisms.

\subsection{Radio Mach number estimates}
For the DSA model, there is a relation between the radio injection spectral index $\alpha_{\rm inj}$ and the Mach number $\mathcal{M}$ of the shock \citep[e.g.][]{drury83,blandford+eichler87}:

\begin{equation}\label{eq:mach_number}
\mathcal{M} = \sqrt{\frac{2\alpha_{\rm inj} + 3}{2\alpha_{\rm inj} -1}} \, .
\end{equation}
One usually assumes a standard power-law energy distribution of relativistic electrons just after acceleration\footnote{$\frac{dN(E)}{dE} \propto E^{-\delta_{\rm inj}}$, where $\delta_{\rm inj}=1-2\alpha_{\rm inj}$}. For ``stationary conditions'', with the lifetime of the shock and the electron diffusion time being much longer than the electron cooling time, the integrated spectral index $\alpha_{\rm int}$ is steeper than the injection index by 0.5 \citep{kardashev62}:

\begin{equation}\label{eq:inj_spix}
\alpha_{\rm inj} = 0.5 + \alpha_{\rm int}\, .
\end{equation}

However, the assumption that the timescale on which the shock properties change is much longer than the electron cooling time does not necessarily have to hold for relics \citep{kang15}. For example, for a spherically expanding shock, \citeauthor{kang15} established that the slow decline of the injected particle flux over time could increase the low frequency radio emission downstream away from the shock. This means that $\alpha_{\rm inj}$ calculated by Eq. \ref{eq:inj_spix} can lead to significant errors in the derived Mach number (i.e. 0.2 units in $\alpha$).

Therefore, a more accurate way to compute the Mach number from the radio spectral index, would be to directly measure $\alpha_{\rm inj}$ at the shock front and not use $\alpha_{\rm int}$. 
At the shock front, where the particles have recently been (re-)accelerated, the injection spectral index should be ``flat'', while it should steepen in the downstream region because of synchrotron and IC energy losses. However, directly measuring  $\alpha_{\rm inj}$, requires (i) highly-resolved maps of the downstream cooling region, to avoid the mixing of different electron populations and (ii) minimum projection effects, which means that the merger should have to occur in, or close to, the plane of the sky. Numerical simulations \citep[e.g.][]{vanweeren+11, kang+12} have indicated that for CIZA2242 the projection effects are probably small, at least for the northern relic, with a merger axis angle $|i|\lesssim10^\circ$. Furthermore, the mixing of emission from regions with different spectral ages can be largely avoided thanks to our high-resolution images, as described in Sect. \ref{sec:spix}.

\subsection{Spectral index profiles and color-color diagrams}
To investigate possible differences in the spectral index properties at the shock and the energy losses in the post-shock region, we analyzed the spectral index profile across the radio relics. We extracted flux densities across the width of the relics in narrow annuli spaced by the beam size.
To avoid mixing of emission from regions with different amounts of aging, we used the highest resolution images available for all frequencies, i.e. $5^{\prime\prime}$. 
The flux densities were fitted with a first order polynomial. Following \cite{vanweeren+12}, we did not include the calibration uncertainties (15\% on $S_{\rm LOFAR}$, 5\% on $S_{\rm GMRT}$ and 2.5\% on $S_{\rm VLA}$) in the estimation of the flux density uncertainties, as this would shift measurements in different annuli/regions in the same way.

Another way to investigate the spectral shapes is by means of so-called color-color (cc) diagrams \citep{katz-stone+93, rudnick+katz-stone96, rudnick01}. These diagrams emphasize spectral curvature, since they represent a comparison between spectral indices calculated at low- and high-frequency ranges. In our case, we plot $\alpha_{\rm 150 MHz}^{\rm 610 MHz}$ on the {\it x}-axis and $\alpha_{\rm 1.5 GHz}^{\rm 3.0 GHz}$ on the {\it y}-axis (see bottom panel in Fig. \ref{fig:spix_profileRN}). Moreover, cc-diagrams  are particularly useful to discriminate between different theoretical synchrotron spectral models, and to give constraints on the injection spectrum.

The time evolution of the CR electron energy distribution depends on two parameters:

\begin{equation}\label{eq:sync_ic_losses}
\frac{dE}{dt} = -(\varepsilon_{\rm sync} + \varepsilon_{\rm IC})E^2 \, ,
\end{equation}
where $\varepsilon_{\rm IC}$ is the contribution of the IC energy losses, 
(i.e. $\varepsilon_{\rm IC} \propto B_{\rm CMB}^2$, where $B_{\rm CMB}=3.25(1+z)^2~\mu$G), while $\varepsilon_{\rm sync}$ is the contribution of the synchrotron energy losses, whose parametrization depends on the aging model assumed.
Generally, a JP \citep{jaffe+perola73} aging model is used, which involves 
a single burst of particle acceleration, and a continues isotropization of the angle between the magnetic field and the electron velocity vectors (the so-called pitch angle) on a time-scale shorter than the radiative timescale\footnote{Another aging model is the KP \citep{kardashev62, pacholczyk70} one, but it assumes a constant pitch angle in time.}. Here, the synchrotron losses are described by the $\varepsilon_{\rm sync} \propto B^2$ relation. 
Other models include continuous particle acceleration \citep[CI;][]{pacholczyk70} or a modification to the JP model with a finite period of electron acceleration \citep[KGJP;][]{komissarov+gubanov94}.

One characteristic that makes  cc-diagrams particularly useful is that the shape of the different models only depends on the injection spectral index, while it is independent of the magnetic field value, adiabatic compression/expansion and radiation losses. 

\begin{figure*}%[t]
\centering
\includegraphics[width=\textwidth]{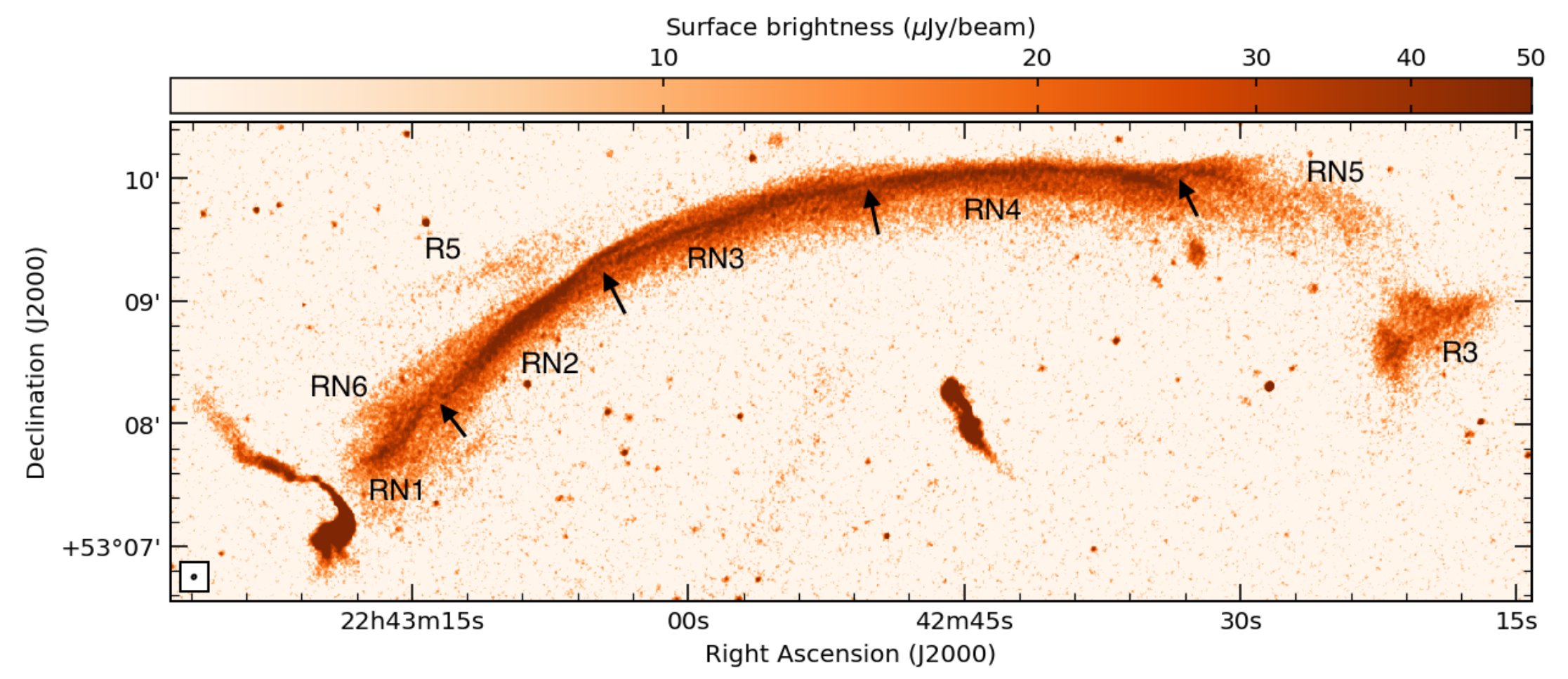}
\caption{Full resolution ($2.1^{\prime\prime}\times1.8^{\prime\prime}$) L-band image zooming in the northern relic. The black arrows indicate the points where the relic breaks into several separate  sheets.}\label{fig:zoomRN_fullLband}
\end{figure*}

\subsection{Northern Relic}\label{sec:RN}
Our high-resolution images reveal that the northern relic 
contains filamentary substructures. We labeled these sheets as RN1, RN2, RN3, RN4, RN5 and RN6 in Fig.~\ref{fig:zoomRN_fullLband}. To avoid mixing of different electron populations, we obtained the spectral index profiles in several sub-sectors (RN1--RN5). We define each sector such that it does not include regions where two sheets overlap, or contain compact sources. 
Despite their different locations, RN1 to RN5 display similar $\alpha_{\rm inj}$ values at the shock, and similar trends in the downstream region (see Table \ref{tab:spix_profiles}).

The injection spectral index for the RN was estimated by considering a narrow region (with a width identical to the beam-size, i.e. the combined red boxes in Fig.~\ref{fig:spix_RN}, bottom panel) following the length of the relic. From this region, we obtain $\alpha_{\rm inj}=-0.86\pm0.05$, corresponding to a Mach number of $\mathcal{M_{\rm N}}=2.58\pm0.17$ (Eq.~\ref{eq:mach_number}). The injection spectral indices of the different filaments in the northern relic, and the corresponding Mach numbers, are reported in Table~\ref{tab:spix_profiles}. We find a good agreement with the X-ray Mach number estimate \citep[$\mathcal{M_{\rm N}}=2.7^{+0.7}_{-0.4}$][]{akamatsu+15}
 and with the previous radio studies ($\mathcal{M_{\rm N}}=2.9^{+0.10}_{-0.13}$, \citealp{stroe+14}; $\mathcal{M_{\rm N}}=2.7^{+0.6}_{-0.3}$, \citealp{hoang+17}).

\begin{table}[h!]
\caption{Spectral indices and Mach numbers for the relics in CIZA2242. A comparison with literature values is shown in columns~5, 6  (radio) and~7 (X-ray).}
\begin{center}
\resizebox{0.75\textwidth}{!}{\begin{minipage}{\textwidth}
\begin{tabular}{lcccccc}
\hline
\hline%\noalign{\smallskip}
Region & res.  & $\alpha_{\rm inj}$ & $\mathcal{M_{\rm radio}}^\star$ & $\mathcal{M_{\rm radio}}^\dagger$ & $\mathcal{M_{\rm radio}}^\ddagger$ & $\mathcal{M_{\rm X-ray}}^\diamond$\\
& ($^{\prime\prime}$) \\ %\times^{\prime\prime}
%& ($^{\prime\prime}$) \\
\hline%\noalign{\smallskip}
RN   & 5 & $-0.86\pm0.05$ & $2.58\pm0.17$ & $2.9^{+0.10}_{-0.13}$ & $2.7^{+0.6}_{-0.3}$ & $2.7^{+0.7}_{-0.4}$ \\
RN1 & 5 & $-0.89\pm0.05$ & $2.47\pm0.14$ \\
RN2 & 5 & $-0.90\pm0.03$ & $2.46\pm0.07$ \\
RN3 & 5 & $-0.89\pm0.04$ & $2.47\pm0.10$ \\
RN4 & 5 & $-0.81\pm0.04$ & $2.72\pm0.16$ \\
RN5 & 5 & $-0.89\pm0.10$ & $2.48\pm0.38$ \\
RN6 & 5 & $-0.84\pm0.16$ & $2.62\pm0.86$\\
RS & 10 & $-1.09\pm0.05$ & $2.10\pm0.08$ & $2.8^{+0.19**}_{-0.19}$ & $1.9^{+0.3**}_{-0.2}$ & $1.7^{+0.4}_{-0.3}$ \\ 
R1    & 10 & $-0.82\pm0.02$ & $2.69\pm0.06$ & 				     & $2.4^{+0.5}_{-0.3}$ & $2.5^{+0.6\ddagger}_{-0.2}$ \\
R4    & 10 & $-0.86\pm0.04$ & $2.57\pm0.12$\\
R5*	& 10 & $-1.06\pm0.21$ & $2.13\pm0.55$ \\
\hline%\noalign{\smallskip}
\end{tabular}
\end{minipage}}
\end{center}
{Note: $^\star$ this work; $^\dagger$ \cite{stroe+14,stroe+13}, for the north and south relics, respectively; $^\ddagger$ \cite{hoang+17}; $^\diamond$ \cite{akamatsu+15}. * Source is only visible in our new deep VLA images, hence we calculated the spectral index (and derived the Mach numbers) only between 1.5 and 3.0~GHz flux density measurements. ** Obtained at different resolutions: $18^{\prime\prime}$ \cite{stroe+13} and $25^{\prime\prime}$ \citep{hoang+17}.}		
\label{tab:spix_profiles}
\end{table}

To investigate possible variations across the length of RN, we calculated the spectral index in individual beam-sized regions at the shock front (red boxes in Fig.~\ref{fig:spix_RN} bottom panel). The results are shown in the top panel in Fig.~\ref{fig:spix_RN} 
and suggest that the eastern part ($\lesssim1$ Mpc) of RN is slightly steeper than the western one. However, given the uncertainties on the spectral indices of the two relic sides, i.e. $\alpha_{\rm inj,\lesssim1Mpc}=-0.89\pm0.05$ and $\alpha_{\rm inj,\gtrsim1Mpc}=-0.81\pm0.08$, this difference is not significant.  Significant spectral index variation are visible on smaller scales (i.e. $\sim 200$~kpc): we measure a flattening of the spectral index around 1600~kpc, and a steepening around 400 and 1000~kpc (see Fig. \ref{fig:RN_avg_spix}). Mach number variations, or different aging trends with the broken-shaped shock surfaces, different amounts of mixing, and magnetic field variation are possible explanations (see Sect.~\ref{sec:filaments}).

\subsubsection{Spectral curvature}\label{sec:curvature}
\begin{figure}%[h!]
\centering
{\includegraphics[width=0.51\textwidth]{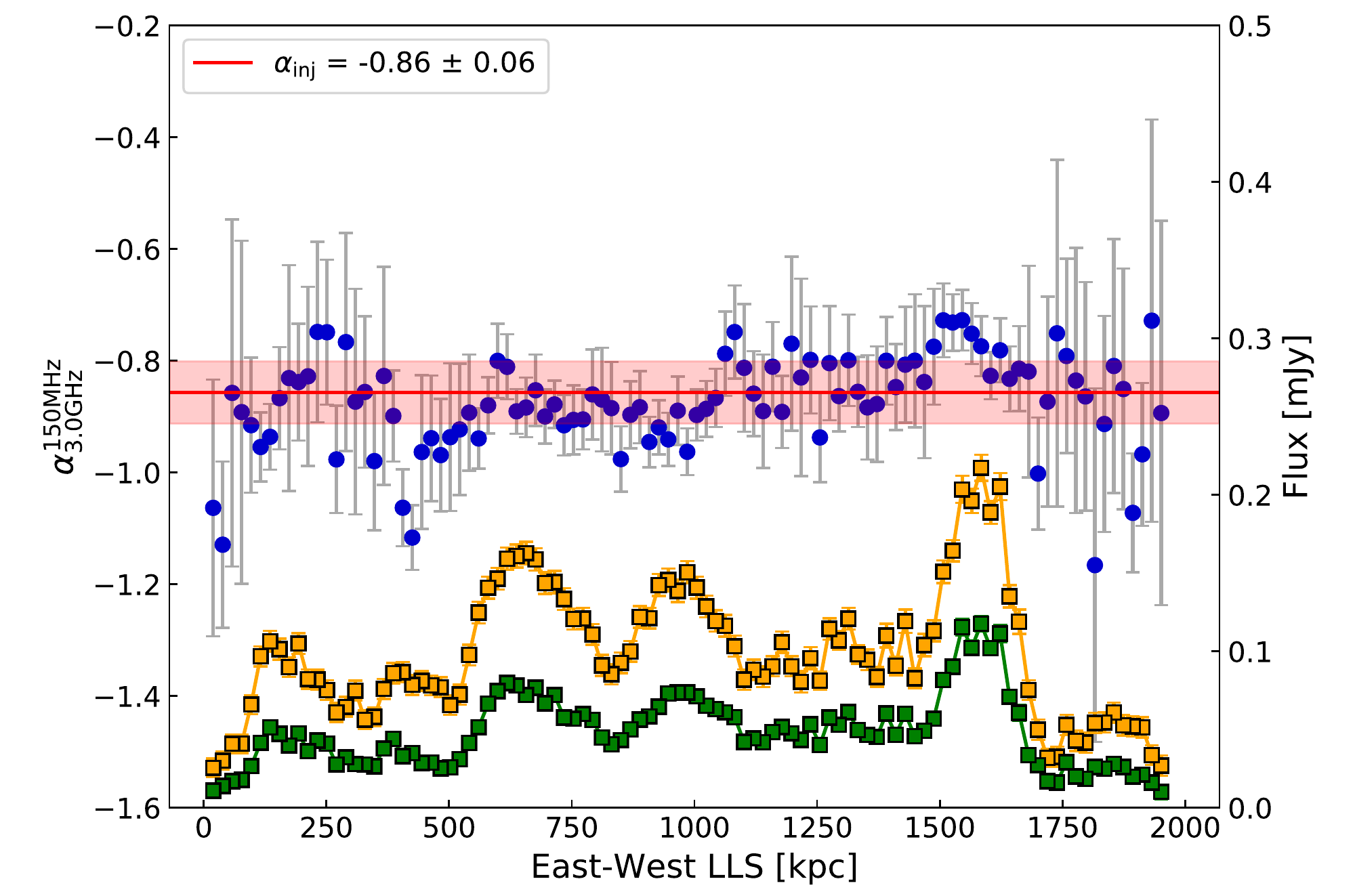}}\\
{\includegraphics[width=0.47\textwidth]{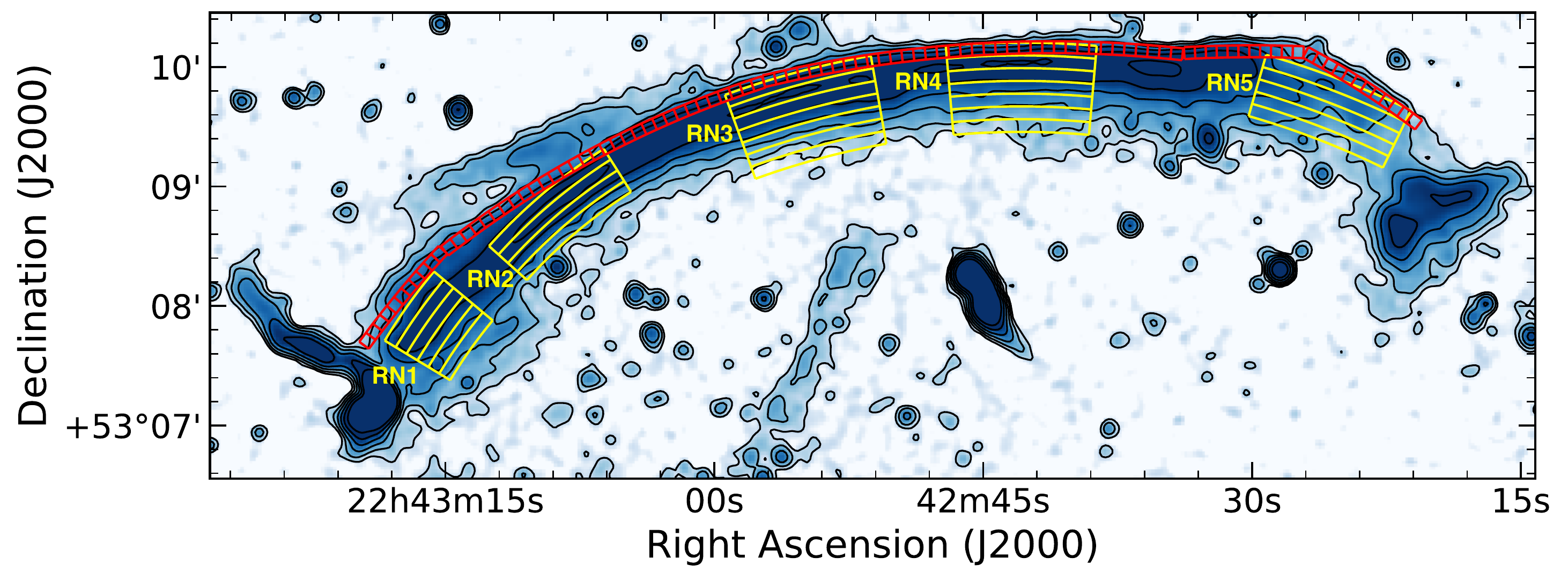}}
\caption{Top panel: East-west spectral index profile (blue circles) obtained with a first order polynomial fit and VLA flux densities at 1.5 and 3.0~GHz (orange and green squares, respectively) at $5^{\prime\prime}$-resolution of the northern relic. The filled red region represents the standard deviation associated with the injection spectral index (red solid line). Bottom panel: 1--4~GHz $5^{\prime\prime}$ resolution image of the northern relic. The red boxes shown the beam-sized region where the spectral indices were extracted. The yellow sectors represent the regions where we extracted the spectral index and curvature profiles and the color-color diagram.}\label{fig:spix_RN}
\end{figure}

\begin{figure}%[h!]
\centering
{\includegraphics[width=0.5\textwidth]{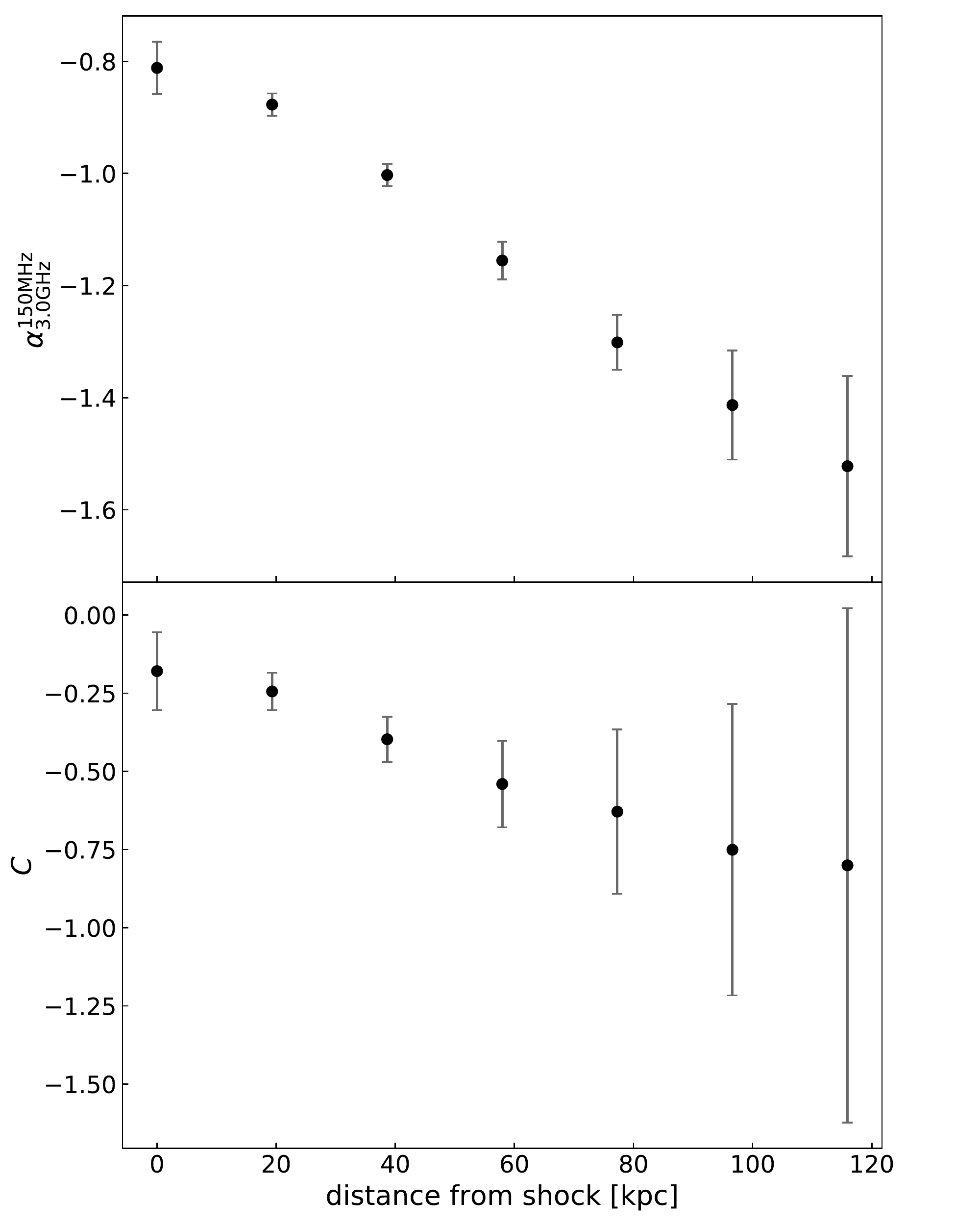}}
{\includegraphics[width=0.5\textwidth]{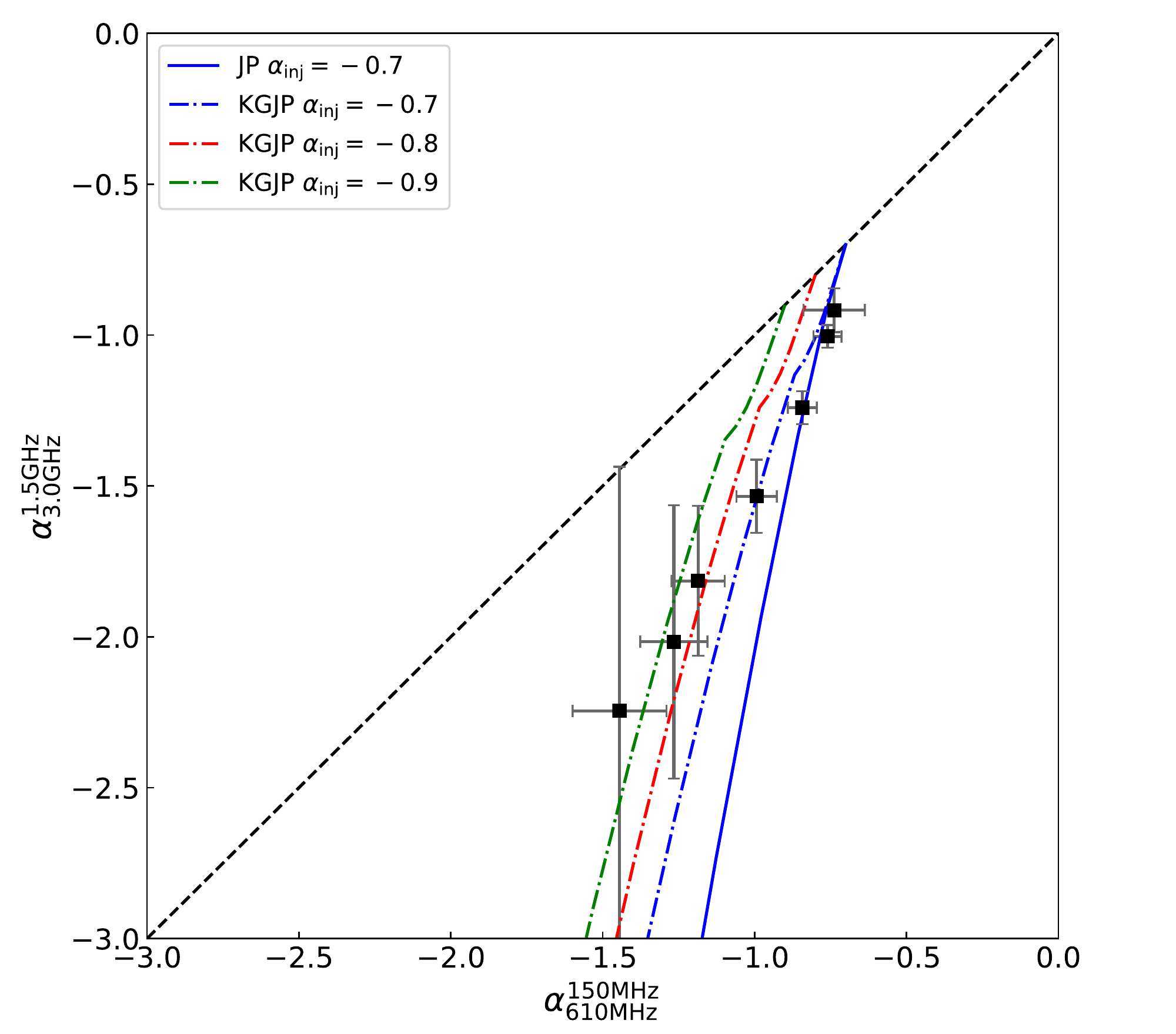}}
\caption{Spectral index profile (top panel), curvature profile (middle panel) and color-color diagram (bottom panel) towards the cluster center obtained via a first order polynomial fits to the flux density measurements of the RN4 filament of the northern relic at $5^{\prime\prime}$ resolution. The color-color diagram shows the JP (blue solid lines) and the KGJP (dot dashed lines) aging models obtained for different injection spectral indices ($\alpha_{\rm inj} = -0.7$ in blue, $\alpha_{\rm inj} = -0.8$ in red, $\alpha_{\rm inj} = -0.9$ in green).}\label{fig:spix_profileRN}
\end{figure}

In Fig.~\ref{fig:spix_profileRN} we display the spectral index and the curvature profiles, and the color-color diagram for the RN4 filament (see the bottom panel of Fig.~\ref{fig:spix_RN} for the sector placement), obtained with a first order polynomial fit. The steepening of the spectral index in the post-shock region (top panel in Fig.~\ref{fig:spix_profileRN}) qualitatively agrees with synchrotron and IC energy losses. The non-negligible curvature in each annulus, which was also detected by \cite{stroe+13}, is better seen in the middle panel in Fig.~\ref{fig:spix_profileRN}, where the curvature profile is shown. All the points in the plot lie below the $C=0$ line, and the convexity of the spectrum ($C<0$, see Eq. \ref{eq:curv}) increases further towards the cluster center. 
In the cc-diagram we also show the JP and KGJP (solid and dash-dotted lines, respectively) aging models. The $\alpha_{\rm 150 MHz}^{\rm 610 MHz}=\alpha_{\rm 1.5 GHz}^{\rm 3.0 GHz}$ line (black dashed) represents a power-law spectral shape, as there are no differences between the spectral index calculated in the low- and the high-frequency range. Our data lie between the $\alpha_{\rm inj}=-0.7$ and  $\alpha_{\rm inj}=-0.9$ curves, assuming a KGJP model with a particle injection time of $0.6\times10^8$~yr for a magnetic field of $B=7~\mu$G.  
The JP model with an injection spectral index of $-0.7$ fits well only for the first three annuli, which correspond to a linear size of about 60 kpc. A possible, at least partly, explanation can be that, as one progresses further downstream, the emission becomes progressively more affected by projection effects, or by actual mixing of different electron populations. Mixing of different electron populations would indeed generate a spectrum closer to the power-law shape, generating the discrepancy with the models (solid and dot-dashed lines in the bottom panel in Fig. \ref{fig:spix_profileRN}). 
This result is not completely a surprise since the northern relic is almost 2~Mpc long and it is likely that it is also  characterized by extended emission in the third dimension. This would easily led to regions with slightly different amounts of spectral aging being contained in a single annulus. Moreover, we note that this effect is stronger  further downstream towards the cluster center. A similar result has also been seen for the Toothbrush cluster \citep{vanweeren+12}. We describe the effects of the projection in Sect. \ref{sec:proj}.

\label{fig:RN_avg_spix}

Interestingly, the injection spectral index directly obtained from the spectral index map is slightly different from the one obtained via the cc-diagram ($\alpha_{\rm inj,2D map}\approx-0.8$ while $\alpha_{\rm inj,model}=-0.7$). The first point in our color-color diagram, at the relic's outer edge, already shows a small amount of spectral curvature indicating that even at this location there is already some mixing of emission with different spectral ages, likely due to projection effects. Extrapolation in the cc-diagram to the $\alpha_{\rm{low}} = \alpha_{\rm{high}}$ line might therefore provide a more reliable estimate than simply taking the flattest spectral index from a map.  This discrepancy between the two methods, shows the limitation of the direct estimation of the injection spectral index from a radio map. 
In this sense, we can only assert that the injection spectral index, estimated from the maps, is equal to or steeper than the real injection spectral index.

\subsubsection{3D model}\label{sec:proj}
When computing the spectral model in Section \ref{sec:curvature}, it was assumed that the relic is caused by a planar shock front perfectly aligned with the line of sight. The actual shock front in a merging cluster, which causes a radio relic, has certainly a much more complex geometry. A better approximation, although still very simplistic, is the assumption that the shock front is shaped like a spherical cap with uniform Mach number.

\begin{figure*}
\centering
\includegraphics[width=\textwidth]{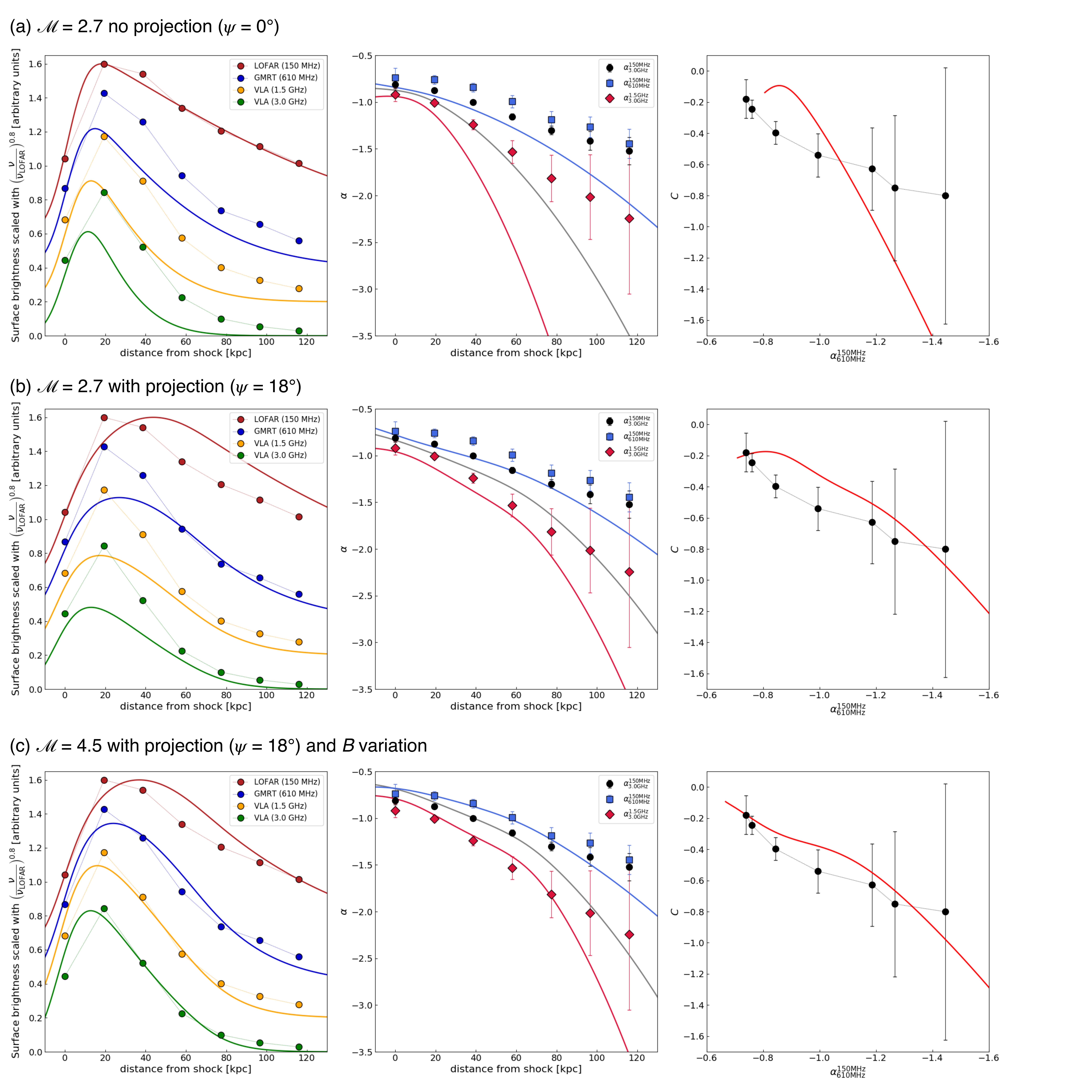}
\caption{Left panels: 150 MHz LOFAR, 610 MHz GMRT, 1.5 GHz VLA and 3.0 GHz VLA flux profiles (red, blue, orange and green solid lines, respectively) 
convolved with the resolution of the observation, i.e. $5^{\prime\prime}$; filled circles display the flux density measurements from RN4 with the same color-code of the solid lines. Central panels: observed and modeled spectral index profiles calculated between 150 MHz and 3.0 GHz (black filled circles and solid lines), 150 and 610 MHz (blue squares and solid lines) and 1.5 GHz and 3.0 GHz (red diamonds and solid lines). Right panels: observed (black filled circles) and modeled (red solid line) curvature profile. The three rows show the different results assuming different Mach numbers ($\mathcal{M}$), spherical projection ($\psi$) and magnetic field variation ($B_0$ and $\log\sigma$). All the models assume a curvature radius $R=1.5$ Mpc and a downstream temperature $T=7$ keV.}\label{fig:3d_models}
\end{figure*}

Before we assess the spherically-shaped shock front, we model the profile of a plane shock, i.e. without including any projection effects. 
We compute the radio emission in the downstream region as described by \cite{hoeft+bruggen07}. The model assumes a power-law spectrum for the electron energy distribution at the location of injection. A proper average over pitch angles was recently included (Hoeft et al. in prep.) to best represent the fast electron velocity isotropization according to  the JP model. The resulting model profiles are convolved with the observation resolution, i.e. $5^{\prime\prime}$. The top row in Figure \ref{fig:3d_models} shows the profiles for our four observing frequencies (left panel), and the spectral index (central panel) and curvature (right panel) profiles assuming a shock aligned with the line of sight ($\psi=0^\circ$), a Mach number of $\mathcal{M}=2.7$ (i.e. $\alpha_{\rm int}=-0.82$) according to the X-ray analysis \citep[see][]{akamatsu+15} and spectral index profile (see Tab. \ref{tab:spix_profiles}), and a homogeneous magnetic field of $B=3~\mu$G. 
As a result, the peak of the flux density profiles is shifted towards to the downstream region of the shock, even though the injection is located at the rising flank of the profiles. 
Although the flux profile at 150 MHz is matched well, this model fails to describe the profiles at the other frequencies, resulting in a much steeper spectral index profile with respect to that observed (central top panel in Fig. \ref{fig:3d_models}). 

A possible fix  might be made by introducing projection effects, which mix the emission coming from the outermost region and what appears to be in the downstream region in the 2D plane. To create a toy model, we adopt a curvature radius of the shock of $R= 1.5$ Mpc and an opening angle of $2\psi=36^\circ$. This implies that, in projection, there is injection up to about 60 kpc in the downstream direction. The middle row in Figure \ref{fig:3d_models} shows the resulting profiles. Evidently, the flux profiles (left panel) get wider and the spectral index profile (central panel) is less steep, at least up to about 70 kpc. However, the profiles are now too much smeared out, showing only a weak peak at about 20 kpc. Moreover, the spectral index profile is still generally too steep.  

As argued in the previous Section, the injection spectral index estimated from the RN4 profile (see top panel in Fig. \ref{fig:spix_profileRN}) is a limit, and the real one can be flatter. To assess the implications, we adopt a quite flat injection index, namely $\alpha_{\rm inj}=-0.6$ ($\mathcal{M}=4.5$)\footnote{Such choice comes from Eq. \ref{eq:inj_spix}, where $\alpha_{\rm int}=-1.1$ given the 150 MHz and 3.0 GHz integrated fluxes \citep[1550,][and 56 mJy, respectively]{hoang+17}.}. 
Moreover, as it has been pointed out for the ``Toothbrush'' cluster \citep{rajpurohit+17}, since we expect that the magnetic field is not homogeneous for a projected Mpc-size shock front, we also included a log-normal distribution for the magnetic field $B$ in the downstream region \citep[see Eq. 7 in][]{rajpurohit+17}. 
The best match of our observed and modeled profiles is given by a magnetic field strength of $B_0=1~\mu$G and scatter $\log\sigma = 1.0$. The resulting profiles are shown in the bottom row in Figure \ref{fig:3d_models}. The flux densities now match reasonably well and the shapes of the profiles are similar to the observed ones. Although, there are still clear deviations for distances larger than 80~kpc. 
The discrepancy between the toy model and the observed profiles might be attributed to a much more complex shape of the shock front, injection spectral index and efficiency variations across the shock front, and a more complex magnetic field distribution including large scale modes. However, our study, combined with the recent results by \cite{rajpurohit+17} on the ``Toothbrush'' relic, seems to support the importance of the combination of projection effects and magnetic field variation in the downstream region to explain the cooling of the electrons.

\subsubsection{Origin of the filaments}\label{sec:filaments}
The existence of filaments/sheets in the northern relic is not completely unexpected, since numerical simulations show that shock surfaces are complex-shaped structures \citep[e.g.][]{vazza+12, skillman+13} and the shock Mach numbers are not constant. In combination with a highly non-linear shock acceleration efficiency \citep[e.g.][]{hoeft+bruggen07} this would lead to morphologically complex radio relics, with the radio emission primarily tracing localized regions with higher Mach numbers.

Another possibility is that the sheets trace magnetic structures. 
If they are produced by strands of strong magnetic fields this would imply strong magnetic fields with coherence lengths of Mpc in cluster outskirts. This would constitute a strong constraint on magnetogenesis models. 
A rough estimate of the amount of magnetic field ($B$) variation necessary to explain the radio flux density variations along the relic's length (top panel in Fig. \ref{fig:spix_RN}) is given by \citep[Eq. 4 in][]{katz-stone+93}:

\begin{equation}\label{eq:mag_var}
\frac{S_1}{S_2} = \left ( \frac{B_1}{B_2} \right )^{1-\alpha} \, ,
\end{equation}
where $\alpha$ is, in our case, the injection spectral index, $-0.86$ (see Tab. \ref{tab:spix_profiles}).  According to Eq.~\ref{eq:mag_var}, to explain a flux variation of $\sim55\%$ (i.e. between the flux density peak at $\sim1600$ kpc and the plateau between $\sim$1100--1400 kpc\footnote{The flux density peak at $\sim1600$ kpc is located in the RN4 filament, while the plateau partly covers both RN3 and RN4. We chose these regions because they represent the strongest flux variation.}, i.e. $S_2\approx0.22$ and $S_1\approx0.1$ mJy at 1.4 GHz, see top panel in Fig. \ref{fig:spix_RN}), we need a magnetic field variation of about $30\%$.
The possible influence of the magnetic field on the radio emission can be addressed via polarization measurements. A Faraday Rotation Measure analysis for CIZA2242 will be presented in an upcoming paper.

\begin{figure}
\centering
{\includegraphics[width=0.51\textwidth]{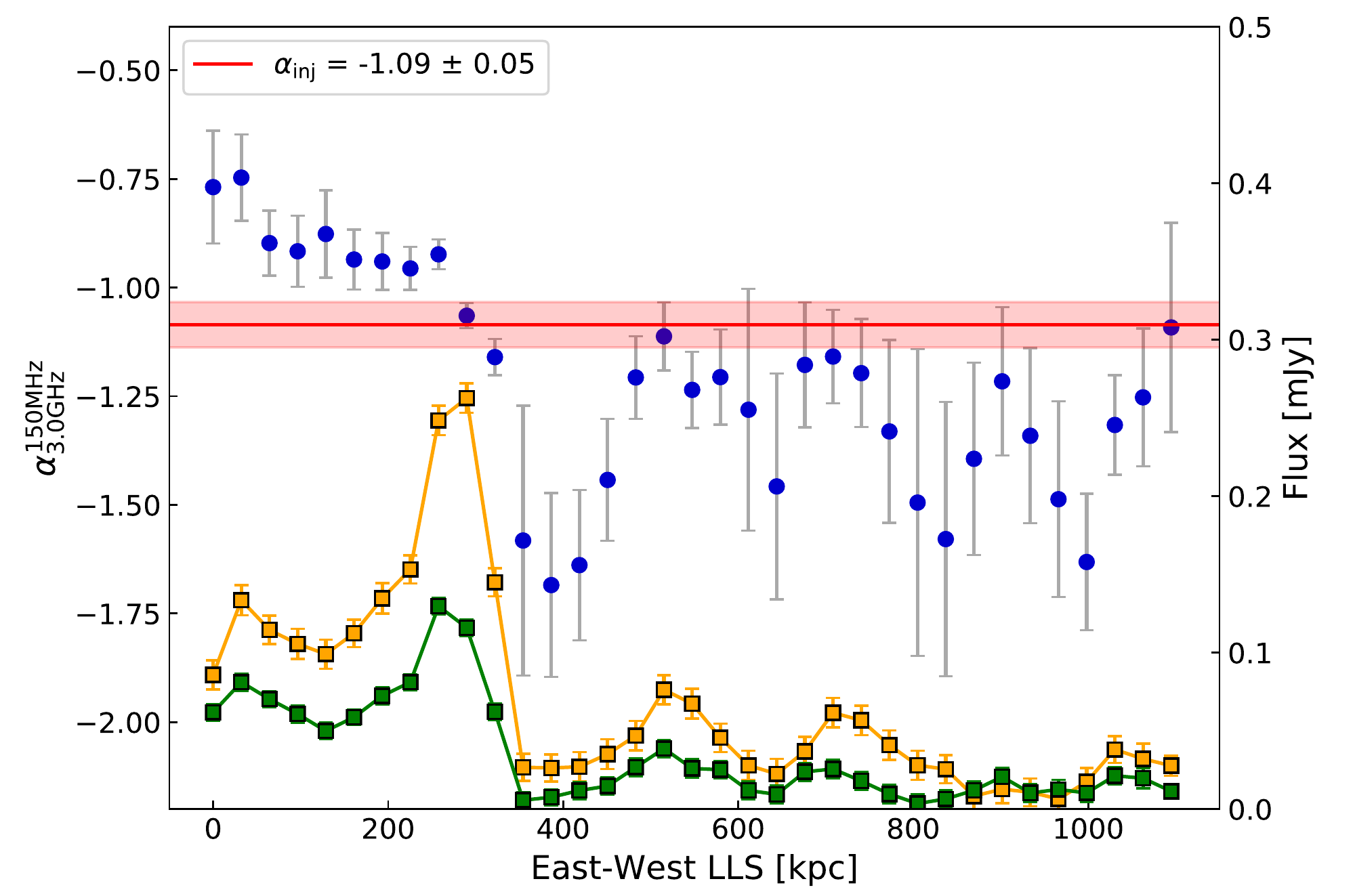}}
{\includegraphics[width=0.47\textwidth]{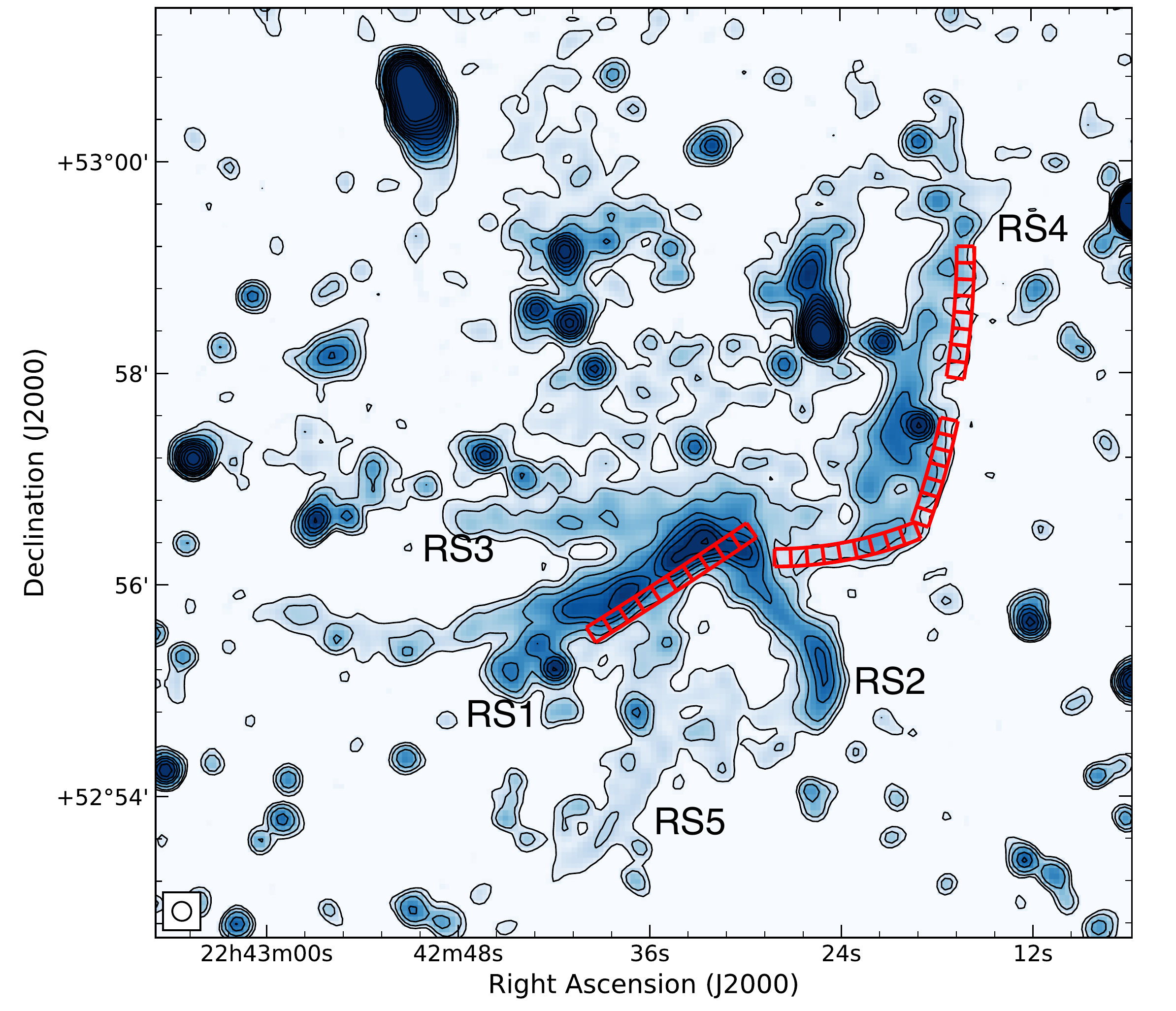}}
\hspace{0mm}
\caption{East-west spectral index profile (blue circles) obtained with a first order polynomial fit and VLA flux densities at 1.5 and 3.0~GHz (orange and green squares, respectively) at $10^{\prime\prime}$-resolution of the southern relic. The filled red region represents the standard deviation associated with the injection spectral index (red solid line). Bottom panel: 1--4~GHz $10^{\prime\prime}$ resolution image of the southern relic. The red boxes shown the beam-sized region where the spectral indices were extracted.}\label{fig:spix_RS}
\end{figure} 

\begin{figure}
\centering
\includegraphics[width=0.5\textwidth]{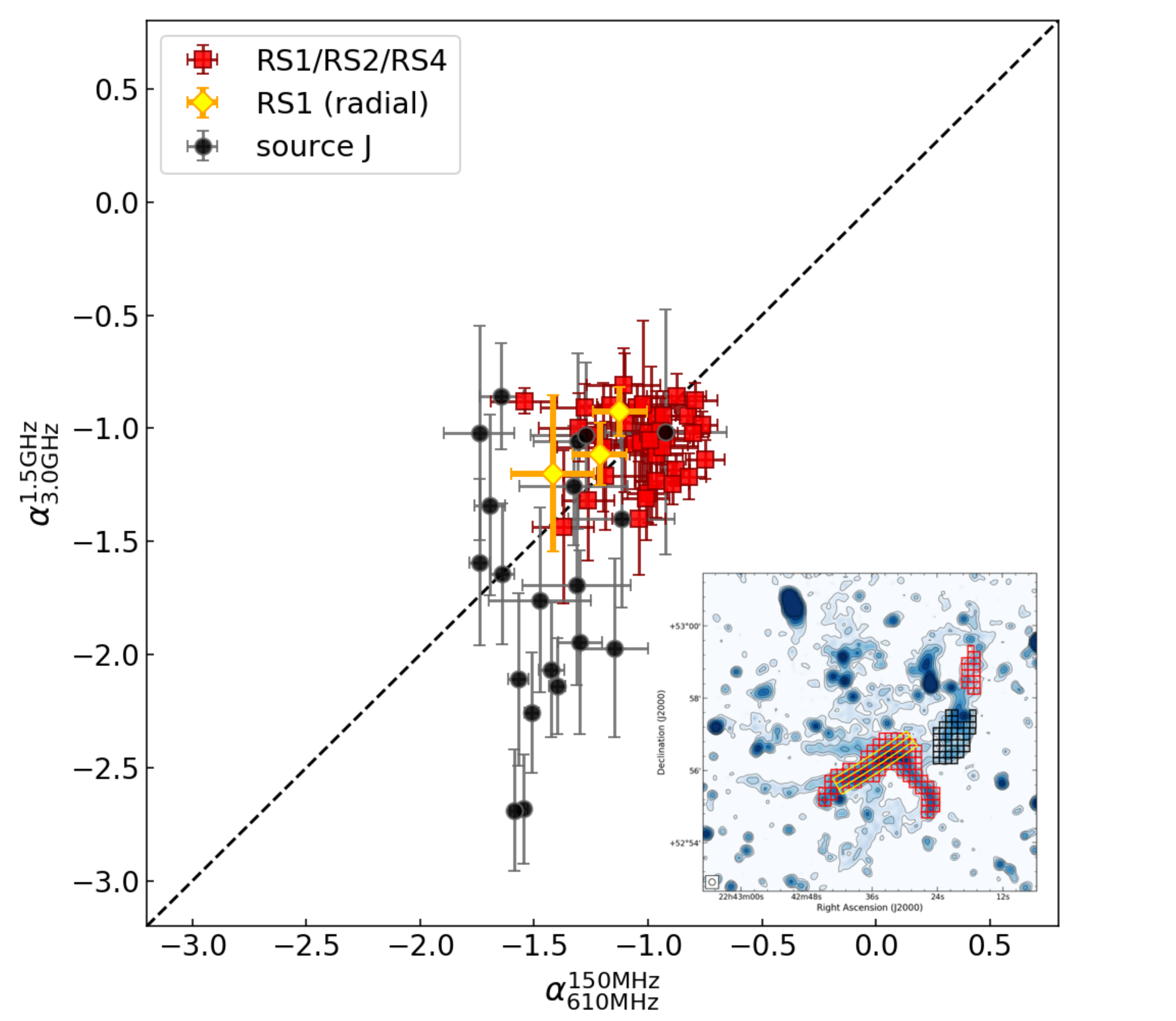}
\caption{Color-color diagram of the southern region of CIZA2242, using the flux densities extracted in  $10^{\prime\prime}\times10^{\prime\prime}$ boxes as shown in the inset in the bottom right corner (same images as the bottom panel in Fig. \ref{fig:spix_RS}). Red squares are obtained from the area covered by the RS1, RS2 and RS4 ``arms'' of the southern relic. The black dots come from the lobe of source~J. The yellow diamonds represent the values obtained from the radial binning (yellow annuli in the inset).}\label{fig:cc_RS}
\end{figure}

Another possibility to generate non-uniform radio emission is in the re-acceleration scenario. In this case, brightness variations might reflect initial variations in the distribution of fossil radio plasma. A non-uniform distribution of fossil plasma is to be expected considering the complex morphologies of tailed radio galaxies in clusters.

Observational evidences of filaments in radio relics have been found in instances, for example in A\,2256 \citep{owen+14} and in 1RXS\,J0603.3+4214 \citep[i.e. the ``Toothbrush'' cluster,][]{rajpurohit+17}. However, with the current data in hand we cannot distinguish between the various scenarios above. It is also possible that the sheet-like, filamentary morphology is due to a combination of two or more effects.

\begin{figure*}
\centering
\includegraphics[width=\textwidth]{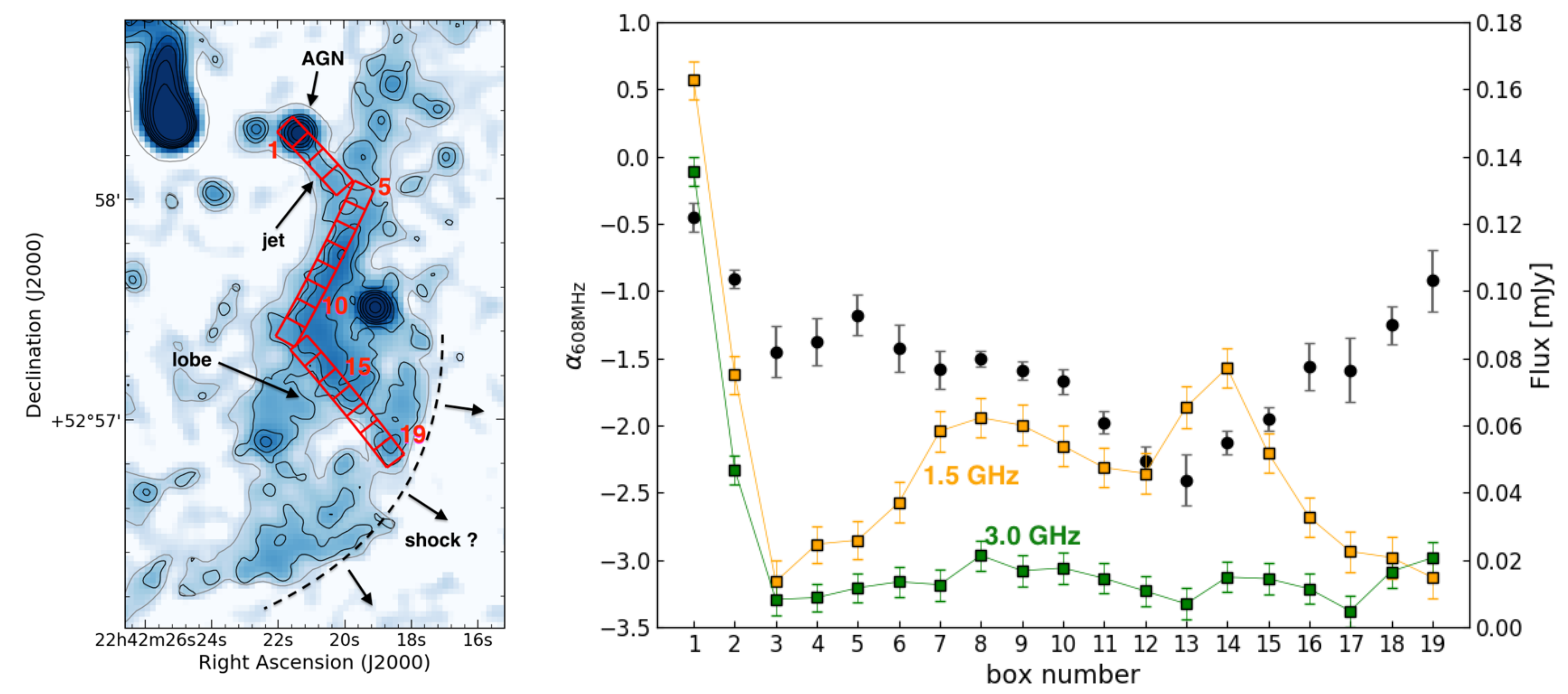}
\caption{Left panel: 1--4~GHz $5^{\prime\prime}$ resolution image of source J. We overlay the $5^{\prime\prime}$ (gray) and $2.5^{\prime\prime}$ (black) resolution radio contours, at $3\sigma$ and $3\sigma\times\sqrt{(1, 4, 16, \ldots)}$, respectively. Right panel: Spectral index profile at 608 MHz, derived by a second order polynomial fit (Eq. \ref{eq:second_order}), across source~J obtained from the red boxes in the left panel. The yellow and green squares trace the VLA L- (1.5~GHz) and S-band (3.0~GHz) flux densities across the radio tail, respectively.}\label{fig:spixJ}
\end{figure*}

\subsection{Southern Relic}\label{sec:RS}
Our low-resolution images ($>10^{\prime\prime}$) show that the southern relic has a complex morphology. Unlike RN, which is characterized by a single arc-like shape at low resolution, this relic is formed by five ``arms'' (see Fig.~\ref{fig:labels}). This complex morphology does not allow us to produce a clear spectral index profile. 
Thus, we only calculated the radio Mach number from the injection spectral index, following the same procedure as carried out for RN, avoiding the compact sources. We did that at $10^{\prime\prime}$ resolution, a compromise between resolution and SNR.
We follow the relic from RS1 to RS4, including the lobe of source~J (see the red boxes in the bottom panel in  Fig.~\ref{fig:spix_RS}). From the region given by the combination of these boxes, we obtain an injection spectral index of $\alpha_{\rm inj}=-1.09\pm0.05$, which corresponds to a Mach number of $\mathcal{M}=2.10\pm0.08$.
However, a clear variation of spectral index values from east to west is seen (top panel in Fig.~\ref{fig:spix_RS}). We measure flatter $\alpha$ values at $\rm{LLS_{EW}}\lesssim350$~kpc ($\alpha_{\rm inj,\lesssim350 kpc}=-0.93\pm0.08$ and $\mathcal{M}=2.36\pm0.18$), and steeper values at  $\rm{LLS_{EW}}\gtrsim350$~kpc ($\alpha_{\rm inj,\gtrsim350 kpc}=-1.36\pm0.08$ and $\mathcal{M}=1.19\pm0.05$). We notice that the steepest spectral index values are coincident with the location of source~J. Around 400 kpc, we measure the steepest spectral index values, coincident with the end of the tail of source~J (see Fig.~\ref{fig:spix25}). 

X-ray observations performed with {\it Suzaku} \citep{akamatsu+15} found a Mach number for RS of $1.7^{+0.4}_{-0.5}$. Although this value is comparable within the uncertainties to our radio spectral index derived Mach number, the injection spectral index is affected by the presence of the embedded source~J, which is characterized by a very steep spectral index (see Fig.~\ref{fig:spix25}). The actual injection spectral index might thus be somewhat flatter.

The spectral index map shows a spectral index steepening from the relic outskirts towards the cluster center (Fig. \ref{fig:spix_RS5}). 
To investigate whether this region is characterized by curved spectra, like RN, we created a cc-diagram using the flux densities extracted in $10^{\prime\prime}\times10^{\prime\prime}$ boxes (i.e. the beam size) covering RS1, RS2, RS4 and the lobe of source J (see the red and black boxes, respectively, in the inset in the bottom right corner in Fig. \ref{fig:cc_RS}). To keep the values with the best SNR, we selected only flux densities above a threshold of $4\sigma_{\rm rms,\nu}$ (where $\sigma_{\rm rms,\nu}$ the noise of each radio map, see Tab. \ref{tab:images}). Moreover, we also created a cc-diagram dividing the RS1 ``arm'' in $10^{\prime\prime}$-wide annuli (yellow sector in the same inset).
Interestingly, we found that all the values for the RS1, RS2 and RS4 regions (red squares in Fig. \ref{fig:cc_RS}), as well as the radial distribution (yellow diamonds in the same Figure), fall along the $\alpha_{\rm 150 MHz}^{\rm 610 MHz}=\alpha_{\rm 1.5 GHz}^{\rm 3.0 GHz}$ line. 
This is somewhat surprising result as curved spectra would be expected due to aging. These results suggest that that (i) at each location, there is an actual power law which represents the steady-state solution to the local shock strength, and that shock strength varies throughout the region, or (ii) there is at each location a very broad range of magnetic fields and/or electron loss histories which smear out the underlying spectra, and make them close to power-law. It remains an open question, though, why different power laws at different locations are present for this relic. 

\subsection{A case for fossil plasma re-acceleration?}\label{sec:fossil_reacc}
There are several issues with the DSA model for radio relics. For some relics, the required fraction of energy flux through the shock surface  that needs to be transfered to CR electrons to explain the observed radio power is too large \citep[e.g.][]{vazza+14,vanweeren+16}. The absence of relic emission at some shocks \citep[e.g.][]{shimwell+14} is also puzzling. In addition, in a few cases, discrepancies have been found between the Mach numbers measured from X-ray observations and those calculated from the radio spectral index \citep[e.g.][]{vanweeren+16}. The re-acceleration of fossil electrons would address some of these problems.

CIZA2242 represents an interesting case to study the interplay between the diffuse radio emission and the tailed radio galaxies.
We propose that Source~J, which is completely embedded in the southern relic RS, presents a possible case of shock re-acceleration because: (i) there is a morphological connection between the tailed radio galaxy, source~J, and the relic. We find that source J consists of an active radio core (J1) and a single jet that ends up in a lobe-like structure (right panel in Fig.~\ref{fig:spix_tails}(e)); (ii) A {\it Suzaku} analysis revealed the presence of a temperature jump roughly at the location of RS \citep{akamatsu+15}; finally, (iii) the spectral index map displays a steepening from the radio core (J1) through the radio lobe and then a flattening at the cluster shock (right panel in Fig.~\ref{fig:spix_tails}(e)). Also, the cc-diagram for the lobe of source J (black dots in Fig. \ref{fig:cc_RS}) shows values both on and below the $\alpha_{\rm 150 MHz}^{\rm 610 MHz}=\alpha_{\rm 1.5 GHz}^{\rm 3.0 GHz}$ line. Radio galaxy's lobes are indeed expected to have curved spectra, described by a JP model (or modification thereof), since electrons experiment aging going further from the AGN. Then if the shock passes through the lobe it could re-accelerate or re-energize the electrons. In this case, electrons with steeper spectra than the ``critical'' spectral index associated with the shock will be flatten out and ``moved'' towards the power law line.
Such flattening has been observed recently in the Abell\,3411-3412 system \citep{vanweeren+17a}, MACS\,J0717.5+3745 \citep{vanweeren+17b} and Abell\,1033 \citep{degasperin+17}.

\begin{figure}
\centering
{\includegraphics[width=0.5\textwidth]{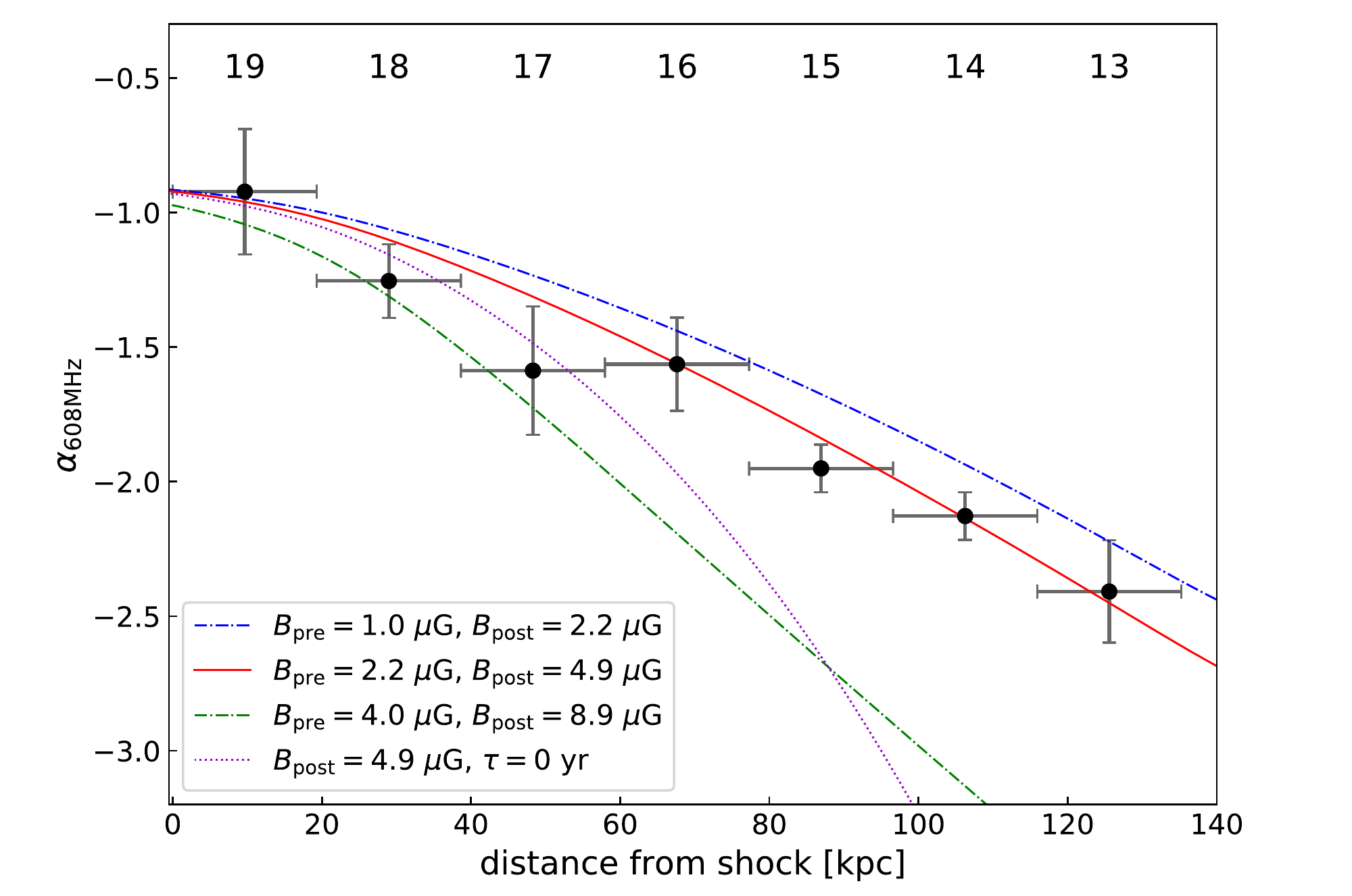}}
{\includegraphics[width=0.5\textwidth]{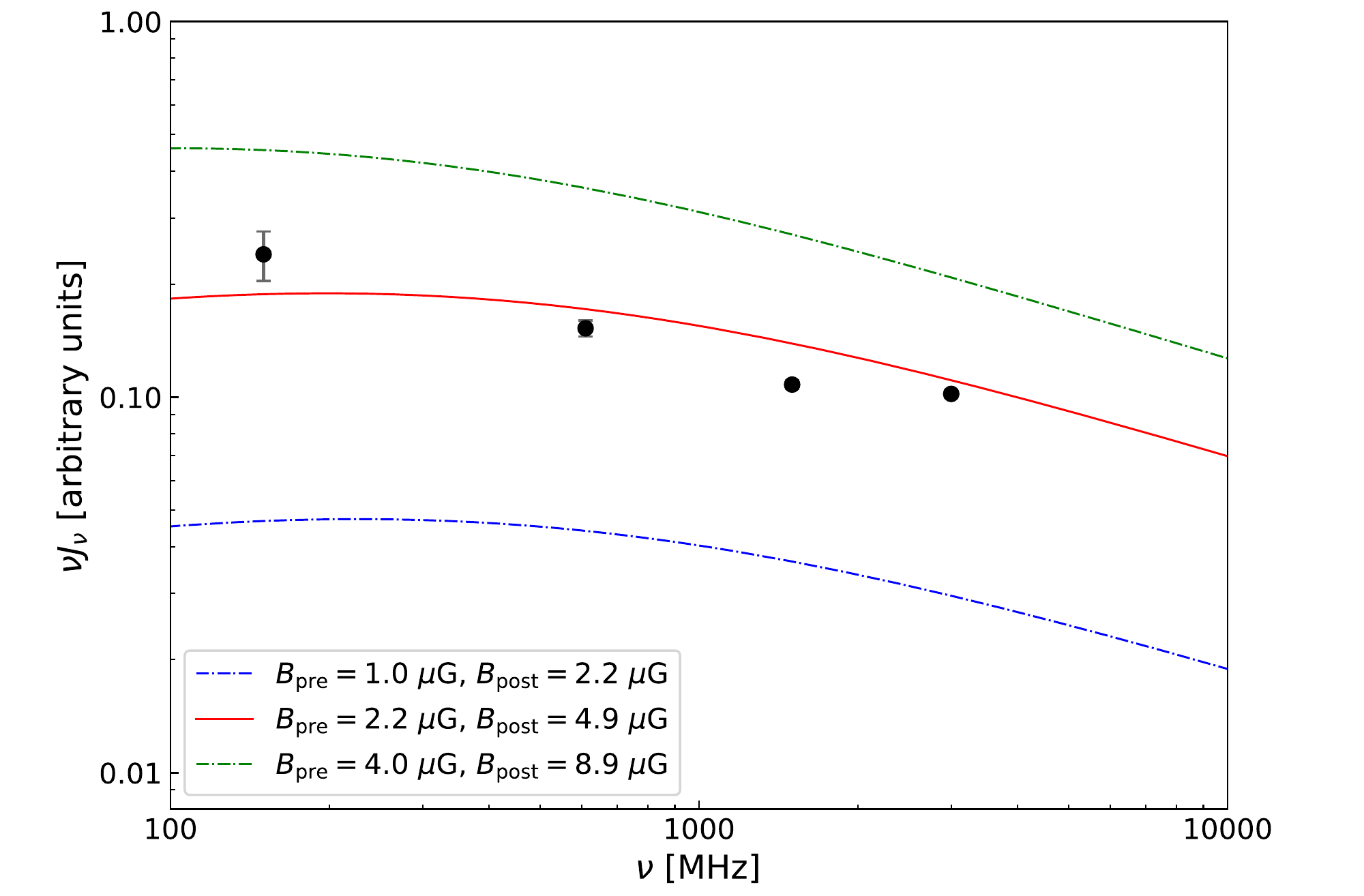}}
\caption{Top panel: Spectral index profile calculated at 608 MHz with a second order polynomial fit (Eq. \ref{eq:second_order}) from the shock position to the region just before the re-acceleration (boxes from 19 to 13 in the left panel in Fig. \ref{fig:spixJ}). Bottom panel: Volume-integrated spectrum of the shock region, obtained by assuming a constant pre-shock electron density distribution.
Lines display models for different pre- and post-shock magnetic fields as indicated in the legend of the figures.}\label{fig:spixJ_zoom}
\end{figure}

\subsubsection{Modeling the spectral index change with re-acceleration}

Following \cite{kang+17}, we attempt to model the observed spectral index change from the lobe of source~J to the relic RS\footnote{See \cite{kang+ryu15} for the geometry of the cloud containing the fossil electrons.}. We assume a region of fossil plasma that is run over up by a spherically expanding shock propagating through the cluster outskirts. To obtain the spectral index distribution across the relic, we  extract the flux densities at $5^{\prime\prime}$ resolution, following the shape of the tail (left panel in Fig.~\ref{fig:spixJ}). 
We consider an energy spectrum of the fossil electrons following a power-law distribution with exponential cutoff, of the type: 

\begin{equation}\label{eq:agn_reacc}
f(\gamma_e)\propto\gamma_e^{-s}\exp{\bigg [-\bigg ( \frac{\gamma_e}{\gamma_{e,c}} \bigg )^2 \bigg]} \, ,
\smallskip
\end{equation}
where $s= 1-2\alpha$ is the injection index from the AGN and $\gamma_{\rm e,c}$ the cutoff Lorentz factor. We use a typical AGN injection spectral index, i.e. $\alpha=-0.75$ \citep[e.g.][]{mullin+08}, corresponding to $s=2.5$. The spectral index where the tail meets the relic (box 13), i.e. $\alpha=-2.4$ at 608 MHz (right panel in Fig.~\ref{fig:spixJ}), gives a cutoff Lorentz factor of $\gamma_{\rm e,c}=4.2\times10^3$. This parameter mimics the steepening of the fossil plasma along the radio tail, and is expected to change with distance from the AGN. We set the post-shock temperature to $9.0\pm0.6$ keV, based on the {\it Suzaku} data \citep{akamatsu+15}, and assume an ``extension angle'' of $\psi=12^\circ$ \citep{vanweeren+17a}. We also included post-shock turbulent re-acceleration with $\tau_0=10^8$ yrs \citep[for details see][]{kang+17}, which prevent the faster steepening behind the shock (purple dotted line in the top panel of Fig. \ref{fig:spixJ_zoom}). The pre-shock magnetic field was assumed to be 2.2~$\mu$G, similar to the magnetic field in the lobes of radio galaxies, leading to a post-shock value of 4.9~$\mu$G.

The post-shock profile of the radio spectral index depends mainly on the shock Mach number and the magnetic field strength (top panel Fig. \ref{fig:spixJ_zoom}). 
Our best-fit model reproduces the spectral change from box 13 (i.e. just before the re-acceleration) to box 19 (i.e. immediately after re-acceleration behind the shock) by assuming a sonic Mach number of $\mathcal{M}_s=2.4$, corresponding to a pre-shock temperature of 3.4 keV\footnote{$\frac{T_{\rm post}}{T_{\rm pre}} = \frac{5\mathcal{M}^4+14\mathcal{M}^2-3}{16\mathcal{M}^2}$, see \citep{landau+lifshitz59}.}. Such Mach number is chosen to match the radio spectral index $\alpha_{\rm 608~MHz} \approx -0.9$ at the relic edge (box 19 in Fig. \ref{fig:spixJ}). This value is not consistent with the {\it Suzaku} analysis (i.e. $\mathcal{M}=1.7^{+0.4}_{-0.3}$ and $T_{\rm pre}=5.1^{+1.5}_{-1.2}$ keV, respectively), although the Mach number in the latter case could be underestimated because of instrumental limitations. 
It is also true, though, that since the fossil electrons are expected to cool due to Coulomb and synchrotron/IC losses advecting away from the AGN core, assuming a constant value for $\gamma_{\rm e,c}$ is a simplification.
However, $\gamma_{e,c}$ would not change significantly over $\sim100~kpc$ scales, i.e. the length of the re-acceleration region, since the radiative cooling time scale for electrons with $\gamma < 4.2\times10^3$ is $t_{\rm cool} > 2.2\times 10^8$~yr, which is much longer than the shock crossing time (i.e. $\approx 6.7\times 10^7$~yr). Additionally, it has been shown that changes in the $\gamma_{\rm e,c}$ value do not affect significantly the energy spectrum of post-shock electrons, since the re-acceleration process ``erases'' most of the spectral information from the seed fossil particles \citep[see also Supplementary Figure 10 in][]{vanweeren+17a}. 
Last but not least, the complex radio tail morphology with a superposition of electron populations could also be the reason of the $\mathcal{M}_{\rm radio}$/$\mathcal{M}_{\rm X-ray}$ discrepancy.

For the re-acceleration, the radio flux density profile across the relic's width and the integrated spectrum also depend on the spatial variation of fossil electrons. 
Generally, the radio emission is expected to be the brightest at the approximate shock location, while it fades downstream of the shock because of energy losses. Instead, we find an  opposite trend, with the highest fluxes decreasing towards the shock front (see left panel Fig.~\ref{fig:spixJ}, from box 13 to 19). 
However, the model can reproduce reasonably well the observed volume-integrated spectrum (red solid line in the bottom panel of Fig.~\ref{fig:spixJ_zoom}). 
Likely, the shock downstream region is ``contaminated'' by radio plasma from the lobe of source~J, which is embedded in the relic. The failure to accurately predict the flux profiles in the shock downstream region highlights the difficulties of modeling complex AGN lobe emission and its interactions with a shock, in addition to particle re-acceleration in the same general region.

\subsection{The relation between tailed radio galaxies and diffuse cluster sources}
As discussed above, RS, or at least part of RS, seems to be related to the tailed radio source~J. However, this could not be the unique case in CIZA2242, which contains a number of complex shaped relics (R1, R2, I, R3, and RS) and tailed radio sources (sources C, D, F, and H). In this section, we speculate that some of the other diffuse relics are also related to AGN fossil plasma that has been revived, either by re-acceleration or solely by adiabatic compression.

The toroidal morphology of R2, similarly to the ones found in \cite{slee+01}, and the large attached filament I (see Fig. \ref{fig:low_res_comb}) could be the result of a fossil plasma lobe/tail that has been compressed by a merger shock wave. Source G could be another example of an old radio galaxy lobe which has been revived by a merger shock. Additional evidence of revived fossil plasma is provided by the several tailed radio sources whose tails suddenly brighten (e.g. source H), suggesting that some re-energizing of the lobe electron population due to adiabatic compression and/or re-acceleration \citep[e.g.][]{pfrommer+jones11,cuciti+17,degasperin+17} might have taken place, and implying we are witnessing the birth of other radio relics or radio ``phoenices''. Moreover, recent numerical simulations \citep{jones+17} show that in the presence of a ``cross wind'', resulting from a passing shock, jet plasma could be revived and displaced, producing the ``bottle-neck'' morphology we see for the radio tails C, D and F.

CIZA2242 thus seems to display the entire range of evolutionary stages of AGN fossil plasma and relic formation, from classical NAT/HT sources (E and K1), to currently active tailed radio sources with signs of revived lobes (C, D, F), active tailed source connected to a relic (J, RS), patch of fossil plasma (G), and large Mpc-sized complex shaped diffuse sources (RS, R2-I).
Whether fossil plasma and re-acceleration are required to produce the main northern relic remains an open question though. Despite the fact the formation scenario for the northern relic remains unclear, our observations do imply that to produce realistic models of diffuse cluster radio emission, fossil plasma will need to be included in simulations.

\section{Conclusions}\label{sec:concl}
We have presented deep 1--4 GHz VLA observations of the galaxy cluster CIZA\,J2242.8+5301 ($z=0.1921$). This cluster is one of the best examples of a merging system, hosting two Mpc-size relics and several other diffuse radio sources. Our images reached a resolution of $2.1^{\prime\prime}\times1.8^{\prime\prime}$ and $0.8^{\prime\prime}\times0.6^{\prime\prime}$ and a noise level of $3.8~\mu$Jy~beam$^{-1}$ and $2.7~\mu$Jy~beam$^{-1}$ at 1.5 and 3.0~GHz respectively. These observations were combined with existing GMRT 610 MHz and LOFAR 150 MHz data \citep{vanweeren+10,stroe+13,hoang+17} to carry out a radio continuum and spectral study of the cluster. 
The main results of our study are summarized below:

\begin{itemize}
\item[$\bullet$] The high-resolution images reveal that the northern relic (RN) is not a continuous structure, but is broken up into several filaments with a length of $\sim 200-600$ kpc each.
Moreover, we detect and characterize additional radio emission north of RN, labeled as R5.
\medskip

\item[$\bullet$] In agreement with previous studies, we find a trend of North-South spectral steepening for the northern relic, indicative of IC and synchrotron energy losses in the downstream region of a shock. We measure an injection spectral index at the shock front from the radio map of $-0.86\pm0.05$, corresponding to a Mach number of $\mathcal{M}=2.58\pm0.17$. Although this value is consistent with the Mach number estimated from X-ray studies, a color-color diagram shows a small amount of spectral curvature at the relic's northern edge indicating mixing of emission with different spectral ages, possibly due to projection effects. This suggests that the true injection spectral index is slightly flatter ($\alpha\sim-0.7$). 
\medskip

\item[$\bullet$] The low-resolution VLA images reveal that the southern relic (RS) has a complex morphology, characterized by the presence of five ``arms''. The origin of this complex morphology is unclear, but it might be partly explained by the presence of AGN fossil plasma that has been revived by the passage of a shock wave. Interestingly, the color-color diagram revealed no curvature in the downstream region of the southern relic, which can be explained either by a sum of power-law spectra with different $\alpha_{\rm inj}$ at different locations or smearing of broad range of magnetic fields and/or electron loss histories. 
\medskip

\item[$\bullet$] We identify source~J, a tailed radio galaxy embedded in RS, as possible source of fossil radio plasma. Our spectral index maps suggest that old plasma from the lobe of source~J is re-accelerated. We attempted to model the flattening of the spectral index from the lobe to the relic RS. We find that a $\mathcal{M}=2.4$ is compatible with the observed spectral index change. 
\medskip

\item[$\bullet$] The cluster contains several tailed radio galaxies, which show disturbed morphologies likely due to the interaction between the radio plasma, ICM, and the merger event.  In addition, the cluster contains a number of complex diffuse radio sources that seem to be related to plasma from tailed radio sources (source~G, RS, I, R2). The range in morphologies of tailed-radio sources, strong signs of interactions with the ICM, and presence of nearby complex-shaped relics suggest that fossil plasma from cluster radio galaxies plays an important role in the formation of (at least some) diffuse radio sources. To produce realistic models of diffuse cluster radio emission, this fossil plasma will need to be included in simulations.
\medskip

\end{itemize}

\begin{acknowledgements}
{\it Acknowledgements:} 
We thank the anonymous referee for useful comments which have improved the quality of the manuscript. 
This paper is partly based on data obtained with the International LOFAR Telescope (ILT). LOFAR \citep{vanhaarlem+13} is the Low Frequency Array designed and constructed by ASTRON. It has facilities in several countries, that are owned by various parties (each with their own funding sources), and that are collectively operated by the ILT foundation under a joint scientific policy. Portions of this work were performed under the auspices of the U.S. Department of Energy by Lawrence Livermore National Laboratory under Contract DE-AC52-07NA27344. 
RJvW and HJAR acknowledge support from the ERC Advanced Investigator programme NewClusters 321271.  RJvW acknowledges support of the VIDI research programme with project number 639.042.729, which is financed by the Netherlands Organisation for Scientific Research (NWO). HK and DR were supported by the NRF of Korea through grants 2016R1A5A1013277, 2017R1A2A1A05071429, and  2017R1D1A1A09000567. We thank the staff of the GMRT that made these observations possible. GMRT is run by the National Centre for Radio Astrophysics of the Tata Institute of Fundamental Research. The NRAO is a facility of the National Science Foundation operated under cooperative agreement by Associated Universities, Inc. Support for this work was provided by the National Aeronautics and Space Administration through {\it Chandra} Award Number GO6-17113X issued by the {\it Chandra} X-ray Observatory Center, which is operated by the Smithsonian Astrophysical Observatory for and on behalf of the National Aeronautics Space Administration under contract NAS8-03060. This research made use of APLpy, an open-source plotting package for Python \citep{2012ascl.soft08017R}.
\end{acknowledgements}

\appendix
\renewcommand\thefigure{\thesection.\arabic{figure}}

\section{Additional radio and optical images}
\setcounter{figure}{0} 

\begin{figure*}%[h!]
\centering
\includegraphics[width=1.\textwidth]{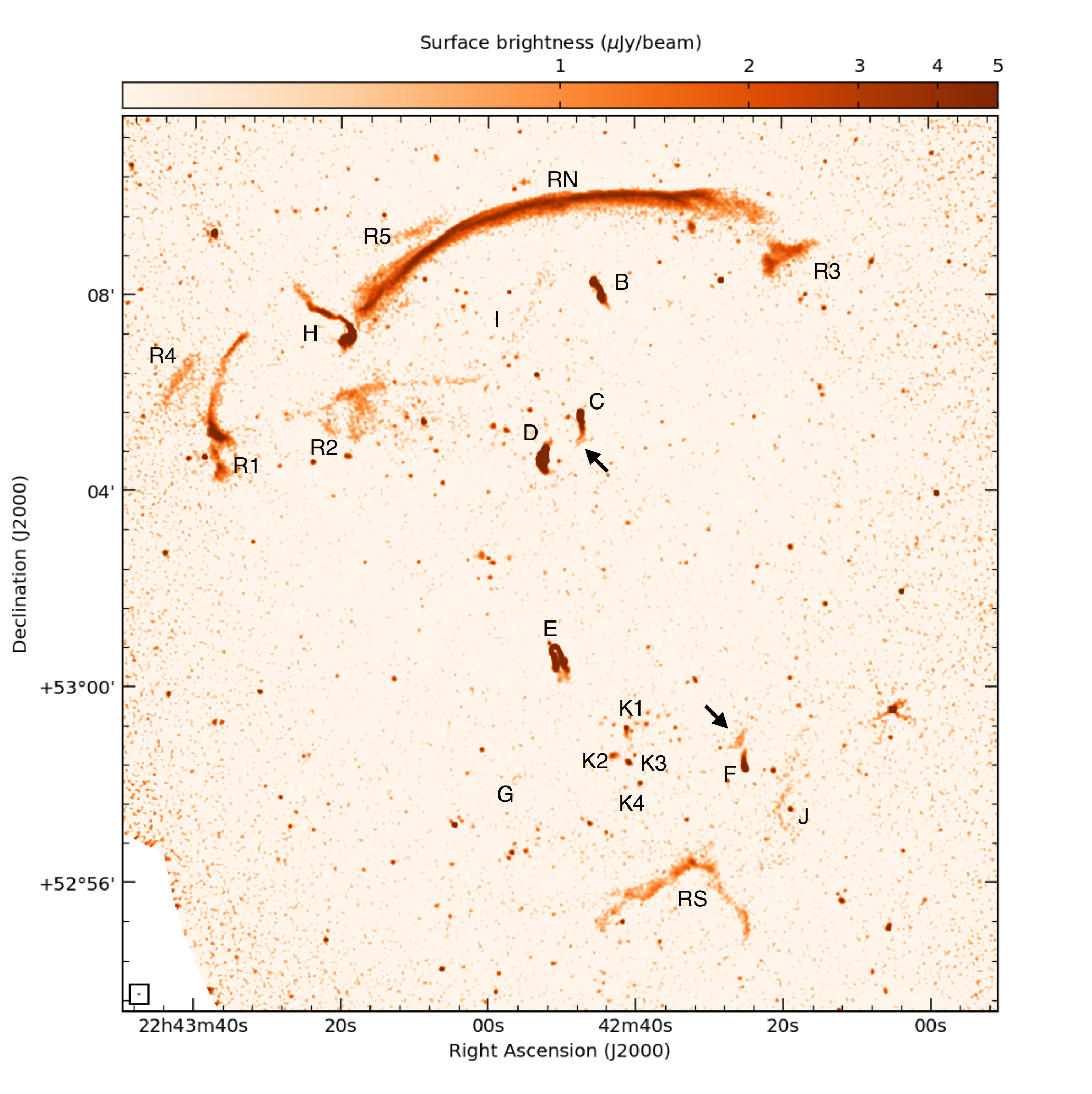}
\caption{S-band (3.0 GHz) VLA high-resolution image of CIZA2242. To aid the visibility of the sources, we smoothed the image with a 3-pixel (1.5\arcsec) Gaussian. The black arrows highlights the ``broken'' nature of the radio tails. The map has a noise of $\sigma_{\rm rms}=2.7~\mu$Jy beam$^{-1}$.
}\label{fig:deep_fullSband}
\end{figure*}

\begin{figure*}
\centering
\includegraphics[width=1.\textwidth]{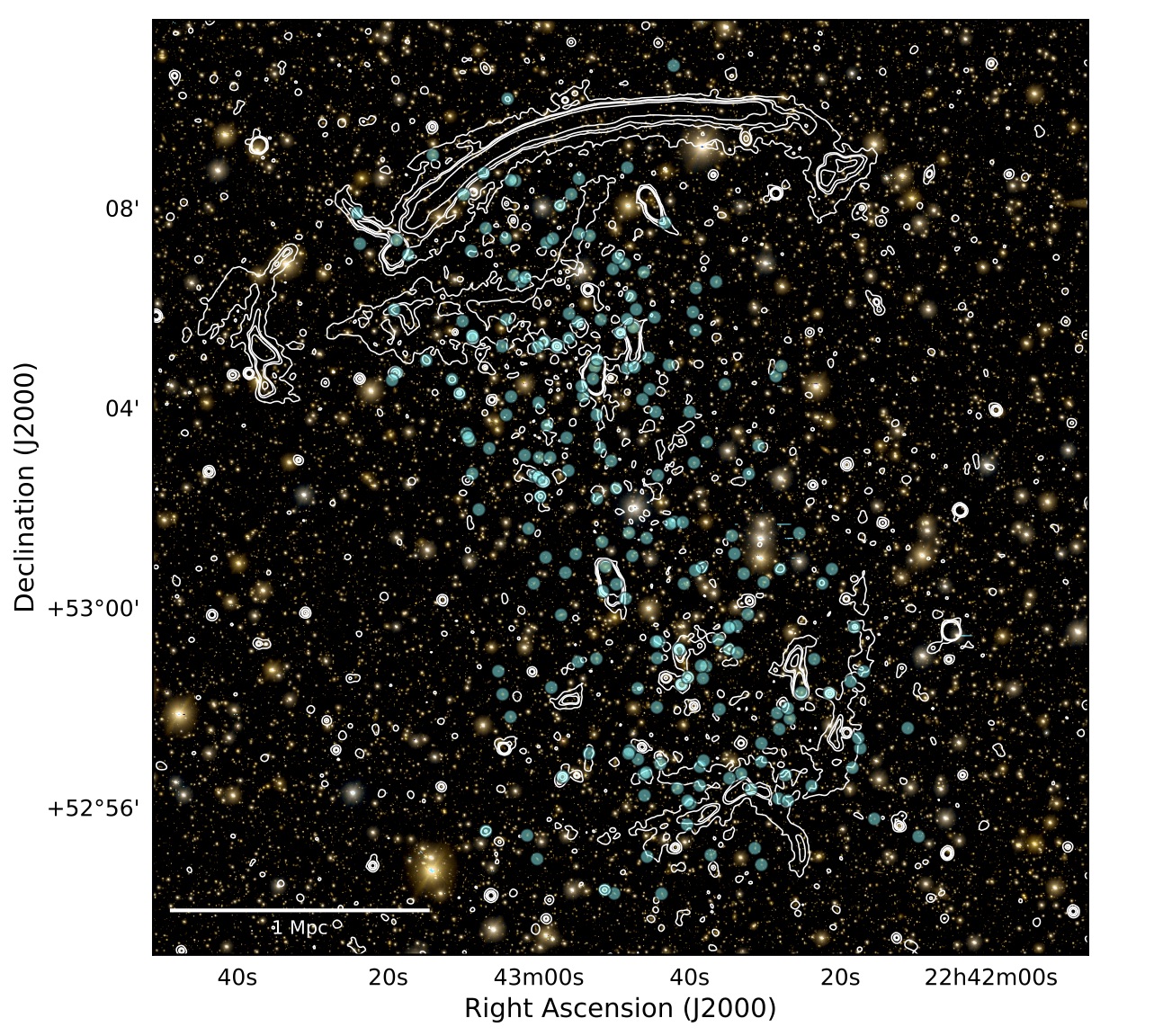}
\caption{Subaru $gri$ composite optical image \citep{dawson+15,jee+15} with the 1--2 GHz radio contours ($5^{\prime\prime}$ resolution) overlaid. Light-blue dots indicate the spectroscopically confirmed cluster galaxies, with $0.17\leq z \leq 0.2$ \citep{dawson+15}.}\label{fig:opt_galaxies}
\end{figure*}

\section{Additional spectral index and spectral index uncertainty maps}
\setcounter{figure}{0}    

\begin{figure*}[h!]
\centering
{\includegraphics[width=0.47\textwidth]{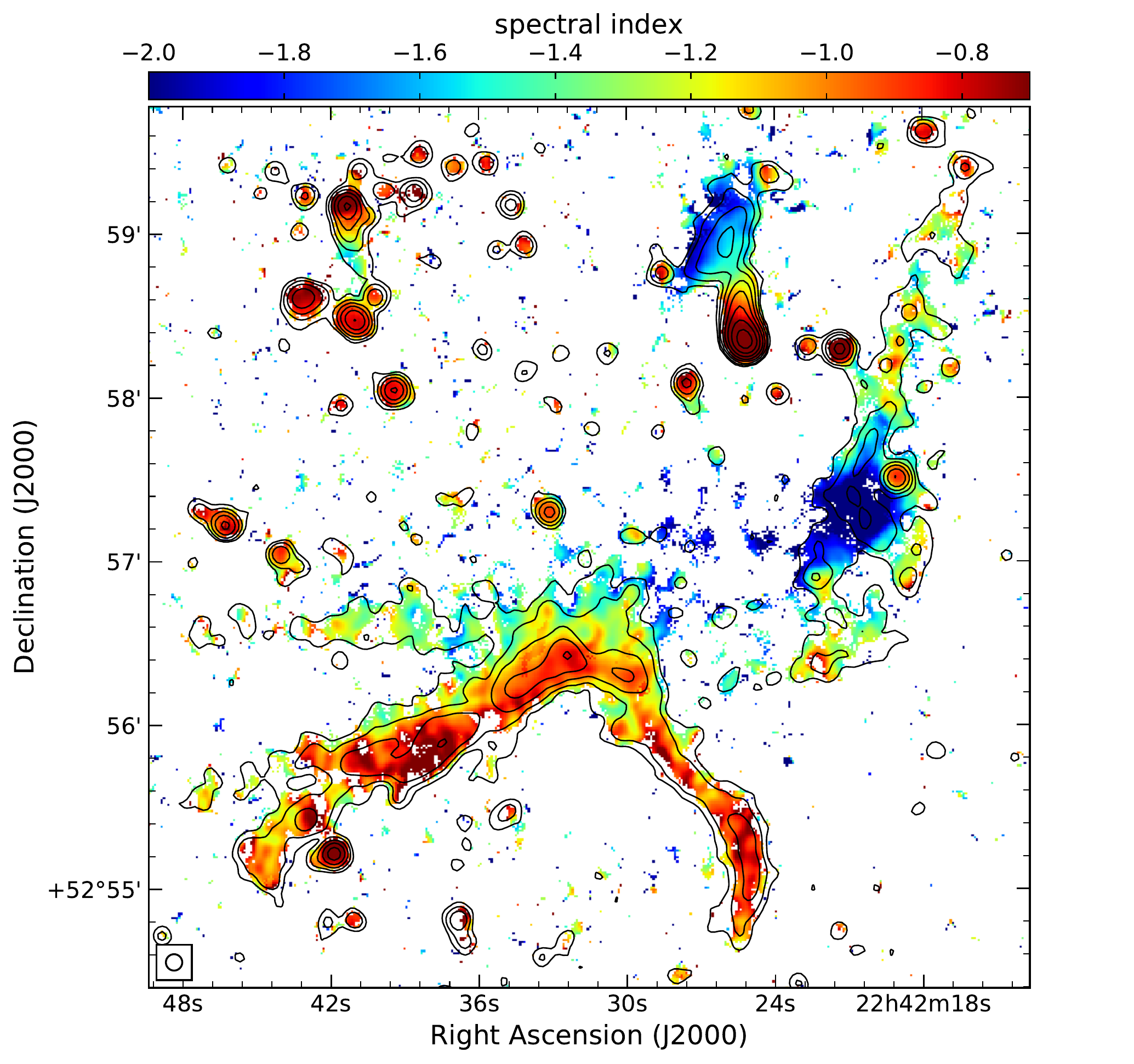}}
\hspace{0mm}
{\includegraphics[width=0.47\textwidth]{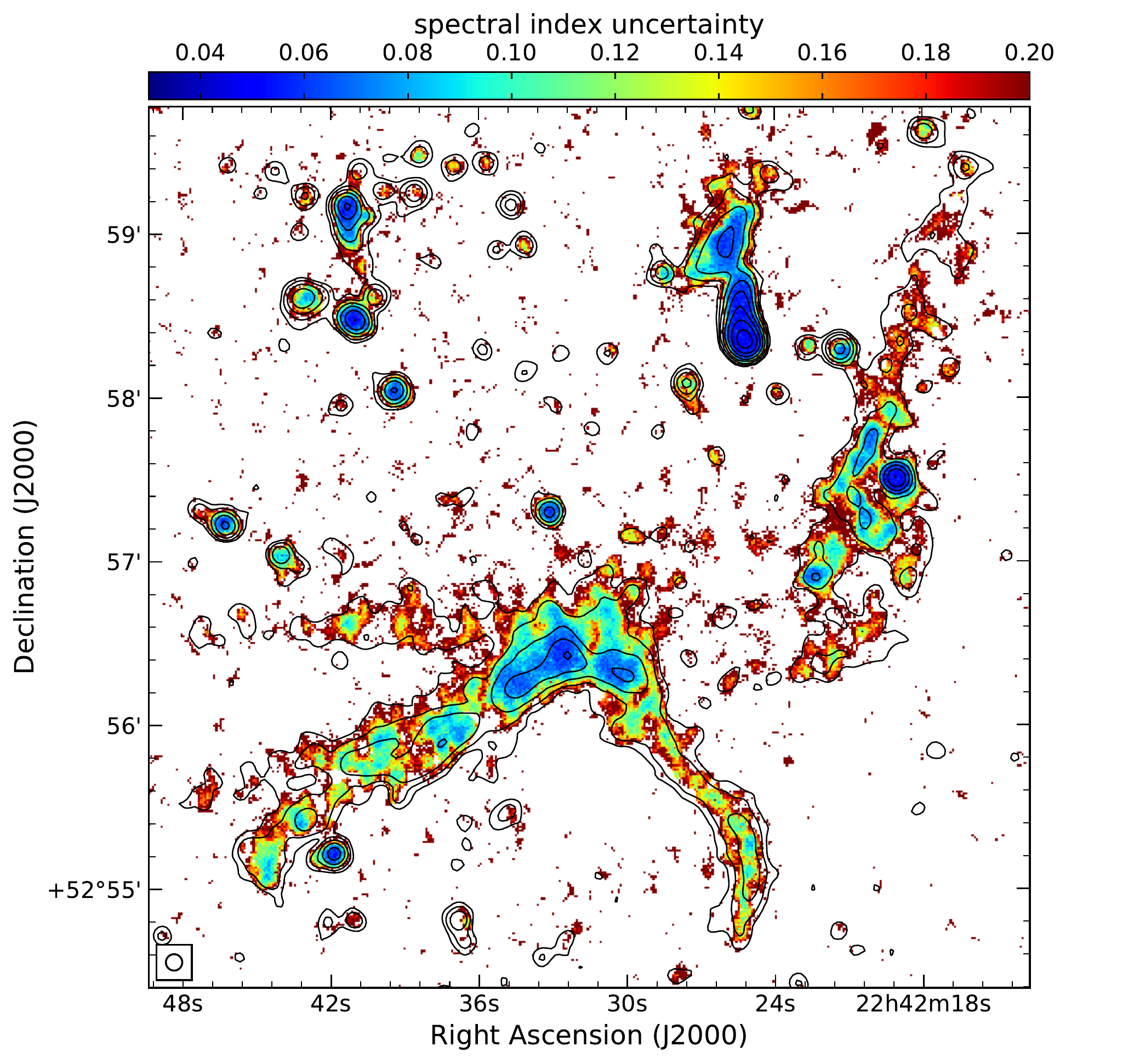}}
\caption{$5^{\prime\prime}$ spectral index map (left panel) and the correspondent uncertainty map (right panel) of the southern relic obtained between 145 MHz, 610 MHz, 1.5 GHz and 3.0 GHz. The radio contours are from the combined L- and S-band observation, leveled in the same way as the top right panel in Fig.~\ref{fig:low_res_comb}.}\label{fig:spix_RS5}
\end{figure*}

\begin{figure}[h!]
\centering
{\includegraphics[width=0.7\textwidth]{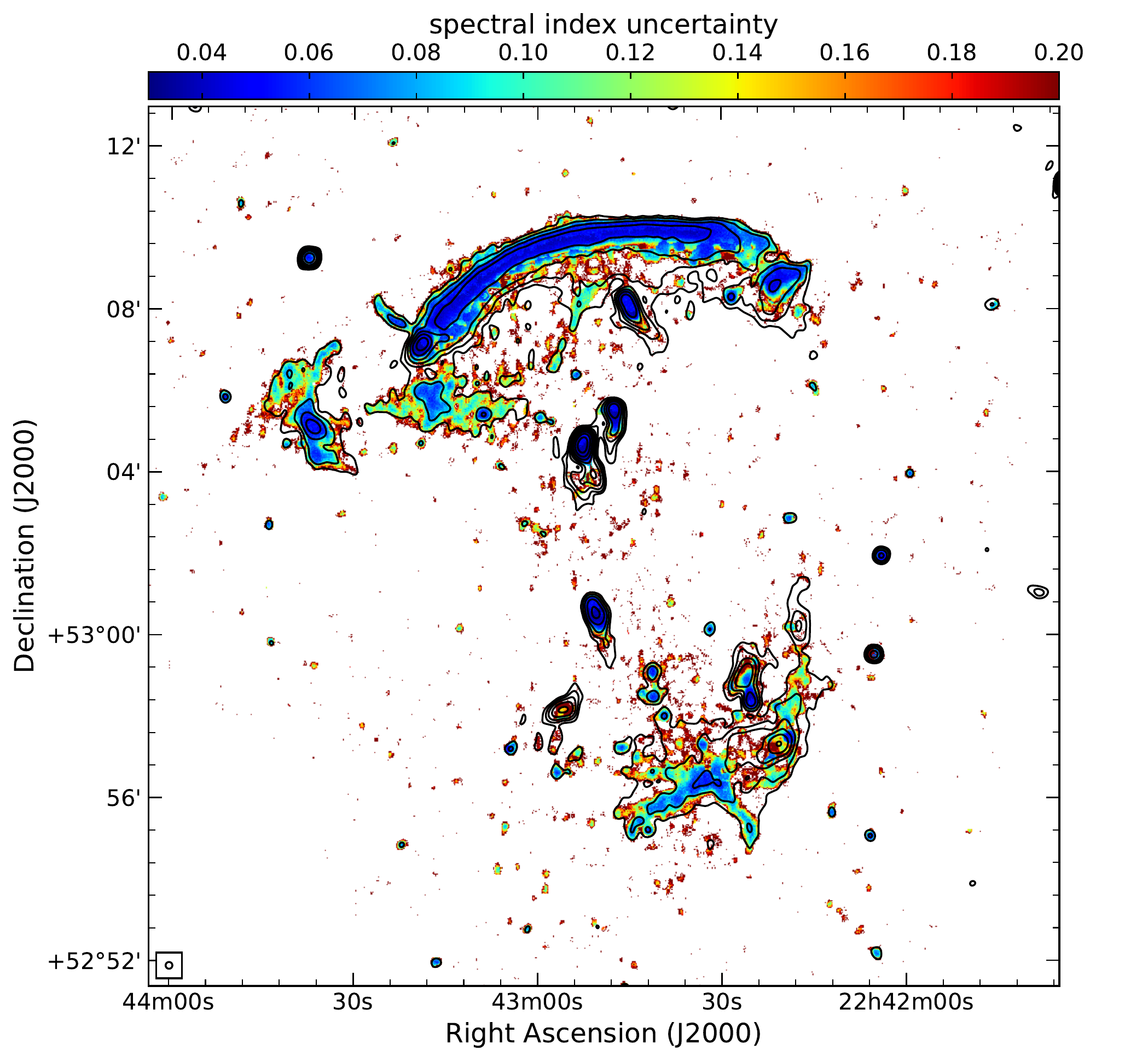}}
\caption{The correspondent spectral index uncertainty map of Fig.~\ref{fig:spix10}.}\label{fig:spix_error}
\end{figure}

\begin{figure*}
\centering
\includegraphics[width=0.302\textwidth]{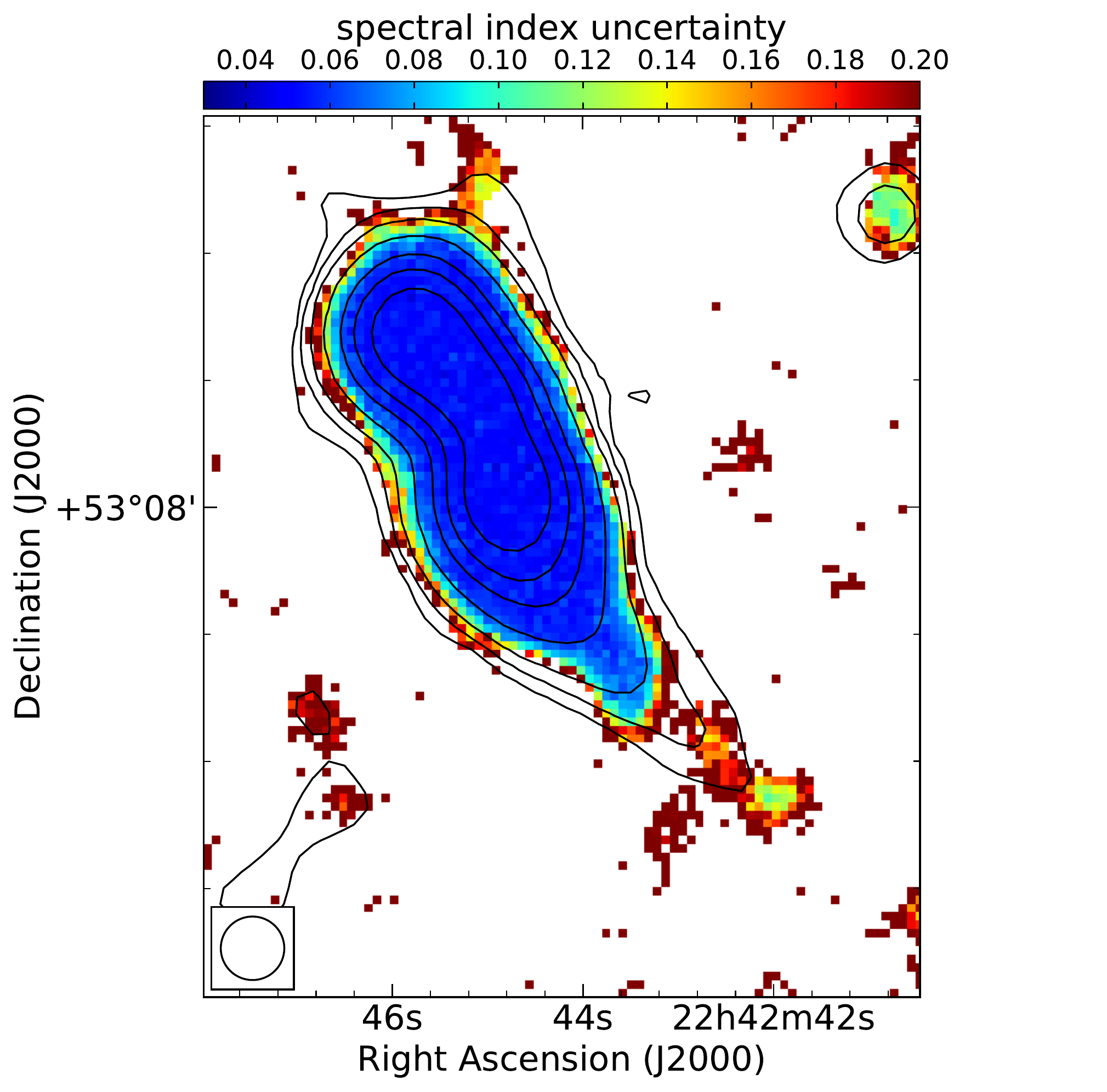}
\includegraphics[width=0.305\textwidth]{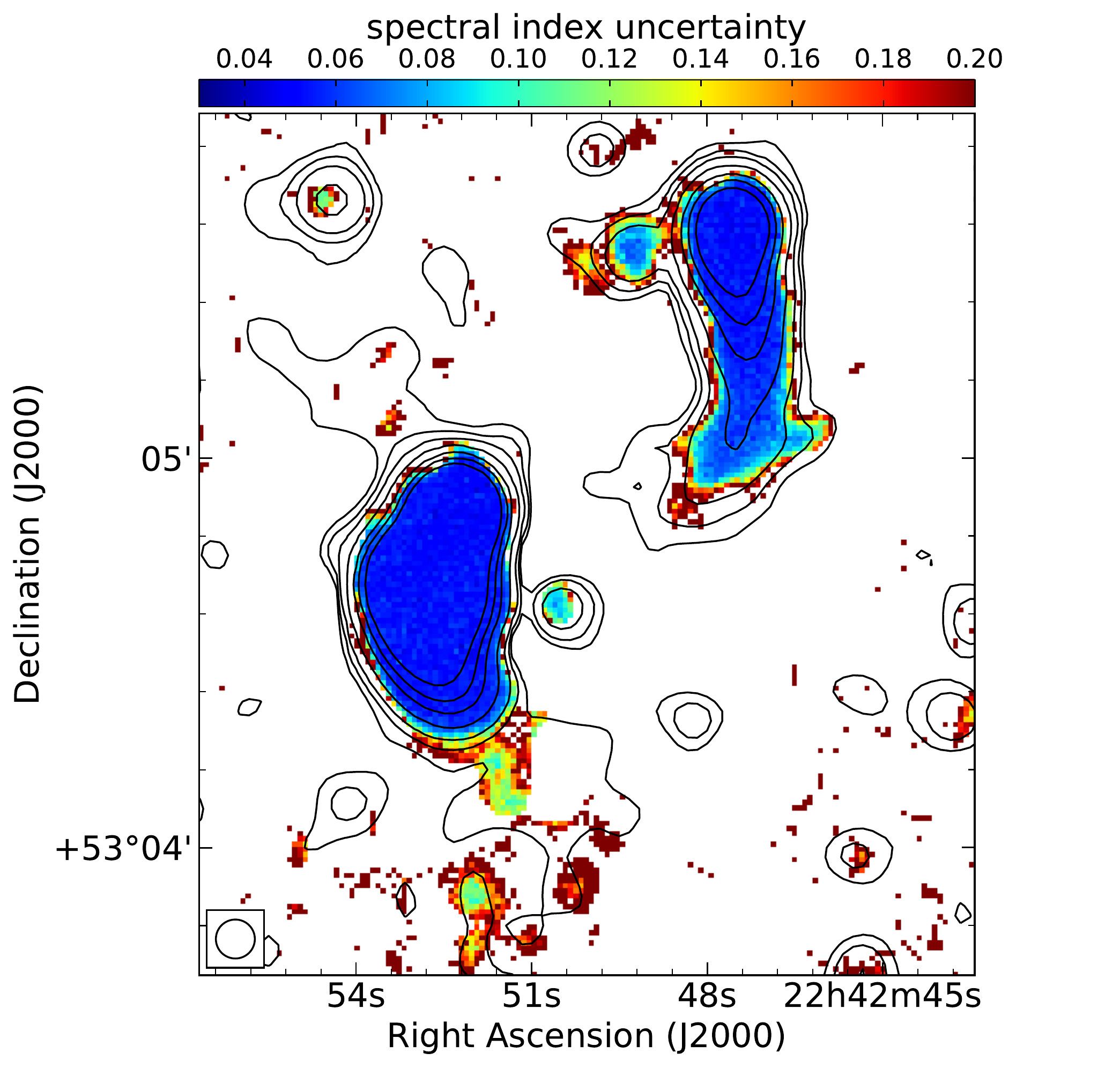}
\includegraphics[width=0.3\textwidth]{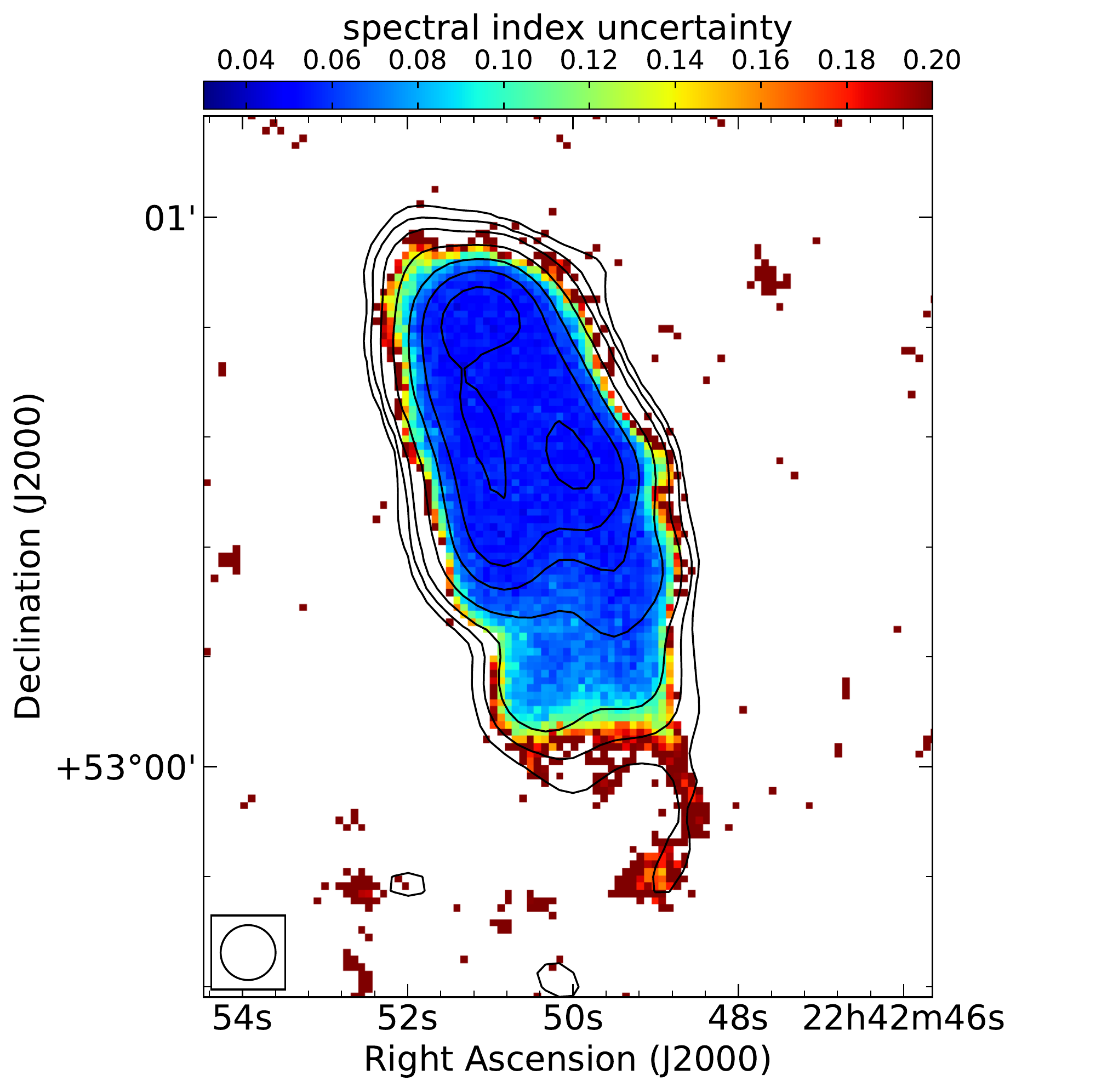}\\
\medskip
\includegraphics[width=0.3\textwidth]{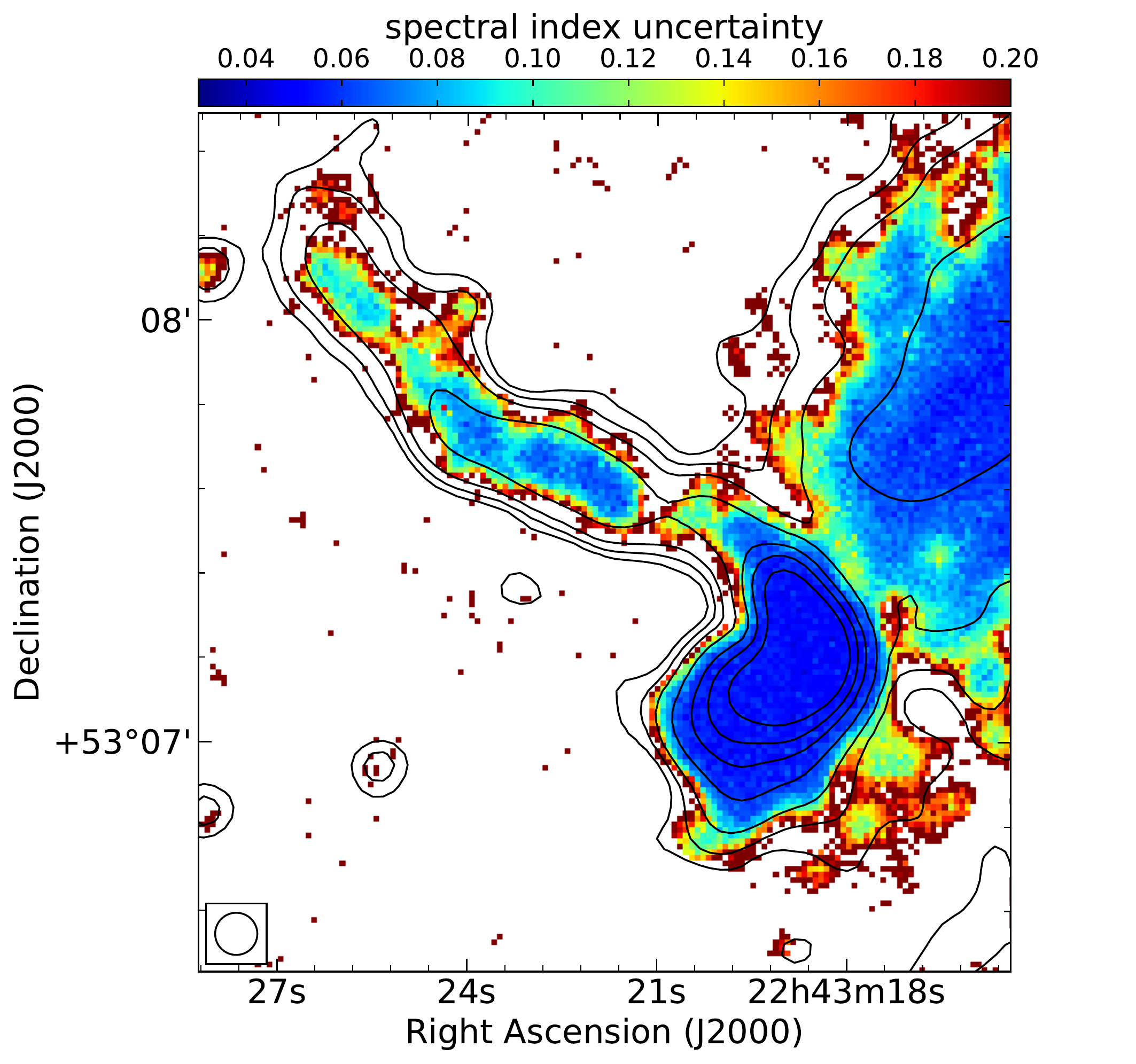}
\includegraphics[width=0.29\textwidth]{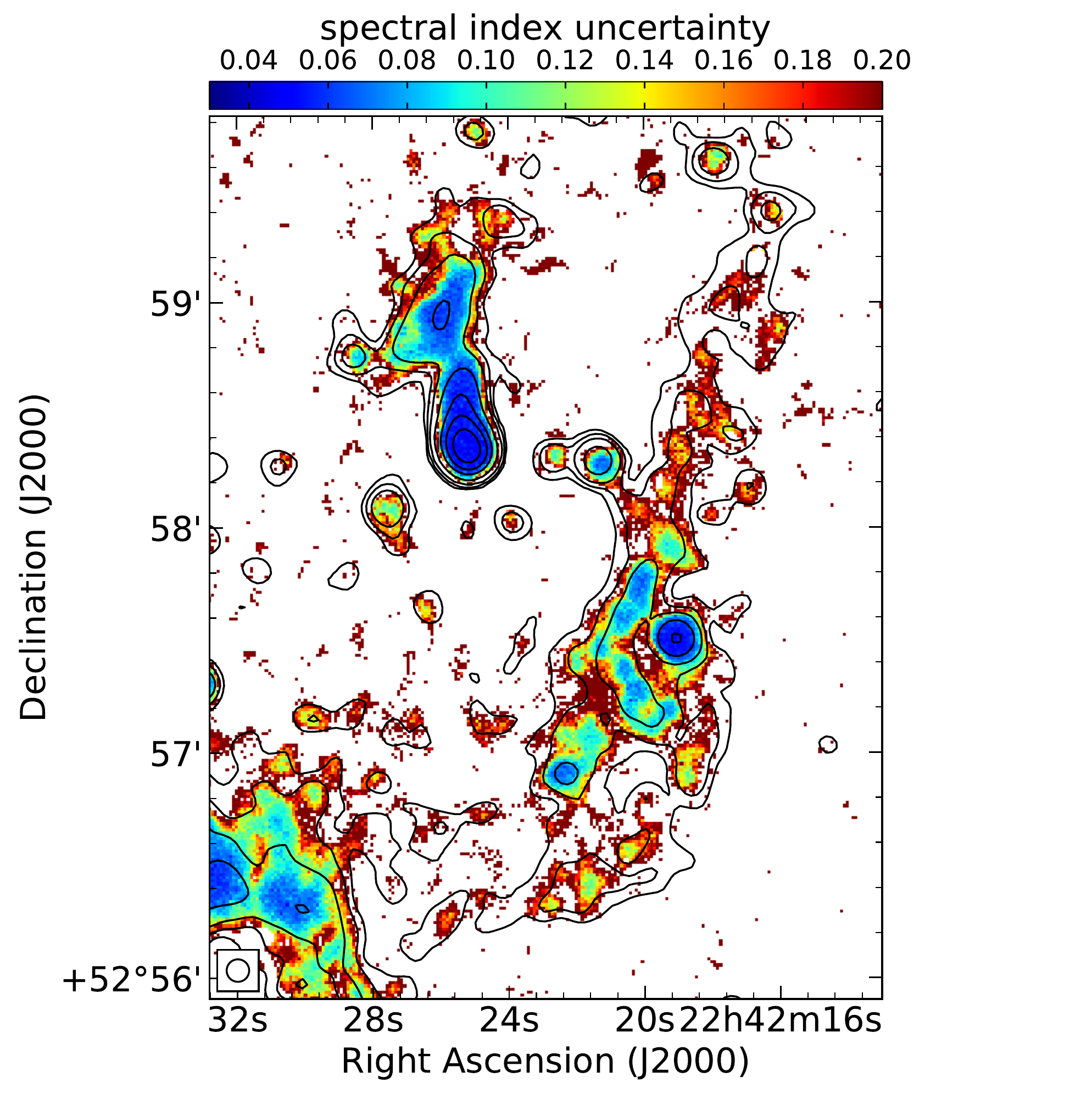}
\includegraphics[width=0.3\textwidth]{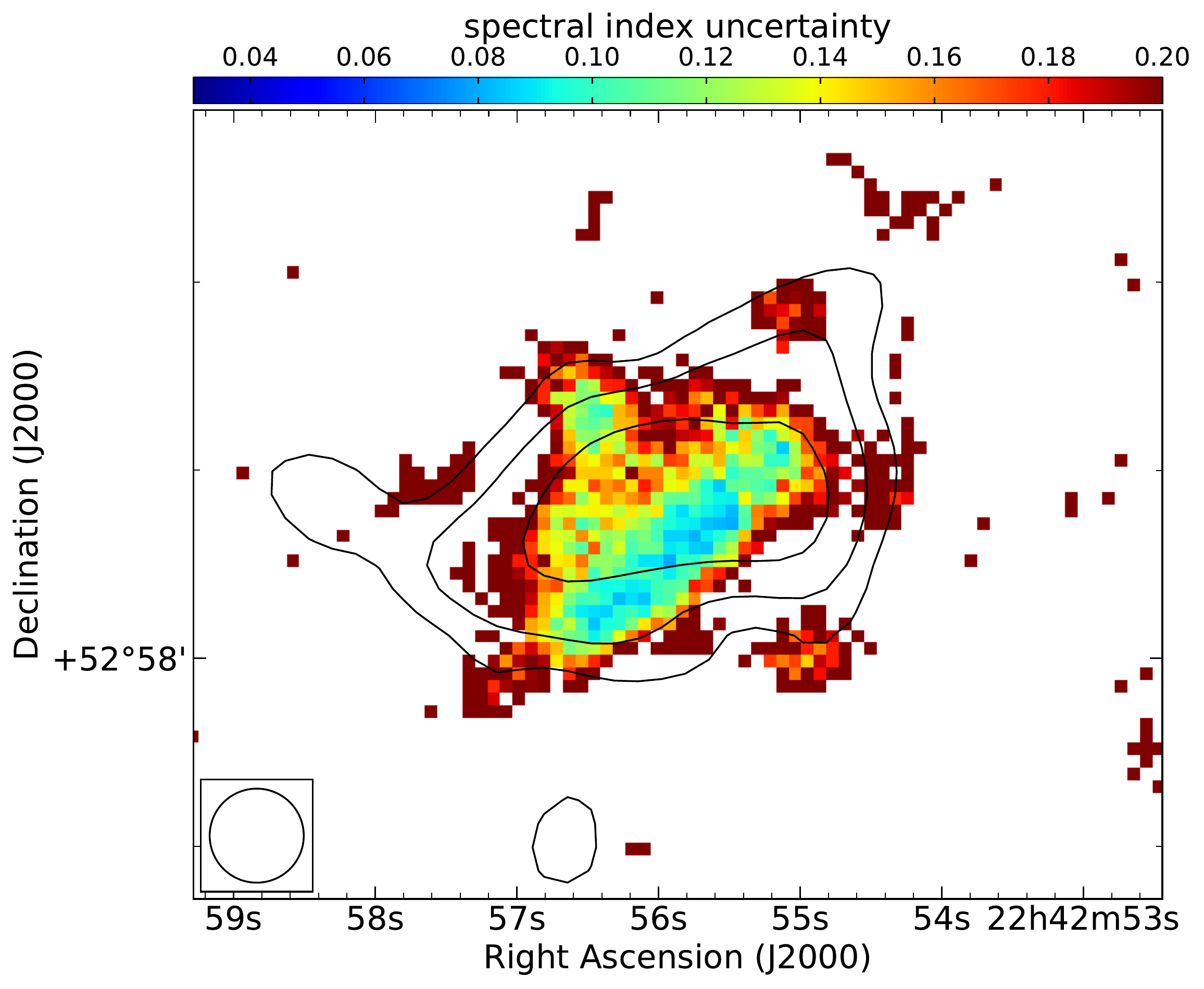}
\includegraphics[width=0.3\textwidth]{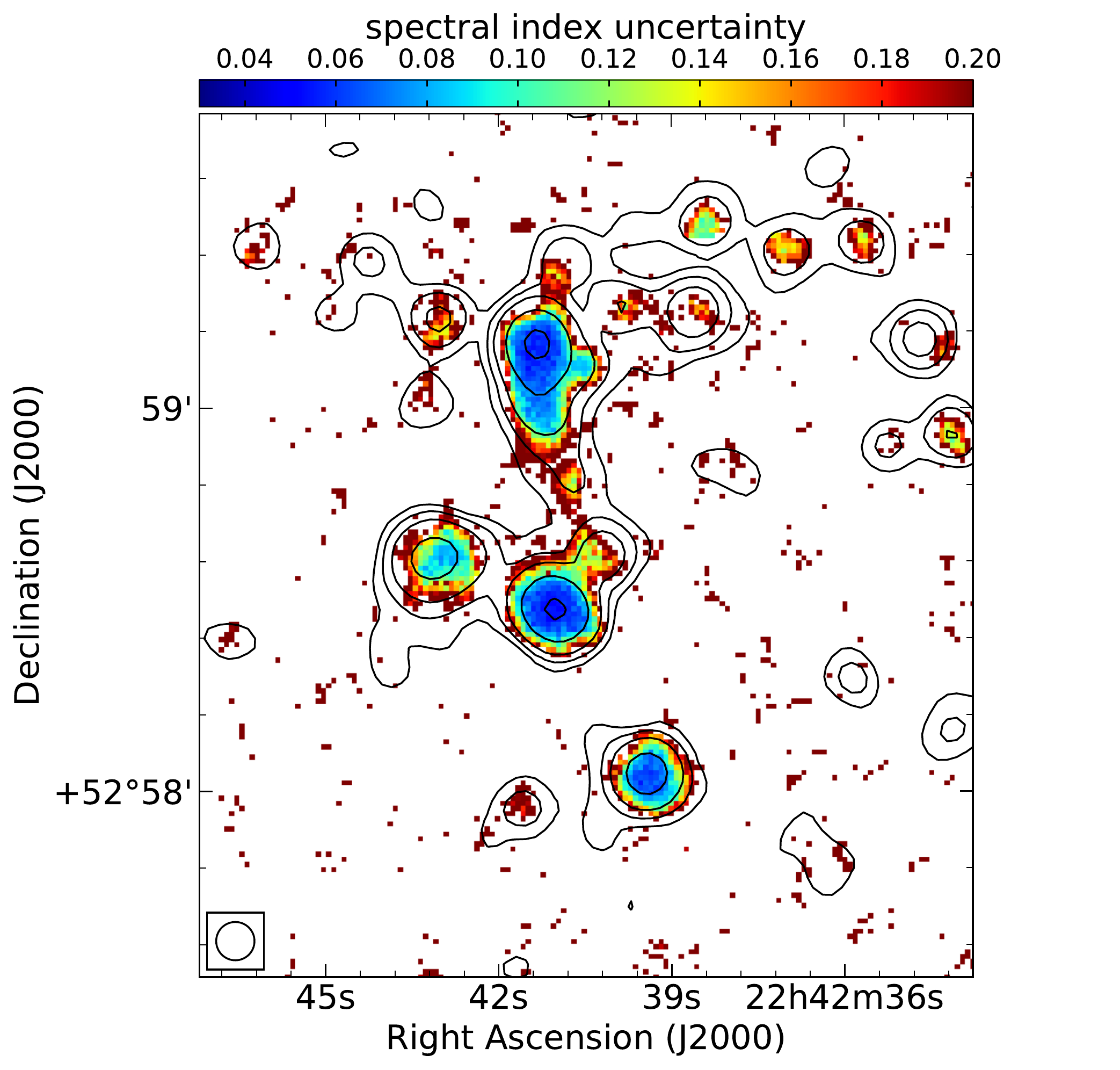}
\caption{The correspondent spectral index uncertainty maps of the left panels in Fig.~\ref{fig:spix_tails}. From the top left to the bottom right: source B, C and D, E, H, F and J, G and K1-4.}\label{fig:spix_tails_err}
\end{figure*}

 \bibliography{biblio.bib}
\end{document}